\documentclass{pasa}%

\title[Galaxies at $z >$ 6]{Observational Searches for Star-Forming Galaxies at $z >$ 6}
\author[Steven L.\ Finkelstein]{Steven L.\ Finkelstein$^1$\\
\affil{$^1$The University of Texas at Austin, Department of Astronomy,
  Austin, TX 78712}}%
\jid{PASA}
\doi{10.1017/pas.\the\year.xxx}
\jyear{\the\year}

\usepackage[authoryear]{natbib}
\usepackage{psfrag}
\usepackage{amsmath}
\usepackage{mathrsfs}
\bibpunct{(}{)}{;}{a}{}{,}
\newcommand{\sol}{$_{\odot}$}
\newcommand{\apj}{ApJ}
\newcommand{\apjl}{ApJL}
\newcommand{\apjs}{ApJS}
\newcommand{\aj}{AJ}
\newcommand{\mnras}{MNRAS}
\newcommand{\physrep}{Phys.\ Rep.}
\newcommand{\nat}{Nature}
\newcommand{\aap}{A\&A}
\newcommand{\pasp}{PASP}
\newcommand{\pasj}{PASJ}
\newcommand{\pasa}{PASA}
\newcommand{\araa}{ARA\&A}
\newcommand{\aapr}{A\&AR}
\newcommand{\ssr}{Space Sci.\ Rev.}
\def\arcs{\hbox{$^{\prime\prime}$}}

\begin{document}%
\begin{abstract}
Although the universe at redshifts greater than six represents only
the first one billion years ($<$10\%) of cosmic time, the dense nature of the early universe
led to vigorous galaxy formation and evolution activity which we are only now starting to
piece together.  Technological improvements have, over only the past
decade, allowed large samples of galaxies at such high redshifts to be
collected, providing a glimpse into the epoch of formation of the first stars and
galaxies.  A wide
variety of observational techniques have led to the discovery of
thousands of galaxy candidates at $z >$ 6, with spectroscopically
confirmed galaxies out to nearly $z =$ 9.  Using these large samples,
we have begun to gain a physical insight into the processes inherent
in galaxy evolution at early times.  In this review, I will discuss i)
the selection techniques for finding distant galaxies, including a
summary of previous and ongoing ground and space-based searches, and spectroscopic followup efforts,
ii) insights into galaxy evolution gleaned from measures such as the
rest-frame ultraviolet luminosity function, the stellar mass function,
and galaxy star-formation rates, and iii) the effect of galaxies on their
surrounding environment, including the chemical enrichment of the
universe, and the reionization of the intergalactic medium.  Finally, I conclude with prospects for future observational
study of the distant universe, using a bevy of
new state-of-the-art facilities coming online over the next decade and
beyond.

\end{abstract}
\maketitle%
\section{INTRODUCTION }
\label{sec:intro}

\subsection{Probing our origins}
A central feature of humanity is our inherent curiosity about our
origins and those of the world around us.  These two desires reach a crux with
astronomy, where by studying the universe around us we also peer back
into our own genesis.  It is thus no surprise that an understanding of
the emergence of our own Milky Way galaxy has long been a highly pursued field in
astronomy.  A number of observational probes into our
Milky Way's beginnings are available, from studying the stellar
populations in our Galaxy (or nearby galaxies), stellar archaeology,
and even primordial abundances in our Solar System.

The finite speed of light is a key property of nature that by delaying our perception of
distant objects allows us to glimpse deep into the history of our
Universe.  A complementary approach is thus to directly study the likely progenitors
of the Milky Way by peering back in time \citep[e.g.,][]{papovich15}.  
Over the past several decades, such studies have arrived at the prevailing theory that today's galaxies
formed via the process of hierarchical merging, where smaller galaxies
combine over time to form larger galaxies \citep[e.g.,][]{searle78,blumenthal84}.
With present-day technology, we can now peer
back to within one billion years of the Big Bang, seeing galaxies as
they were when the universe was less than 10\% of its present-day age.

\subsection{The first galaxies in the Universe}

One key goal in the search for our origins is to uncover the first
galaxies.  In the present day universe, normal galaxies have typical stellar
masses of log(M/M\sol) $\sim$ 10-11 (for reference, the stellar mass
of the Milky Way is $\sim$5 $\times$ 10$^{10}$ M\sol; e.g.,
\citealt{mutch11}).  However, as stellar mass builds
up with time, it is sensible to assume that early galaxies were likely
less massive.  To understand how small these first galaxies might be,
we need to turn to simulations of the early universe to explore
predictions for the first luminous objects.

The universe at a time $\sim$10$^{8}$ years after the Big Bang ($z
\sim$ 30) was a much different environment than today.  The era of
recombination had just
ended, and the cosmic microwave background (CMB) was filling the universe at
a balmy temperature of $\sim$ 85 K.  At that time, baryonic
matter, freed from its coupling with radiation, had begun to
fall into the gravitational potentials formed by the previously collapsed
dark matter halos.  Simulations show that the first stars in the universe
formed in dark matter halos with masses of $\sim$10$^{6}$ M\sol\
(known as mini-halos; \citealt{couchman86,tegmark97}).  Stars forming in these halos were
much different than those in the local universe, as the chemical
composition in the universe prior to the onset of any star formation
lacked any metals.  As metals are a primary coolant in the local
interstellar medium (ISM), gas in these minihalos must cool through
different channels to reach the density threshold for star formation \citep[e.g.,][]{galli98}.
Even lacking metals, gas can still cool through hydrogen atomic line
emission, but only in halos where the virial temperature is $>$10$^4$
K.  These ``atomic cooling halos'' have virial masses of
$\gtrsim$ 10$^8$ M\sol, and are thus more massive than the likely host halos
of the first stars.

In the absence of more advanced chemistry, small amounts of molecular
hydrogen (H$_2$) were able to form in these minihalos, and these dense
gas clouds were the sites of the formation of the first stars.
Lacking the ability to cool to temperatures as low as present day
stars, simulations have shown that the first stars were likely much more
massive, with characteristic masses from $\sim$10 M\sol\ up to $>$ 100
M\sol\ \citep[e.g.,][]{bromm04,glover05}.  These stars consisted solely of hydrogen and helium,
and are thus referred to as Population III stars, compared to the
metal-poor Population II, or metal-rich Population I stars.

However, these first objects were not galaxies, as the first
simulations of such objects showed that most minihalos would form but
a single Population III star \citep[e.g.,][]{bromm04}.  Subsequent simulations have
shown that due to a variety of feedback effects (Lyman-Werner
radiation, photoionization, X-ray heating, etc.), the collapsing gas
may fragment and form a small star cluster.  The most advanced
\textit{ab initio} simulations show this fragmentation, but they also
show that some of these protostars merge back together \citep{greif11}.  Given the
computational cost involved in these latter simulations, they are not
yet able to run until the stars ignite, thus it is
unknown whether the final result is a single highly massive star, or a
cluster of more moderate-mass stars.  However, even in the latter
case it is likely that the initial mass function has a higher characteristic mass than in the
present day universe \citep[e.g.,][]{safranek-shrader14}.

When the one (or more) massive stars in these mini-halos reach the end
of their life and explode as a supernova, the energy injected may be enough to heat and expel much of the
remaining gas, and suppress further star formation.  Eventually, the now-enriched gas will fall back in, cool, and begin forming
Population II stars.  By this time, the dark matter halos have likely
grown to be in the $\sim$ 10$^{8}$ M\sol\ range, with the subsequent
forming stellar masses likely $\gtrsim$10$^6$ M\sol.  These ``first galaxies'' will
ultimately be observable with the \textit{James Webb Space Telescope},
and today, we can observe their direct descendants with stellar masses
of $\sim$10$^{7-8}$ M\sol\ with deep \textit{Hubble Space Telescope}
({\it HST})
surveys.  The nature of these earliest observable galaxies is a key
active area of galaxy evolution studies.
 
\subsection{Reionization}

The build up of galaxies in the early universe is deeply intertwined
with the epoch of reionization, when the gas in the intergalactic
medium (IGM), which had been neutral since recombination at $z \sim$
1000, became yet again ionized.  Although the necessary ionizing
photons could in principle come from a variety of astrophysical
sources, the prevailing theory is that galaxies provide the bulk of
the necessary photons
(e.g.,
\citealt{stiavelli04,richards06,robertson10,finkelstein12b,robertson13,finkelstein15,robertson15,bouwens15b},
though see \citealt{giallongo15,madau15} for a possible non-negligible
contribution for accreting super-massive black holes).  
Thus, understanding both the spatial nature and temporal history of
reionization provides a crucial insight into the formation and evolution
of the earliest galaxies in the universe.

Current constraints from the CMB show that the optical depth to electron
scattering along the line-of-sight is consistent with an instantaneous
reionization redshift of 8.8$_{-1.1}^{+1.2}$ \citep{planck15}.
However, reionization was likely a more extended process.
Simulations show that reionization likely started as an inside-out
process where overdense regions first formed large H\,{\sc ii} regions,
which then overlapped in a ``swiss-cheese'' phase, ultimately
ending as an outside-in process, where the last remnants of the
neutral IGM were ionized \citep[e.g.,][]{barkana01,iliev06,alvarez09,finlator09}.
Current constraints from galaxy studies show that reionization likely ended by
$z \sim$ 6, and may have started as early as $z >$ 10 \citep{finkelstein15,robertson15}.

In addition to ionizing the diffuse IGM, reionization likely had
an adverse impact on star formation in the smallest halos.  Those halos
which could not self-shield against the suddenly intense UV background
would have all of their gas heated, unable to continue forming stars.
This has a significant prediction for the faint-end of the
high-redshift luminosity function -- it must truncate at some point.
Current observational constraints place this turnover at M$_\mathrm{UV} >
-$17, while theoretical results show it occurs likely somewhere in the
range $-$13 $<$ M$_\mathrm{UV}$ $< -$10, though some simulations find
that other aspects of galaxy physics may produce a turnover at $-$16 $<$ M$_\mathrm{UV}$ $< -$14 \citep{jaacks13,oshea15}.  Identifying this turnover is
crucial for the use of galaxies as probes of reionization, as the
steep faint-end slopes observed yield an integral of the luminosity
function that can vary significantly depending on the faintest
luminosity considered.  Even {\it JWST} will not probe faint
enough to see these galaxies (although with gravitational lensing it
may be possible), but the burgeoning field of near-field
cosmology aims to use local dwarf galaxies, which may be the
descendants of these quenched systems, to provide further
observational insight \citep[e.g.,][]{brown14,mbk15,graus15}.

With the ability to study the Universe so close to its beginning, it
is natural to ask: when did the first galaxies appear, and what were
their properties?  This review will concern itself with our progress with
answering this question, focusing on observational searches for
galaxies at redshifts greater than six, building on the work of
previous reviews of galaxies at 3 $< z <$ 6 of \citet{stern99} and
\citet{giavalisco02}.  In \S 2 I discuss methods for discovering distant galaxies, while in \S 3 I
highlight recent search results, and in \S 4 I discuss spectroscopic
followup efforts.   In \S 5 and \S 6, I discuss our current understanding of
galaxy evolution at $z >$ 6, while in \S 7 I discuss reionization.  I
conclude in \S 8 by discussing the prospects towards
improving our understanding over the next decade. Throughout this paper, when
relevant, a {\it Planck} cosmology of H$_{0} =$ 67.3,
$\Omega_m =$ 0.315 and $\Omega_{\Lambda} =$ 0.685 \citep{planck15} is assumed.

\section{Selection Techniques for Distant Galaxies }

To understand galaxies in the distant universe, one needs a method to
construct a complete sample of galaxies, with minimal contamination.
The obvious course here is spectroscopy - with a deep, wide
spectroscopic survey, one can construct a galaxy sample with
high-confidence redshifts, particularly when the continuum and/or
multiple emission lines are observed.  This has been accomplished in
the low-redshift universe, by, for example, the CfA Redshift Survey
\citep{huchra83}, the 2dF Galaxy Redshift Survey \citep{colless01}, and the Sloan Digital
Sky Survey \citep[SDSS; e.g.,][]{strauss02}.  While still highly
relevant surveys more than a decade after
their completion, these studies are limited to $z \lesssim$ 0.5.
Future spectroscopic surveys, such as the Hobby Eberly Telescope Dark
Energy Experiment \citep[HETDEX;][]{hill08} and the Dark Energy
Spectroscopic Instrument \citep[DESI;][]{desi14}, will enhance the
spectroscopic discovery space out to $z \sim$ 3.  However, due to the
extreme faintness of distant galaxies, the $z
\sim$ 6 universe is largely presently out of reach for wide-field,
blind spectroscopic surveys.

\subsection{Spectral Break Selection}
 
Succesful studies of the $z >$ 3 universe have thus turned to
broadband photometry.  Although the spectroscopic resolution of
broadband filters is extremely low (R $\sim$ 5 for the SDSS filter
set), photometry can still be used to discern strong spectral
features.  The intrinsic spectra of star-forming galaxies exhibit two
relatively strong spectral breaks.  The first is the Lyman break at
912 \AA, which is the result of the hydrogen ionization edge in
massive stars, combined with the photoelectric absorption of more
energetic photons by neutral gas (H\,{\sc i}) in the interstellar
media (ISM) of galaxies.  The second break due to a combination of
absorption by the higher-order Balmer
series lines down to the Balmer limit at 3646 \AA\ (strongest in
A-type stars), along with absorption from metal lines in lower mass stars, primarily the Ca H
and K lines (3934 and 3969 \AA), strongest in lower-mass, G-type
stars.  Although this so-called ``4000 \AA\ break'' can become strong in galaxies
dominated by older stellar populations, it is typically much weaker
than the Lyman break, which can span an order of magnitude or more in
luminosity density.  

The intergalactic medium adds to
the amplitude of the Lyman break, as neutral gas along the line of
sight (either in the cosmic web, or in the circum-galactic medium of
galaxies) efficiently absorbs any escaping ionizing radiation (with a
rest-frame wavelength less than 912 \AA).  Additionally, the continuum of galaxy
spectra between 912 and 1216 \AA\ will be attenuated by 
Ly$\alpha$ absorption lines in discrete systems along the
line-of-sight, known as the Lyman-$\alpha$ forest.  
This latter effect is redshift-dependent, in that at higher redshift, more opacity is
encountered along the line-of-sight.  By $z \gtrsim$ 5, the region
between the Lyman continuum edge and Ly$\alpha$ is essentially opaque, such
that no flux is received below 1216 \AA, compared to 912 \AA\ at
lower redshifts.  As such, the Lyman break is occasionally referred
to as the Lyman-$\alpha$ break at higher redshifts, although the
mechanism is generically similar.

At $z >$ 3, the Lyman break feature shifts into the optical, and can
thus be accessed from large-aperture, wide-field ground-based
telescopes.  Building on the efforts of \citet{tyson88},
  \citet{guhathakurta90} and
\citet{lilly91}, \citet{steidel93} was
among the first to realize this tremendous opportunity to study
the distant universe.  They used a set of three
filters ($U_{n}$, $G$ and $\mathscr{R}$) devised a set of color criteria to select
galaxies at $z \sim$ 3 in two fields around high-redshift quasars.  
At this redshift, the Lyman break occurs between the $U_n$ and $G$ bands,
thus a red $U_n-G$ color (or, a high $G$/$U_n$ flux ratio) corresponds to
a strong break in the spectral energy distribution (SED) of a given
galaxy between these filters.  While this alone can efficiently find
galaxies with strong Lyman breaks, corresponding to redshifts of
$z \sim$ 3, it may also select lower-redshift galaxies with red
rest-frame optical continua.  Thus a second color is used, $G-\mathscr{R}$,
corresponding to the rest-frame UV redward of the Lyman break at $z
\sim$ 3, with a requirement that this color is relatively blue, to
exclude lower-redshift passive and/or dusty galaxies from
contaminating the sample.  I refer the reader to the review by
\citet{giavalisco02} for further details on Lyman break galaxies at $z
<$ 6.

\begin{figure*}
\begin{center}
\includegraphics[width=3.55in]{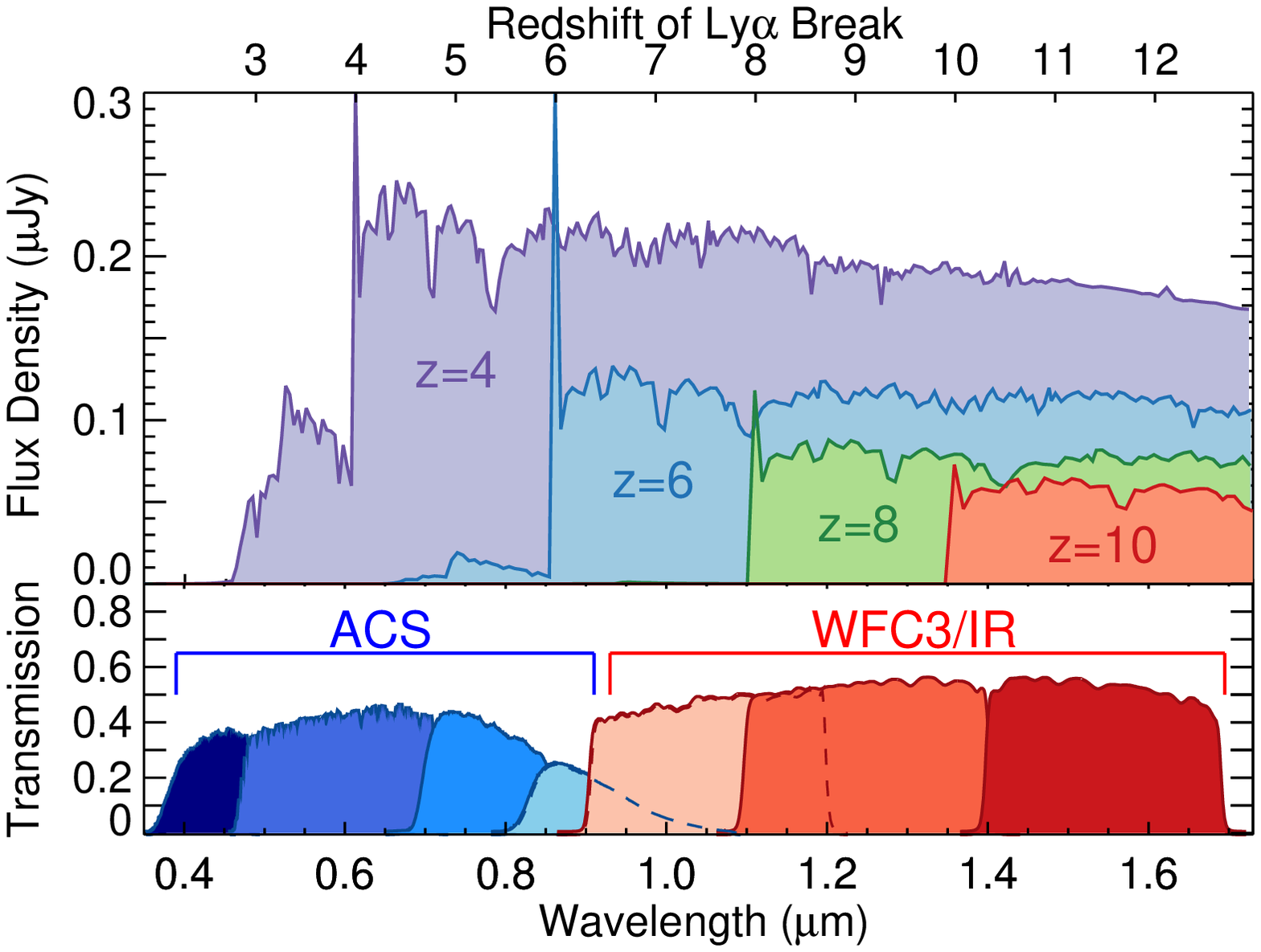}
\hspace{5mm}
\includegraphics[width=2.75in]{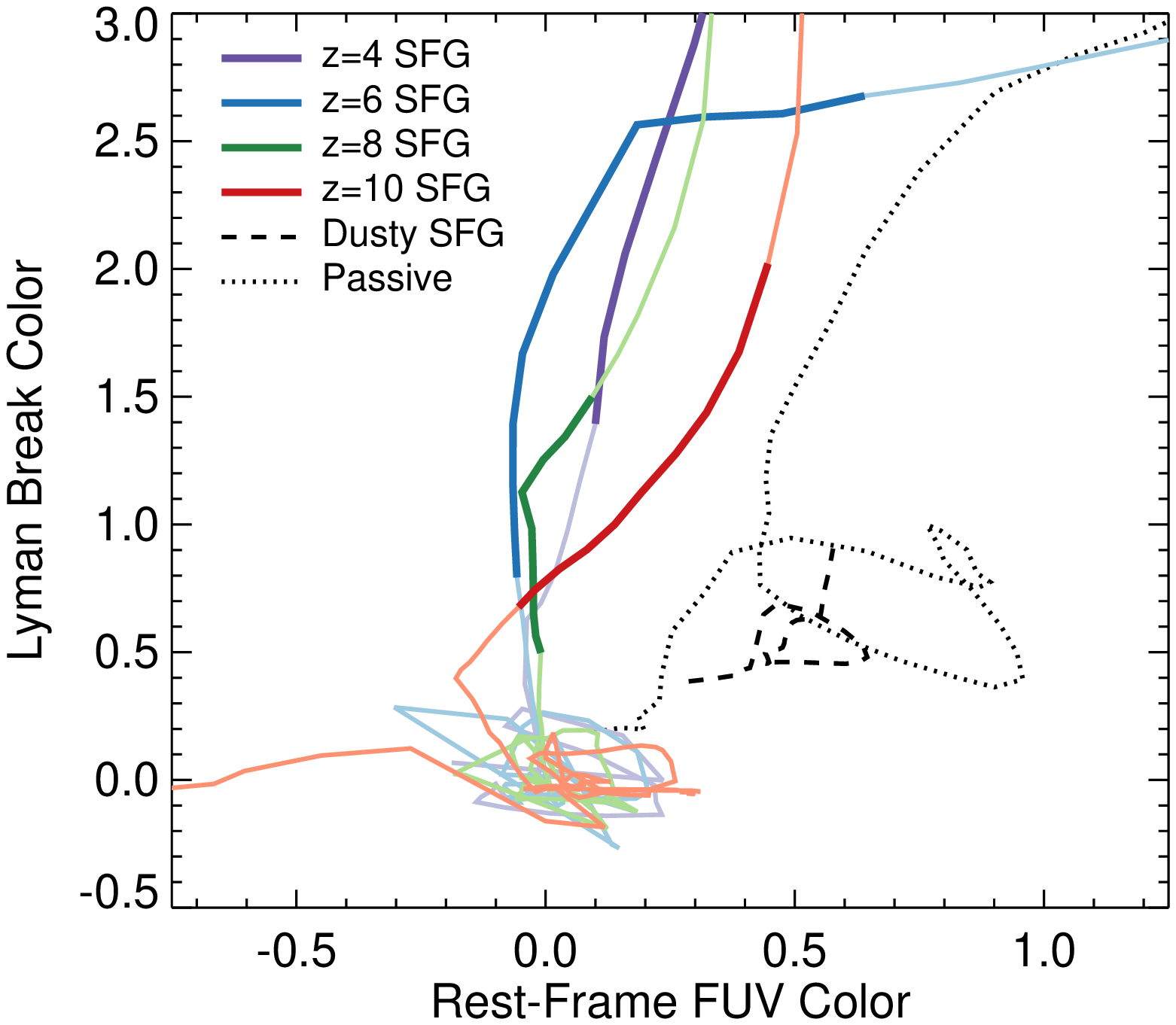}
\includegraphics[width=5.5in]{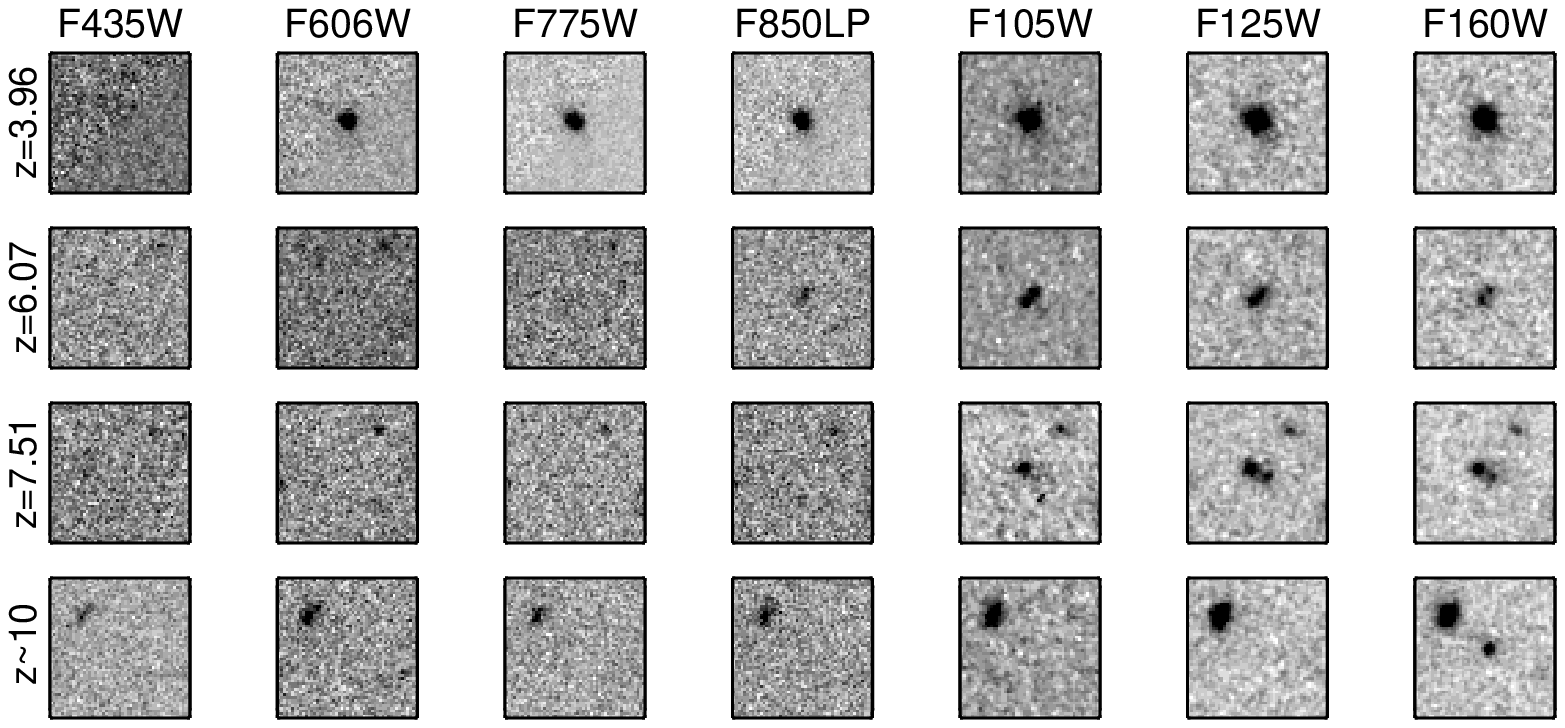}
\caption{Top-Left) Model galaxy spectra at three different redshifts, compared
  to the {\it Hubble Space Telescope} optical (ACS) and near-infrared (WFC3)
  filter set available in the GOODS/CANDELS fields.  The models shown
  have log (M/M\sol) = 9, an age of 10$^8$ yr, and
  $E(B-V)=$0.03.  At $z >$ 7, the
  Lyman break shifts into the near-infrared, rendering such distant
  galaxies literally ``invisible''.  Top-Right) Color-color plot
  showing how   the colors of normal star-forming galaxies (SFGs) at 4 $< z <$ 10
  change with redshift.  For the SFGs,
  the vertical axis represents the Lyman break color:
  $B-V$, $i^{\prime}-z^{\prime}$, $Y-J$ and $J-H$ at $z =$ 4, 6, 8
  and 10, respectively.  The horizontal axis represents the rest-frame
  far-ultraviolet color: $V-i^{\prime}$, $z^{\prime}-Y$, $J-H$ and
  $H-[3.6]$ at $z =$ 4, 6, 8 and 10, respectively.  The dark portion
  of these curves represent when the curve is within $\Delta z \pm$ 0.5
  of the center of the redshift bin.  The dashed and
  dotted lines show the colors of dusty SFGs and passive galaxies from
  0 $< z <$ 5, where for these we plot $z^{\prime}-Y$ versus
  $i^{\prime}-z^{\prime}$ (i.e., showing how these galaxies would
  compare to $z =$ 6 SFG colors).  One can construct a box  in each
  color-color combination which selects the desired high-redshift
  population, and excludes the low-redshift interlopers.  Bottom)
  3\arcs\ stamp images in the seven filters shown in the top-left
  panel centered on example galaxies at $z \sim$ 4, 6, 8 and 10 (the $z \sim$
  4, 6 and 8 galaxies are spectroscopically confirmed, and come from
  the sample of \citealt{finkelstein15}, while the $z \sim$ 10
  candidate galaxy comes from \citealt{bouwens15c}).}
\label{fig:selection}
\end{center}
\end{figure*}

The Lyman break is the primary spectral feature used in nearly all
modern searches for $z >$ 6 galaxies.  In \S 3 below, I will discuss
recent searches with this technique.  Some studies continue to use a
set of color criteria, typically involving three filters,
qualitatively similar to \citet{steidel93}.  However, a more recent
technique is beginning to become commonplace, known as photometric
redshift fitting.  In this technique, one compares the colors of a
photometric sample to a set of template SEDs in \emph{all} available
filters.  The advantage here is that one uses all available
information to discern a redshift.  Additionally, most photometric
redshift tools calculate the redshift probability distribution
function, therefore giving a higher precision on the most likely
redshift, typically with $\Delta$z $\sim \pm$0.2-0.3 at $z >$ 6, compared
to $\Delta$z $\sim \pm$0.5 for the two-color selection technique.  The
disadvantage is that this technique is dependent on the set of SED
templates assumed.  However, at $z \sim$ 6 this worry is alleviated
as the primary spectral feature dominating the photometric redshift
calculation is the Lyman break, thus the effect of different template
choices is likely minimized.  In Figure~\ref{fig:selection}, I show model spectra at
three different redshifts, highlighting how the wavelength of the
Lyman break, and its strength,
changes with redshift, as well as example real galaxies at each of the
redshifts shown.  The right panel of Figure~\ref{fig:selection} also
shows a color-color plot, highlighting how one could use the filter
set from the left panel to perform a Lyman break selection.
 One note for the reader on terminology -
frequently authors use the term ``Lyman break galaxy'' to denote a
galaxy selected via the Lyman break technique.  However, all distant
galaxies, whether they are bright enough to be detected in a continuum
survey or not, will exhibit a Lyman break.  Thus, in this review I will
use the term ``continuum selected star-forming galaxies'' to denote
such objects, which are selected via either Lyman break color-color selection, or
photometric redshift selection.

\subsection{Emission Line Selection}
Another method to select distant galaxies is via strong emission
lines.  At $z >$ 6, the only strong emission line accessible with
current technology is the Lyman-$\alpha$ line, at $\lambda_{rest} =$
1216 \AA.  While a blind spectroscopic survey for this
feature is likely not practical at such high redshifts, this line is strong enough
that it can be discerned with imaging in narrowband filters.  A
galaxy with a line at a particular wavelength covered by a narrowband
filter will appear brighter in that narrowband then in a broadband
filter covering similar wavelengths (as it will have a greater
bandpass-averaged flux in the narrowband).  This particular line was noted
decades ago as a possible signpost for primordial star-formation in
the early universe \citep{partridge67}, thus a variety of studies were
commissioned with the goal of selecting
large samples of Lyman-$\alpha$ emitting galaxies (LAEs).  
Although one of the first narrowband-selected LAEs was discovered by
\citet{djorgovski85}, it wasn't until the advent of large aperture telescopes and/or wide-field
optical imagers that the first large samples of LAEs were discovered
\citep[e.g.][]{cowie98,rhoads00}.

LAEs form a complementary population of galaxies to continuum-selected
star-forming galaxies.  The study of \citet[][and
subsequent studies from that group]{steidel93} typically
restricted their analyses to galaxies with observed optical AB
magnitudes brighter than approximately 25, due to the limited depth
available from ground-based broadband imaging.  
As narrowband selection techniques require only evidence of a large flux ratio
between the narrowband and encompassing broadband filters, continuum
detections are not required, and high
equivalent width (EW) Ly$\alpha$
emission from galaxies with much fainter continuum levels can be detected.  Deep broadband
imaging of ground-based narrowband-selected LAEs shows that they are
indeed on average fainter than continuum-selected galaxies, with continuum magnitudes as
faint as $\sim$28 \citep[for $z =$ 3--5 LAEs, e.g.,][]{ouchi08,finkelstein09c}.
Whether LAEs form a completely separate population, or are simply the
low-mass extension of the more massive continuum-selected galaxy population, remains an
active area of study \citep[e.g.,][]{hashimoto13,nakajima13,song14}.

\subsection{Infrared Selection}

The previously discussed selection methods relied either on the
detection of stellar continuum emission, or of nebular gas emission
(due to photo-ionization from predominantly stellar-produced ionizing photons).  An
alternative method is also available, selecting galaxies based on the
far-infrared emission from UV radiation re-processed by dust grains in
the interstellar medium \citep[e.g.,][]{smail97,hughes98,barger98}.  The advent of the {\it Herschel Space
  Observatory} as well as large-dish ground-based telescopes such as
the James Clerk Maxwell Telescope (JCMT) with fast survey capabilities
have allowed searches for rare, highly luminous dusty
star-forming galaxies.  The redshift distribution of such dusty
star-forming galaxies peaks at $z \sim$ 2--3 (with the exact peak
redshift depending on the selection-wavelength, with redder
wavelengths selecting on average higher-redshift galaxies due to the
redshifting of the dust-emission SED peak).  

Surprisingly, the redshift
distribution extends out to $z >$ 5 \citep[see][and references
therein]{casey14}, implying that vast quantities of dust are produced
in the early universe.  Obtaining spectroscopic redshifts for such
sources is difficult, as Ly$\alpha$ is easily attenuated by dust
\citep[though see, e.g.,][]{barger99,chapman05,capak08}.  However, the advent of the Atacama
Large Millimeter Array (ALMA) now allows spectroscopic confirmation via a
number of sub/millimeter lines, such as those on the CO ladder, or the
[C\,{\sc ii}] 158 $\mu$m fine structure line
\citep[e.g.][]{riechers13}.  Although dust emission has been detected
from galaxies as distant as $z \approx$ 7.5 \citep{watson15}, the vast
majority of known dusty star-forming galaxies lie at $z <$ 6, thus I
will not discuss these studies further in this review.  However, ALMA,
combined with a recent update to the Plateau
du Bure interferometer (NOEMA), and a potential future update to the
Jansky Very Large Array \citep[NGVLA;][]{carilli15}, will further
enable $z >$ 6 millimeter studies, and will soon provide key insight into such distant galaxies.

\section{Surveys for High-Redshift Galaxies}

In this section I discuss the results from recent surveys designed to
discover galaxies at $z \geq$ 6.  I will focus on the surveys and the
galaxy samples, leaving the results from such studies for the
subsequent sections.

\subsection{Broadband Searches for Star-Forming Galaxies at $z =$ 6}

Selecting galaxies at $z =$ 6 via the Lyman break technique requires
imaging at the extreme red end of the optical, in the $z$-band at 0.9 $\mu$m, as
galaxies at such redshifts are not visible at bluer wavelengths.
Additionally, the imaging must be deep enough to detect these
galaxies, as their increased luminosity distance results in an
observed magnitude at $z =$ 6 that is $\sim$2 magnitudes fainter than a
comparably intrinsically bright galaxy at $z =$ 2.  Deep imaging in
the $z$-band is difficult both due to the early lack of red-sensitive
detectors, as well as (from the ground) the numerous night sky
OH emission lines, which set a high background level.  
Although there were earlier examples of $z \approx$ 6 galaxies from,
e.g., the Hubble Deep Field with WFPC2 \citep[e.g.,][]{weymann98},
larger samples of $z =$ 6 star-forming galaxies were not compiled until 
the installation of the red-sensitive
optical Advanced Camera for Surveys (ACS) on {\it HST} in 2002.  
While initial studies using early observations found a handful of $z
\sim$ 6 candidates (\citealt{bouwens03,stanway03,yan03},
including spectroscopic confirmations, e.g., \citealt{bunker03}),
larger samples of more than 100 candidate $z =$ 6 galaxies were soon
compiled using both the Great Observatories Origins Deep Survey
\citep[GOODS;][]{giavalisco04b} and Hubble Ultra Deep Field \citep[HUDF;][]{beckwith06}
datasets \citep{dickinson04,bunker04,giavalisco04,bouwens06}.

Significant progress at $z =$ 6 has also been made from the ground.
Although difficult due to the bright night sky emission, ground-based
surveys can cover much larger areas, and thus provide complementary
results on the bright-end of the galaxy luminosity function, which is
difficult to constrain with {\it HST} due to the small field-of-view
of {\it HST}'s cameras.  Surveys such as the Subaru/XMM-Newton Survey
(SXDS), the UKIRT Infrared Deep Sky Survey (UKIDDS)/Ultra Deep Survey
(UDS), the Canada-France Hawaii Telescope Legacy Survey (CFHTLS), and
the UltraVISTA Survey
have searched areas of the sky from 1-4 deg$^2$ for $z =$ 6 galaxies
\citep[e.g.,][]{kashikawa06b,mclure09,willott13,bowler15}.  As discussed in \S
5, these wide-field surveys are necessary to probe luminosities much
brighter than the characteristic luminosity at $z =$ 6.

One of the major conclusions
from these early studies was that the galaxy population at $z =$ 6
could maintain an ionized IGM only if faint galaxies dominate
the ionizing budget, which required that the luminosity function
must maintain a steep faint-end slope well below L$^{\ast}$
\citep[e.g.,][]{bunker04,yan04}.  Turning to the evolution of the
cosmic star-formation rate (SFR) density, \citet{giavalisco04} found that the
evolution was remarkably flat out to $z =$ 6, such that the rate of
cosmic star formation was similar at $z =$ 6 as at $z =$ 2.  However,
in a combined analysis using data from multiple {\it HST} surveys,
\citet{bouwens07} found that there was a steep drop in the SFR density,
by more than 0.5 dex from $z =$ 2 to 6.  Some of this discrepancy may
be due to the fact that many of these early $z =$ 6 galaxies were only
detected in a single band, making robust samples (and their
completeness corrections) difficult to
construct, as well as hampering the ability to derive a robust
dust-correction, which is necessary for an accurate measure of the SFR
density.

This issue was alleviated with the installation of the Wide Field
Camera 3 (WFC3) on {\it HST} in 2009, which contains both an
ultraviolet/optical camera (WFC3/UVIS), and a near-infrared camera
(WFC3/IR).  Three major surveys were initiated with the infrared
camera to probe high-redshift galaxies.  The Hubble Ultra Deep Field
2009 survey (HUDF09; PI Illingworth) obtained deep imaging
in three near-infrared filters (centered at 1, 1.25 and 1.6 $\mu$m) on
the HUDF as well as two nearby parallel fields,
while the Ultra Deep Field 2012 (UDF12; PI Ellis) survey increased the depth in
these filters in the HUDF, and added a fourth filter at 1.4 $\mu$m.
At the same time, the Cosmic Assembly Near-infrared Deep Extragalactic
Legacy Survey (CANDELS; PIs Faber \& Ferguson) was one of the three
{\it HST} Multi-cycle Treasury Programs awarded in Cycle 18.  CANDELS
observed both GOODS fields in the same three WFC3/IR filters as the HUDF09
survey\footnote[1]{The northern $\sim$25\% of the GOODS-S field had
  already been observed by the WFC3 Early Release Science (ERS)
  program, using F098M as the 1$\mu$m filter rather than F105W;
  here when I refer to the CANDELS imaging in GOODS-S, I refer to the
  combination of the ERS and CANDELS imaging.}, as well as three
additional fields (COSMOS; Extended Groth
Strip/EGS, and Ultradeep Survey/UDS) in the 1.25 and 1.6$\mu$m
filters.  CANDELS also obtained optical imaging with ACS in parallel,
which was particularly useful in the COSMOS, EGS and UDS fields, which
had less archival ACS imaging than the GOODS fields.
Using the combination of these new near-infrared data with the
previously available ACS optical data, these data now allow full
two-color selection of $z =$ 6 galaxies \citep{bouwens15} , as
well as accurate photometric redshifts \citep{finkelstein15}, for a
sample of $\sim$700-800 robust candidates for $z =$ 6 galaxies.

\subsection{Broadband Searches for Star-Forming Galaxies at $z \geq$ 7}

At $z =$ 7, the Lyman break redshifts into the near-infrared, making
deep near-infrared imaging a requirement.
Prior to the advent of WFC3, the availability of the necessary imaging
was scarce.  Efforts were made with NICMOS imaging \citep[e.g.,][]{bouwens10c}, but with a survey
efficiency $>$40$\times$ worse than WFC3 \citep[a combination of depth and
field size;][]{illingworth13}, significant progress in
understanding early galaxy formation was difficult.  The first large,
robust samples of $z >$ 7 galaxies thus did not come about until the
acquisition of the first year of the HUDF09 dataset.  Several papers were published in
the first few months after the initial Year 1 HUDF09 imaging
\citep[e.g.,][]{oesch10,bouwens10a,bunker10,mclure10,finkelstein10}, finding
$\sim$10--20 robust $z =$ 7 candidate galaxies, as well
as 5--10 $z =$ 8 candidate galaxies.  Sample sizes of faint galaxies were 
increased with the completed two-year HUDF09 dataset, as well as the added
depth from the UDF12 program \citep[e.g.,][]{mclure13,schenker13}.  The largest addition in sample size of
$z =$ 7--8 galaxies was made possible by the CANDELS program,
which, in the two GOODS fields alone, provided 5.4 and 3.7$\times$ more galaxies
than the HUDF alone at $z =$ 7 and $z =$ 8, respectively
\citep{finkelstein15}.  A large-area search for $z =$ 8 galaxies was
also made possible by the Brightest of Reionizing Galaxies (BoRG)
program, which used {\it HST} pure
parallel imaging to find an additional $\sim$40 $z =$ 8 galaxy
candidates at random positions in the sky.
Recently, even larger samples of plausible $z =$ 7 and 8
galaxies were obtained by \citet{bouwens15}, who extended the search
to all five CANDELS fields (as well as BoRG), making use of ground-based $Y$-band
imaging in those regions without {\it HST} $Y$-band (i.e., the COSMOS,
EGS and UDS CANDELS fields).  The {\it HST} samples of $z =$ 7 and 8
galaxies now number $\sim$300-500 and $\sim$100-200, respectively.

Even at such great distances, ground-based searches have provided
valuable data, particularly at $z =$ 7, though the bright night sky
background still limits these studies to be restricted to relatively
bright galaxies.  Some of the first robust $z =$ 7
candidates discovered from the ground were published by
\citet{ouchi09}, who used ground-based near-infrared imaging over the
Subaru Deep Field (SDF) and GOODS-N to select 22 $z =$ 7 candidate
galaxies, all brighter than 26th magnitude, including some which are
spectroscopically confirmed \citep{ono12}.  \citet{castellano10}
also used deep ground-based near-infrared imaging, here from the
VLT/HAWK-I instrument, to find 20 candidate $z =$ 7 galaxies brighter
than 26.7.  \citet{tilvi13} took a complementary approach, using
ground-based \emph{medium}-band imaging to select three candidate $z
=$ 7 galaxies from the zFourGE survey.  Although the numbers
discovered in this latter study were small, the higher spectral
resolution afforded by the medium bands allows much more robust
rejection of stellar contaminants, particularly brown dwarfs, which
can mimic the broadband colors of $z \geq$ 6 galaxies (Figure 2).
The most recent, and most constraining, ground-based results come from
\citet{bowler12} and \citet{bowler14}, who used deep, very wide-area
imaging 1.65 deg$^2$ from the UltraVISTA COSMOS 
and the UKIDSS UDS surveys to discover 34 bright $z =$ 7
candidates.  The combination of the very large area with the depth
allowed \citet{bowler14} to have some overlap in luminosity dynamic range with
the {\it HST} studies, which allows more robust joint constraints on the
luminosity function.

Searches at even higher redshifts have been performed, with a number
of studies now publishing candidates for galaxies at $z =$ 9 and 10.
This is exceedingly difficult with {\it HST} alone, as at $z \gtrsim$ 8.8,
galaxies will be detected in only the reddest two WFC3 filters (at 1.4
and 1.6 $\mu$m), while at $z \gtrsim$ 9.3, the Lyman break is already
halfway through the 1.4 $\mu$m filter, rendering many higher redshift
galaxies one-band (1.6 $\mu$m) detections.  
However, initial surveys
did not include observations in the 1.4 $\mu$m band, thus only
one-band detections were possible.  These can be problematic, as
one-band detections can pick up spurious sources such as noise spikes
or oversplit regions of bright galaxies; the possibility of such a
spurious source being detected in two independant images at the same
locations is extremely low (see discussion in \citealt{schmidt14} and \citealt{finkelstein15}).

The first $z \geq$ 9 candidate galaxy published was a single $z \sim$ 10 object found in the HUDF by
\citet{bouwens11c}.  The addition of 1.4 $\mu$m imaging in the HUDF by
the UDF12 program led to further progress, with a handful of two-band
detected $z \sim$ 9 candidate galaxies being discovered
\citep{ellis13,mclure13,oesch13}.  Interestingly, the initial $z \sim$
10 galaxy from \citet{bouwens11c} was not detected in this new 1.4$\mu$m
imaging, implying that if it is truly at high redshift, it must be at
$z \sim$ 12, although there is slight evidence that it may truly be an
emission-line galaxy at z $\sim$ 2 \citep{brammer13}.
Very high redshift galaxies have also been found via lensing from the
CLASH and Hubble Frontier Fields programs, with
candidates as high as $z \sim$ 11, although none have been
spectroscopically confirmed \citep{coe13,zheng12,zitrin14,ishigaki15,mcleod15}.
More recent work by \citet{oesch14}
and \citet{bouwens15} have increased the sample sizes of plausible $z
=$ 9 and 10 candidate galaxies by probing the full CANDELS area.
Although extremely shallow 1.4 $\mu$m imaging is available
(from pre-imaging for the 3D-HST program), these studies leverage the
deep available {\it Spitzer Space Telescope} Infrared Array Camera
(IRAC) imaging in these fields.  These data cover 3.6 and 4.5 $\mu$m,
which encompasses rest-frame 0.3--0.4 $\mu$m at $z =$ 9 and 10,
and thus can potentially provide a second detection filter (though
this is limited by the shallower depth and much broader point-spread
function of the IRAC imaging).  The latest results come from
\citet{bouwens15c}, which combine the results from \citet{oesch14} and
\citet{bouwens15} with new candidates from additional 1 $\mu$m imaging
over selected galaxy candidates, finding a total sample of $\sim$15 and 6
robust candidate galaxies at $z =$ 9 and 10, respectively.

\subsection{Narrowband Searches for Star-Forming Galaxies at $z \geq$ 6}
 
There has also been an intensive effort to discover galaxies on the
basis of strong Ly$\alpha$ emission with narrowband imaging surveys
at $z >$ 6.  These have been primarily ground-based, as the narrow
redshift window probed combined with the small-area {\it HST} cameras
renders space-based narrowband imaging inefficient.  The narrowband
technique has proven highly efficient at discovering large samples of
LAEs at $z =$ 3--6
\citep[e.g.,][]{cowie98,rhoads00,gawiser06b,ouchi08,finkelstein09c},
thus clearly an extension to higher redshift is warranted, though
surveys at $z >$ 6 are restricted in redshift to wavelengths clear of
night sky emission lines.  The most
complete survey for $z >$ 6 LAEs comes from \citet{ouchi10}, who used
the wide-area SuprimeCam instrument on the Subaru Telescope to discover
$>$200 LAEs at $z =$ 6.6 over a square degree in the SXDS field.
\citet{mathee15} have recently increased the area searched for LAEs at
$z =$ 6.6 to five deg$^2$ over the UDS, SSA22 and COSMOS fields,
finding 135 relatively bright LAEs.

Moving to higher redshift has proven difficult, as the quantum
efficiency of even red-sensitive CCDs is declining.  Nonetheless,
\citet{ota10} imaged 0.25 deg$^2$ of the SXDS with SuprimeCam with a filter centered
at 9730 \AA, finding three candidate LAEs at $z = $ 7.0.
\citet{hibon11} used the IMACS optical camera on the Magellan
telescope to find six candidate $z =$ 6.96 LAEs in the COSMOS field,
while \citet{hibon12}
found eight candidate LAEs at $z =$ 7.02 in the SXDS with a 9755 \AA\
narrowband filter
on SuprimeCam.  To observe LAEs at $z >$ 7 requires moving to the
near-infrared, which has only recently been possible due to the advent
of wide-format near-infrared cameras, such as NEWFIRM on the Kitt Peak
4m Mayall telescope.  An additional complication is the increasing
presence of night sky emission lines, which leaves few open wavelength
windows, and drives many to use even narrower filters to mitigate the night sky emission
as much as possible.  One such window is at 1.06 $\mu$m, which
corresponds to Ly$\alpha$ redshifted to $z =$ 7.7.
At $z =$ 7.7, \citet{hibon10} used WIRCam on the CFHT to
discover seven candidate LAEs, \citet{krug12} used
NEWFIRM on the Kitt Peak 4m to discover four candidate LAEs, and
\citet{tilvi10}, also using NEWFIRM, found four additional candidate LAEs.
However, the majority of these candidate LAEs remain undetected in
accompanying broadband imaging, due primarily to the difficulty of
obtaining deep broadband imaging in the near-infrared from the
ground, and most also remain spectroscopically unconfirmed
\citep[e.g.,][]{faisst14}, though see \citet{rhoads12} for one exception.  Thus, the validity of the bulk of these sources is in
question, and requires either deep (likely space-based) broadband
imaging, or spectroscopic followup.
Although the depths of these $z \gtrsim$ 7 studies vary, a relatively
common conclusion is that the LAE luminosity function is likely
evolving very strongly at $z >$ 7 compared to that at lower
redshifts.  We will discuss the physical implications of this perceived lack of
strong Ly$\alpha$ emission in \S 7.

\subsection{Searches for Non-Starforming Galaxies at High Redshift}

The previous sub-sections focused on searches for star-forming galaxies at
high redshift, via either rest-frame ultraviolet (UV) continuum
emission from massive stars, or Ly$\alpha$ emission from H\,{\sc ii}
regions surrounding such stars.  In the local universe, such a search
technique would be extremely biased, as it would miss passively
evolving galaxies.  An ongoing debate is whether such a bias exists
at very high redshift.  It is conceivable that so close to the Big
Bang, galaxies have not had time to quench and stop forming stars, and
thus current surveys are highly complete.  However, observational
evidence for this is lacking, as the detection of passive galaxies
with only optical and near-infrared imaging is difficult.  Although
no robust passive galaxies have yet been discovered at $z >$ 6
(\citealt{mobasher05}, but see \citealt{chary07}), the first robust samples of
handfuls of passive
galaxies at $z >$ 3 have only recently been compiled with
state-of-the-art near-infrared imaging
surveys, relying either on photometric selection via the
Balmer break, or full photometric redshift analyses
\citep[e.g.,][]{muzzin13,spitler14,nayerri14,stefanon13}.  However, no
robust passive galaxies have yet been discovered at $z >$ 6.  If such
galaxies exist, their discovery should be possible with very deep
infrared imaging with {\it JWST}, allowing selection based on the
rest-frame optical emission from lower-mass stars.  

A large population of such galaxies at
$z >$ 6 is not likely, as they would exist at a time $<$1 Gyr
removed from the Big Bang.  For example, a $z =$ 6 galaxy which formed
log (M/M\sol) =
10 in a single burst at $z =$ 20 would have a magnitude of 29 and 26
at 1.6 and 3.6 $\mu$m, respectively.  Such a galaxy would be
detectable in the HUDF presently.  The lack of such galaxies places an
upper limit on the abundance, although one needs to be cautious as
these types of objects may not be selected by some selection
techniques, and it is possible that they are presently mis-identified
as foreground interlopers.

\begin{figure}
\begin{center}
\includegraphics[width=\columnwidth]{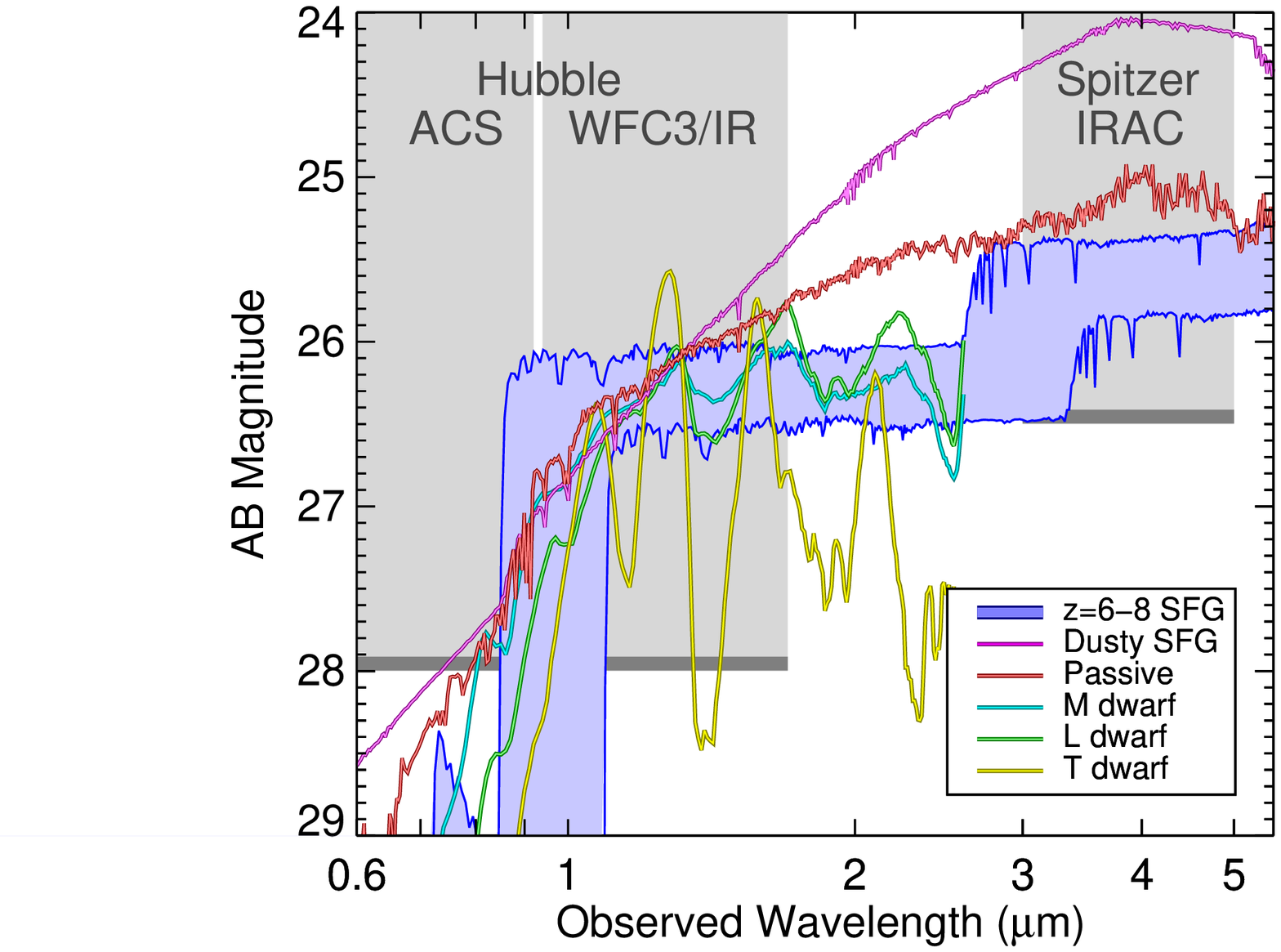}
\caption{A comparison of the SEDs of star-forming galaxies at high
  redshift with possible lower-redshift contaminants.  The
  blue shaded region shows model spectra of high-redshift star-forming
  galaxies (SFGs), with $z
=$ 6 as the upper bound, and $z =$ 8 as the lower bound (both models have
log(M/M\sol) $=$ 9.7, an age of 300 Myr, and A$_\mathrm{V}$=0.4).  The purple and red
lines show a dusty star-forming and a passive galaxy, respectively,
both at $z =$ 1.3.  Cyan, green and yellow curves denote M, L and
T dwarf star empirical near-infrared spectra, taken as the weighted mean of M, L
and T dwarf standards from the SpeX Prism Spectral
Libraries.  The gray shaded regions denote the wavelengths covered by the
HST ACS, WFC3/IR and Spitzer/IRAC imaging used in space-based searches
for $z >$ 6 galaxies, with the lower bound denoting the magnitude
depths at these wavelengths in the CANDELS Deep survey.  All
contaminants shown would likely satisfy a Lyman break criterion for a
$z >$ 6 galaxy, as they would not be detected in typical optical
imaging.  However, colors at redder observed wavelengths can begin to
distinguish between true high-redshift galaxies and low-redshift
contaminants, though this can be difficult when working with low
signal-to-noise data.}
\vspace{-6mm}
\label{fig:contamination}
\end{center}
\end{figure}

\subsection{Contamination}

 All of the studies discussed above select galaxy \emph{candidates},
 meaning that their derived SEDs are consistent with them lying at a
 high redshift, but the vast majority have not had their precise redshifts
 measured with spectroscopy.  I will discuss spectroscopic efforts in
 the following section, but here I discuss the possible sources of
 contamination.   In Figure~\ref{fig:contamination} I show the SEDs of true high redshift galaxies,
along with the plausible contaminating sources discussed below.

For continuum-selected galaxies, the most common contaminants are
lower-redshift dusty galaxies, lower-redshift passively evolving
galaxies, and stars.  Low-redshift dusty galaxies can contaminate as
they would be observed to have very red colors near the anticipated
Lyman break of a true high redshift galaxy.  Similarly, a
lower-redshift passively evolving galaxy can contaminate if the 4000
\AA\ break is mistaken for the Lyman break (at $z =$ 6 and 8 the redshifts of such contaminants
would be $z \sim$ 1.1 and 1.7, respectively).  Both types of
contamination can happen, as depending on the depth of
the imaging used, these contaminants may not be detected in
the bluer of the filters used to constrain the Lyman break, and
detected in the redder of the filters.  However, both types of galaxies should
be rejected as they will also have red colors in the filter
combination just redward of the desired Lyman break, while true high
redshift galaxies are likely bluer.  Additionally, extremely dusty
galaxies may be detected in mid/far-infrared imaging, which is
typically much too shallow to detect true high-redshift galaxies.
Photometric redshift analysis techniques typically show that the redder the galaxy, the more
probability density shifts from a high redshift solution to a low redshift
solution, reflecting the decreased likelihood that the object in
question is a truly red high redshift galaxy.  
For galaxies that are very blue, it is
trivial to rule out any possibility of either a dusty or passive
low-redshift interloper, but there is usually a non-zero chance of
such contamination among the redder galaxies in a given sample.
Contamination estimates from such objects are typically low at $<$10\%
\citep[e.g.,][]{bouwens15,finkelstein15}, though the difficulty of
spectroscopically identifying such interlopers makes it difficult to
empirically measure this contamination rate.

Stellar contamination is typically handled differently, as many
studies search for and remove stellar
contaminants after the construction of the initial galaxy sample
\citep[e.g.,][]{mclure06,bowler12,bowler14,bouwens15,finkelstein15}.  At
$z >$ 6, the colors of M, L and T (brown) dwarf stars can match the
colors of candidate galaxies due to the cool surface temperatures of
these objects.  With {\it HST} imaging it
can be straightforward to remove the brighter stellar contaminants as the
brighter candidate galaxies are all resolved, while stars remain
point sources.  However, this works less well for fainter galaxies, as
near the detection limit it can be difficult to robustly tell whether
a given object is resolved.  This is not a major problem for {\it HST}
studies, as at $J >$ 27, the expected surface density of such
contaminating stars in the observed fields is low \citep{finkelstein15,ryan15}.  

The more major concern is at
intermediate magnitudes, $J =$ 25--26, where the numbers of candidates
are small, yet it can be difficult to robustly discern if an object is 
spatially resolved.  To alleviate this issue, for any objects which
may be unresolved one can examine whether its observed colors are
consistent with any potential contaminating stellar sources.  For this
to be possible, one needs to ensure that the photometric bands
available can robustly delineate between stellar sources and true
high-redshift galaxies; as discussed in \citet{finkelstein15}, this
requires imaging at 1 $\mu$m when working at $z =$ 6--8 \citep[see
also][for a discussion of the utility of medium bands]{tilvi13}.  Using a
combination of object colors and spatial extent, it is likely that
space-based studies are relatively free of stellar contamination.
This may be more of a problem with ground-based studies, though with
excellent seeing even bright $z >$ 6 galaxies can be resolved from the
ground \citep[e.g.,][]{bowler14}.  Future surveys must be cognizant of
the possibility of stellar contamination, and choose their filter set
wisely to enable rejection of such contaminants.

\begin{figure}
\begin{center}
\includegraphics[width=\columnwidth]{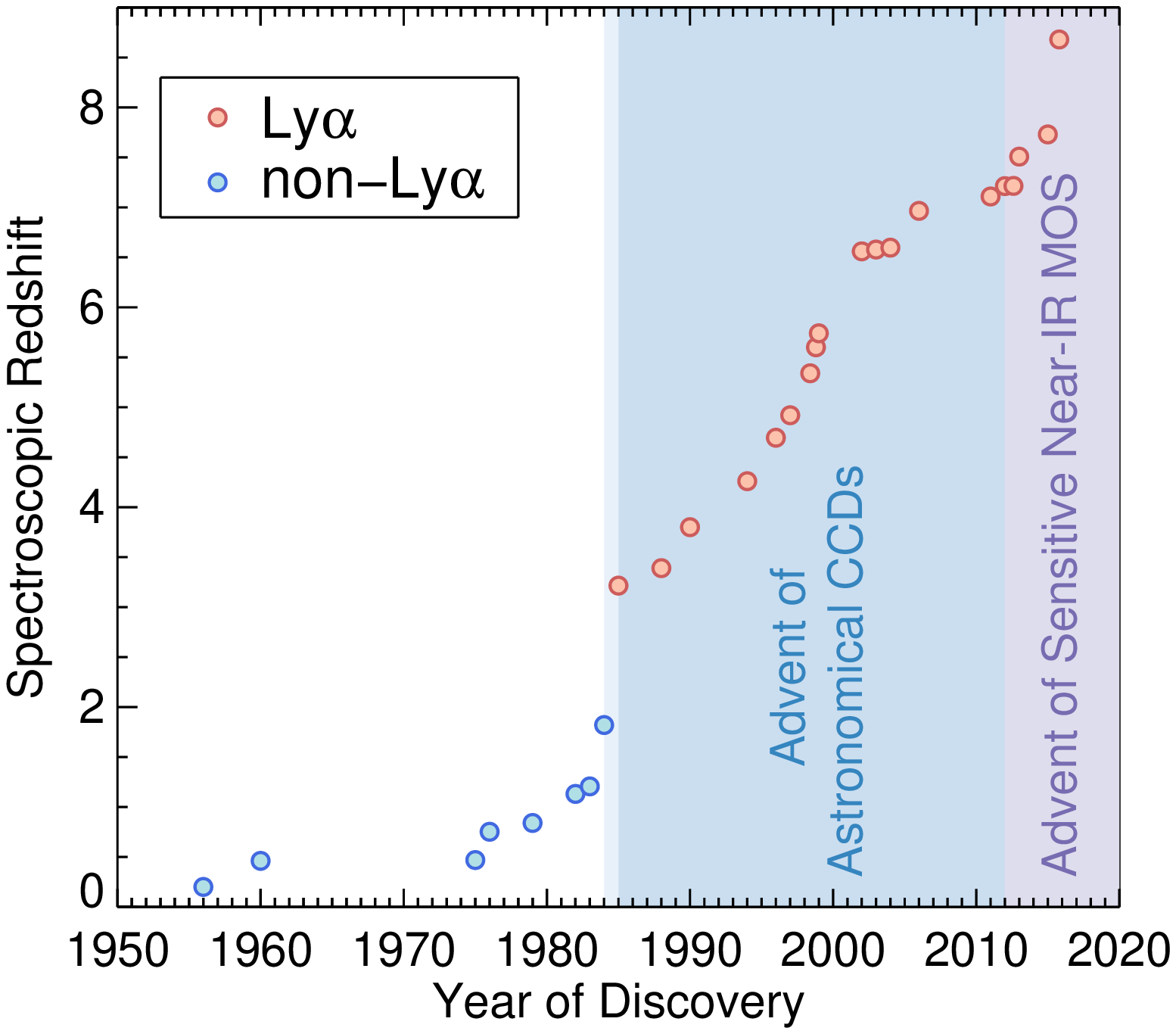}
\caption{The highest redshift spectroscopically confirmed galaxy
  plotted versus the year of discovery.  There are currently only three
  galaxies with robust spectroscopic redshifts at $z >$ 7.5: $z =$
  7.51 from \citet{finkelstein13}, $z =$ 7.73 from
  \citet{oesch15} and $z =$ 8.68 from \citet{zitrin15}.  Data prior to 1999 were taken
from the review of \citet{stern99}, with the references listed
therein.  Objects at later times come from
\citet{hu02,kodaira03,taniguchi05,iye06,vanzella11,ono12,shibuya12,finkelstein13,oesch15,zitrin15}.
The shaded regions denote roughly the time when CCDs became widely
used, as well as when MOSFIRE (the first highly-sensitive
near-infrared multi-object spectrograph) was commissioned on Keck.  Major jumps
in the most-distant redshift are seen to correspond with these
technological advancements.}
\vspace{-8mm}
\label{fig:specz}
\end{center}
\end{figure}

\section{Spectroscopy of $z >$ 6 Galaxies}

While photometric selection is estimated to have a relatively low
contamination rate, it is imperative to followup a representative
fraction of a high-redshift galaxy sample with spectroscopy, to
both measure the true redshift distribution, as well as to empirically
weed out contaminants.  In this section, I discuss recent efforts to
spectroscopically confirm the redshifts of galaxies selected to be at
$z >$ 6.  Figure~\ref{fig:specz} highlights the redshift of the most
distant spectroscopically-confirmed galaxy as a function of the year of discovery.

The most widely used tool for the measurement of spectroscopic
redshifts for distant star-forming galaxies is the Ly$\alpha$ emission line,
with a rest-frame vacuum wavelength of 1215.67 \AA.  While at $z <$ 4,
confirmation via interstellar medium absorption lines is possible \citep[e.g.,][]{steidel99,vanzella09}, the
faint nature of more distant galaxies renders it nearly impossible to
obtain the signal-to-noise necessary on the continuum emission to
detect such features.  Emission lines are thus necessary, and
at $z >$ 3, Ly$\alpha$ shifts into the optical, while strong
rest-frame optical lines, such as [O\,{\sc iii}]
$\lambda\lambda$4959,5007 and H$\alpha\ \lambda$6563 shift into the
mid-infrared at $z >$ 4, where we do not presently have sensitive spectroscopic
capabilities.  Additionally, Ly$\alpha$ has proven to be relatively common amongst
star-forming galaxies at $z >$ 3.  Examining a sample of $\sim$800
galaxies at $z \sim$ 3, \citet{shapley03} found that 25\%
contained strong Ly$\alpha$ emission (defined as a rest-frame EW $>$
20 \AA), while this fraction increases to $\gtrsim$50\% at $z =$ 6 \citep{stark11}.

At higher redshifts, Ly$\alpha$ is frequently the only observable
feature in an optical (or near-infrared) spectrum of a galaxy.  While
in principle a single line could be a number of possible features, in practice, the nearby
spectral break observed in the photometry (that was used to
select a given galaxy as a candidate) implies that any spectral line
must be in close proximity to such a break.  This leaves Ly$\alpha$
and [O\,{\sc ii}] $\lambda\lambda$3726,3729 as the likely
possibilities (H$\alpha$ and [O\,{\sc iii}], while strong, reside in
relatively flat regions of star-forming galaxy continua).  While most ground-based spectroscopy is performed at
high enough resolution to separate the [O\,{\sc ii}] doublet, the
relative strength of the two lines can vary depending on the physical
conditions in the ISM, thus it is possible only a single line could be
observed.  True Ly$\alpha$ lines are
frequently observed to be asymmetric \citep[e.g.,][]{rhoads03}, with a sharp cutoff on the blue
side and an extended red wing, due to a combination of scattering and
absorption within the galaxy (amplified due to outflows), and
absorption via the IGM.  An observation of a single, asymmetric line
is therefore an unambiguous signature of Ly$\alpha$.  However,
measurement of line asymmetry is only possible with 
signal-to-noise ratios of $>$10, which is not common amongst such distant
objects \citep[e.g.,][]{finkelstein13}.  Lacking an obvious asymmetry, other characteristics need to
be considered.  For example, for very bright galaxies, the sheer strength of the Lyman
break can rule out [O\,{\sc ii}] as a possibility, as the 4000 \AA\
break (which would accompany an [O\,{\sc ii}] line) is more gradual
\citep[see discussion in][]{finkelstein13}.  For fainter galaxies,
with a weaker Lyman break, and no detectable asymmetry, a robust
identification of a given line as Ly$\alpha$ is more difficult, and
redshift identification should thus be handled with care.

\subsection{Spectroscopy at $z =$ 6--6.5}

With the advent of ACS on {\it HST}, the frontier for
spectroscopic confirmations of galaxies moved to $z \sim$ 6.   At
these redshifts Ly$\alpha$ is still accessible with
optical spectrographs and was thus an attractive choice for
spectroscopic confirmation.  However, the extreme distances means that
this line will be extremely faint, thus 8-10m class telescopes were
needed to follow them up spectroscopically.  One of the first studies
to spectroscopically observe Ly$\alpha$ from continuum-selected
star-forming galaxies at $z \approx$ 6 was that of
\citet{dickinson04}, who used serendipitous SNe followup ACS grism spectroscopy to
detect the Lyman continuum break from one galaxy, following it up with
LRIS on Keck to discover Ly$\alpha$ emission at $z =$ 5.8.  At that same time
\citet{stanway04} used the GMOS optical spectrograph on the
Gemini 8.2m telescope to measure the redshifts to three galaxies
discovered in the ACS imaging of the HUDF, at $z =$ 5.8--5.9
(one of these, originally published by \citet{bunker03}, is bright
enough to spectroscopically detect the Lyman break).
\citet{stanway07} continued this survey, confirming the redshifts to
two additional galaxies, at $z =$ 5.9--6.1.  \citet{dow07} added
another six redshifts via Ly$\alpha$ at $z =$ 5.5--6.1.  In a series
of papers by Vanzella et al., a larger sample of confirmed $z \approx$ 6
galaxies was obtained with the FORS2 spectrograph on the 8.2m VLT, culminating in the
spectroscopic confirmation of a
total of 32 $z \sim$ 6 galaxies
\citep{vanzella06,vanzella08,vanzella09}.

Another effort for $z \sim$ 6 spectroscopic followup comes
from \citet{stark10,stark11}, who used the DEIMOS optical spectrograph on Keck
to spectroscopically observe continuum selected star-forming galaxies
at 4 $< z <$ 6.  In particular, \citet{stark11} obtained a very deep
12.5 hr single mask observation with DEIMOS, measuring the redshifts
for 11 galaxies at 5.7 $< z <$ 6.0 via Ly$\alpha$ emission.
\citet{stark11} examined the fraction of galaxies with strong (here
defined as EW $>$ 25 \AA) Ly$\alpha$ emission, finding that for
fainter galaxies (M$_{UV} > -$20.25) it rises
from $\sim$35\% at $z =$ 4, to $\sim$55\% at $z \sim$ 6 (for brighter
galaxies, the fraction rises from $\sim$10\% at $z =$ 4 to $\sim$20\%
at $z =$ 6).  These results imply that galaxies at higher redshifts
have a higher escape fraction of Ly$\alpha$ photons, potentially due
to reduced dust attenuation.  In addition to measuring the redshifts of many galaxies
at $z \sim$ 6, the rising fraction of galaxies with detectable
Ly$\alpha$ emission with increasing redshift implied that Ly$\alpha$
should continue to be a very useful tool at $z >$ 6.5.

{\it HST}  provides an alternative to ground-based spectroscopy,
as ACS has grism spectroscopic capabilities.  In this mode, one obtains
very low resolution spectra for every object in the camera's field.
The main advantage is in the multiplexing.  The primary disadvantage
is the contamination from overlapping sources, though this can be
mitigated by splitting the observations over multiple roll angles.
Due to the very low resolution (the G800L grism on ACS has R $\sim$ 100), only
very strong emission lines can be detected.  However, the very low sky
background affords this mode much greater continuum sensitivity,
particularly when searching for galaxies at $z >$ 6, where the night
sky emission makes continuum detections from the ground problematic.
There have thus been a number of {\it HST} surveys seeking to confirm
galaxy redshifts via a detection of the Lyman continuum break.  This
provides somewhat less precision than an emission line detection, but
if the sharpness of the break can be measured, one can confirm that
the break seen in photometry is indeed the Lyman break (which is
sharp, in contrast to the 4000 \AA\ break which is more extended in
wavelength; see Figure~\ref{fig:contamination}).  

The GRAPES survey (PI Malhotra) obtained deep ACS grism observations over
the HUDF.  \citet{malhotra05}
presented the spectroscopic confirmation of 22 galaxies at 5.5 $< z <$
6.7 from this survey, detecting the continuum
break from galaxies as faint as $z_\mathrm{AB} \sim$ 27.5.  The PEARS survey (PI
Malhotra) extended these observations to cover eight additional
pointings in the GOODS fields, culminating in the spectroscopic
detection of a Lyman break at $z =$ 6.6 (with $z_\mathrm{AB} =$ 26.1).  Ground-based followup with
the Keck 10m telescope showed a Ly$\alpha$ emission line at $z =$
6.57 for this galaxy, confirming its high-redshift nature \citep{rhoads13}.  The WFC3/IR
camera also has grism capability, and there are have been efforts
(though none succesful at this time; e.g., \citealt{pirzkal15}) to
confirm redshifts at $z >$ 7 with {\it HST}.  The in-progress
FIGS (PI Malhotra), CLEAR (PI Papovich) and GLASS \citep[PI Treu;][]{schmidt16} surveys may change this, as the very deep
spectroscopy should detect Ly$\alpha$ emission or possibly continuum
breaks for galaxies out to $z \sim$ 8.

\subsection{Pushing to $z >$ 6.5: A deficit of Ly$\alpha$ emission?}

As the first $z \sim$ 7 galaxy samples began to be compiled with the
initial WFC3/IR surveys and ground-based surveys, confirmation
via Ly$\alpha$ was an obvious next step.  However, this proved more
difficult than thought.  One of the first hints that not all was as
expected came from \citet{fontana10}, who observed seven candidate $z
>$ 6.5 galaxies with FORS2 on the VLT.  Given the expected Ly$\alpha$
EW distribution and the magnitudes of their targeted sample, they
expected to detect three Ly$\alpha$ emission lines at
$\geq$10$\sigma$ significance, yet they found none (Ly$\alpha$
emission from one galaxy was
detected at 7$\sigma$ significance at $z =$ 6.972).  Progress was
still made, as \citet{pentericci11} and \citet{vanzella11} each
reported the confirmation of two galaxies via Ly$\alpha$, at $z \sim$ 6.7 in the
former, and $z =$ 7.0-7.1 in the latter.  Yet, as discussed in
\citet{pentericci11}, the fraction of confirmed galaxies was only
25\%, much less than the $\gtrsim$50\% predicted by \citet{stark11}.

\begin{table*}
\caption{Spectroscopically Confirmed Galaxies at $z >$ 7}
\begin{center}
\begin{tabular*}{0.8\textwidth}{@{}c\x c\x c\x c\x c@{}}
\hline \hline
ID & Ly$\alpha$ Redshift & M$_\mathrm{UV}$ & Rest-Frame Ly$\alpha$ EW (\AA)& Reference\\
\hline
BDF-521 & 7.008 & $-$20.6 & 64$^{+10}_{-9}$& \citet{vanzella11} \\
A1703-zD6 & 7.045 & $-$19.4 & 65 $\pm$ 12 & \citet{schenker12} \\
BDF-3299 & 7.109 & $-$20.6 & 50$^{+11}_{-8}$ & \citet{vanzella11} \\
GN 108036 & 7.213 & $-$21.8 & 33 & \citet{ono12} \\
SXDF-NB1006-2  & 7.215 & $-$22.4 & 15$^{+\inf}_{-5}$ & \citet{shibuya12} \\
z8\_GND\_5296 & 7.508 & $-$21.2 & 8 $\pm$ 1 & \citet{finkelstein13} \\
z7\_GSD\_3811 & 7.664 & $-$21.2 & 16$^{+6}_{-4}$ & \citet{song16} \\
EGS-zs8-1 & 7.730 & $-$22.0 & 21 $\pm$ 4& \citet{oesch15} \\
EGSY8p7 & 8.683 &  $-$22.0 & 28$^{+14}_{-11}$ & \citet{zitrin15} \\
\hline
ULAS J1120+0641 & 7.085 & --- & --- & \citet{mortlock11} \\
GRB 090423 & 8.3 & --- & --- & \citet{tanvir09} \\
\hline \hline
\end{tabular*}\label{tab:specz}
\end{center}
\tabnote{The upper portion of the table contains published redshifts based on
  significantly detected ($>$5$\sigma$) Ly$\alpha$ emission at $z >$
  7.  We include published uncertainties on the equivalent width when available.
Not listed are two additional sources which fall in the 4--5$\sigma$
  significance range, from \citet{schenker14} at $z =$ 7.62, and
  \citet{roberts-borsani15} at $z =$ 7.47.  The bottom portion
  contains the highest redshift spectroscopically confirmed quasar
  (via several emission lines, including Ly$\alpha$), and gamma-ray
  burst (via spectroscopic observations of the Lyman break), respectively.}
\end{table*}

Observations of galaxies in this epoch were also performed by
\citet{ono12} and \citet{schenker12}, confirming Ly$\alpha$-based
redshifts at $z =$ 7-7.2, yet still finding less galaxies that would
be the case if the Ly$\alpha$ EW distribution was unchanged from $z =$ 6.
\citet{pentericci14} recently published an extended sample, reporting Ly$\alpha$
emission from only 12 of 68 targeted sources at $z \gtrsim$ 6.5.  After accounting for
the depth of observations and accurate modeling
of night sky emission, \citet{pentericci14} found the fraction of
faint galaxies with Ly$\alpha$ EW $>$25 \AA\ to be only $\sim$30\%.
This deficit is unlikely to be due to significant contamination, as
\citet{pentericci11} showed a much higher fraction of detectable
Ly$\alpha$ emission from galaxies at $z \sim$ 6 selected in a similar way.

Part of these difficulties may be technological in nature, as at $z >$
6.5, these observations were working at the extreme red end of optical
spectrographs, where the sensitivity begins to be dramatically
reduced.  Until recently, similarly efficient multi-object
spectrographs operating in the near-infrared were not available.  This
changed with the installation of MOSFIRE on the Keck I telescope in 2012.
\citet{finkelstein13} used MOSFIRE to observe $>$40 galaxies at $z =$
7--8 from the CANDELS survey in GOODS-N, obtaining very deep 5 hr
integrations over two configurations (of $\sim$ 20 galaxies each).  A
single emission line was detected, which was found to be Ly$\alpha$
from a galaxy at $z =$ 7.51, the most distant spectroscopic detection
of Ly$\alpha$ at that time.  Accounting for incompleteness due to
wavelength coverage and spectroscopic depth, \citet{finkelstein13}
found that they should have detected Ly$\alpha$ from $\sim$six
galaxies, finding that the Ly$\alpha$ ``deficit'' continues well beyond $z
=$ 7.  Other observations have been performed with MOSFIRE, yet most
have only achieved relatively short exposure times, resulting in
primarily non-detections \citep[e.g.,][]{treu13,schenker14}.
Recently, two new record holders for the most distant spectroscopically
confirmed galaxy at the time of this writing have been found, at $z =$
7.73 by \citet{oesch15} and at $z =$ 8.68 by \citet{zitrin15}, 
both detected with MOSFIRE.  Interestingly, the four highest redshift
galaxies known, at $z =$ 7.5--8.7 (including a recent detection at $z
=$ 7.66 by \citealt{song16}), all appear to have very low
Ly$\alpha$ EWs (of $<$ 30 \AA, respectively) and thus are not similar to the much
higher EW sources frequently seen at $z \lesssim$ 6.
In Table~\ref{tab:specz} I summarize the currently known
spectroscopically-confirmed galaxies at $z
>$ 7.  Of note here are again the relatively low EWs, especially at $z
>$ 7.2, as well as the bright UV magnitudes of the confirmed sources.

While there have been some notable successes in the search for
Ly$\alpha$ emission at $z >$ 7, in general all studies report
Ly$\alpha$ detections from fewer objects than expected, as well as
weak Ly$\alpha$ emission from any detected objects.  It thus appears
that something, either in the physical conditions within the galaxies,
or in the universe around them, is causing either less Ly$\alpha$
photons to be produced, or preventing most of them from making their
way to our telescopes.  I will discuss physical possibilities for this
apparent lack of strong Ly$\alpha$ emission in \S 7.1.3.

\subsection{Alternatives to Ly$\alpha$}

Given the apparent difficulties with detecting Ly$\alpha$ at $z >$
6.5, it is prudent to examine whether other emission lines may be
useful as spectroscopic tracers.  While photometric colors imply these
galaxies likely have strong rest-frame optical emission
\citep[e.g.,][]{finkelstein13,smit14,oesch15}, spectroscopic
observations of for example [O\,{\sc iii}] requires {\it JWST}.
However, there may be weaker rest-frame UV emission lines that can be
observed.  \citet{erb10} published a spectrum of a blue, low-mass
star-forming galaxy at $z \sim$ 2 (called BX418) which possessed
physical characteristics similar to typical galaxies at $z >$ 6.
Among the interesting features in the spectrum of this object was
detectable emission lines of He\,{\sc ii} $\lambda$1640 and C\,{\sc
  iii}] $\lambda\lambda$1907,1909.  \citet{stark14} obtained deep
optical spectroscopy of 17 similarly low-mass $z \sim$ 2 galaxies,
finding nearly ubiquitous detections of C\,{\sc iii}].  The strength
of this emission was on average 10\% that of Ly$\alpha$.  However, at
$z >$ 6.5, most of the Ly$\alpha$ is being attenuated (or scattered);
for example, the strength of Ly$\alpha$ at $z =$ 7.51 observed by
\citet{finkelstein13} was only $\sim$10\% of that expected from the
stellar population of the galaxy.  Thus, exposures deep enough to
detect Ly$\alpha$ may also
detect C\,{\sc iii}] at very high redshift.
\citet{stark15} searched for C\,{\sc iii}] from two galaxies with
known Ly$\alpha$ redshifts at $z >$ 6, obtaining tenuous
$\sim$3$\sigma$ detections of C\,{\sc iii}] at $z =$ 6.029 and $z =$
7.213.  Part of the difficulty at higher redshift is because the C\,{\sc iii}]
doublet becomes resolved (splitting the line flux over more pixels),
yet these possible detections imply this may
be a promising line for future study.  Some progress may be made with
MOSFIRE on Keck, though multi-object spectrographs on the next
generation of telescopes, such as the 25m Giant Magellan Telescope or
the Thirty Meter Telescope,
will have the capability to probe these alternative UV emission
features to very faint levels, probing the redshifts of galaxies out
to $z >$ 10.

\section{The Evolution of the Rest-Frame UV Luminosity Function}

One of the most straightforward, and also useful, measures we can make of
distant galaxies is the measurement of the rest-frame UV luminosity
function.  In this section, I will discuss the usefulness of this
observation and recent measurements in the literature.  I will also
derive a ``reference'' luminosity function, as a Schechter fit to all
recent data points from the literature.
 
\subsection{The Significance of the UV Luminosity Function}
 Distribution functions are an immensely useful quantity to measure as
 they are relatively straightforward to compute in both
 observations and theory, and thus provide a direct means to compare
 the two.  Distribution functions of galaxy luminosities, stellar
 masses, and even velocity dispersions have been measured at a variety
 of redshifts, leading to detailed insight into the physical processes
 inherent in galaxy evolution.  At $z >$ 6, however, we are limited in
 what we can measure.  The rest-frame UV is the wavelength
 regime which can be observed very deeply from the
 ground and with {\it HST}, thus the galaxy rest-frame UV luminosity
 function is the best-studied distribution function at such
 redshifts.  While stellar mass functions are also useful (and will be
 discussed in the next section), it is much more direct to correct the
 UV luminosity function for incompleteness, as the luminosity is a
 direct observable, while the stellar mass is a derived quantity.
 There is a downside to the UV luminosity, in that it is highly
 susceptible to dust attenuation, thus to compare observations to theory simulations
 must add dust attenuation, or observations must attempt to correct
 for this attenuation.

Observations of luminosity functions at lower redshifts have shown
that it typically follows a characteristic shape with a
power-law slope at the faint end and an exponential decline at the
bright-end, transitioning at a characteristic magnitude or luminosity
typically referred to as the ``knee'' of the luminosity function.
Parameterized by \citet{schechter76}, the ``Schechter function''
requires three parameters to describe this shape: the characteristic magnitude or
luminosity at the knee (M$^{\ast}$ or L$^{\ast}$), the power-law slope
at the faint end ($\alpha$), and the characteristic number density
($\phi^{\ast}$) which is a normalization factor which defines the
overal number density of galaxies.  Schechter function
parameterizations for luminosity and magnitudes are given in Equations
1 and 2, respectively (in units of number per magnitude or luminosity
bin, per volume).
\vspace{-2mm}

\begin{equation}
\phi(L) = \phi^{\ast}\left(\frac{L}{L^{\ast}}\right)^{\alpha}
\mathrm{exp} \left(-\frac{L}{L^{\ast}}\right)
\end{equation}
\vspace{-5mm}

\begin{equation}
  \begin{split}
    \phi(M) = 0.4~\mathrm{ln}~(10)\!~\phi^{\ast}\!~10^{-0.4(M-M^{\ast})(\alpha + 1)} \\
    e^{-10^{-0.4(M-M^{\ast})}}
  \end{split}
\end{equation}
\vspace{-2mm}

\begin{figure}[!t]
\begin{center}
\includegraphics[width=\columnwidth]{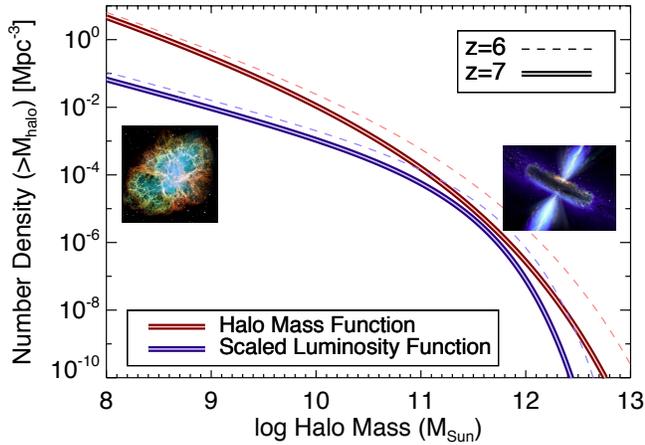}
\caption{The cumulative halo mass function from the Bolshoi
  simulations at $z =$ 6 and 7, shown in red.  In blue, I show the
  cumulative luminosity functions from \citet{finkelstein15}, using
  the relation between stellar mass and UV absolute magnitude from
  \citet{song15}, and scaling by a stellar mass-to-halo mass ratio
  such that the $z =$ 7 functions match at the knee.  Even after this
  scaling, there is still a discrepancy, which is commonly attributed
  to feedback due to supernovae at the faint end, and AGN feedback at
  the bright end (image of the Crab Nebula from \citealt{loll13}).}
\vspace{-12mm}
\label{fig:halomf}
\end{center}
\end{figure}

 A comparison between the shape of the luminosity function and that of
 the underlying halo mass function can provide insight into the
 mechanisms driving galaxy evolution.  A simple toy model may assume
 that the shape of the luminosity function is similar to that of the halo mass
 function, scaling for some constant baryon conversion
 efficiency.  However, as shown in Figure~\ref{fig:halomf}, this is
 not the case.  In this figure, I show the $z =$ 7 luminosity
 function from \citet{finkelstein15}, along with the halo mass
 function at $z =$ 7 from the Bolshoi $\Lambda$CDM simulation
 \citep{klypin11}, measured by \citet{behroozi13b}.  I place the
 luminosity function on this figure by converting from luminosity to
 stellar mass via the liner relation derived by \citet{song15}, and
 scaling vertically such that the two distribution functions touch at
 the knee.  Assuming in this case a ratio of halo mass to stellar mass
 of 80, the number densities match at log (M$_\mathrm{halo}$/M\sol)
 $\sim$ 11.5 (approximately the halo mass of a L$^{\ast}$
 galaxy at this redshift; \citealt{finkelstein15b}), yet the number of both more and
 less massive halos is higher than that of galaxies.  To phrase this
 another way, the conversion of gas into stars in galaxies in both
 more and less massive halos is less efficient.

Such differences have been observed at all redshifts where robust
luminosity functions exist, and a number of physical mechanisms have
been proposed for this observation.  One mechanism that is currently
actively debated is that of feedback.  Models which invoke feedback,
typically due to supernovae (primarily at the faint-end), stellar
radiative feedback, and (primarily at the bright end) accreting
supermassive black hole/active galactic nucleus
(AGN) feedback \citep[see discussion of these processes in
the review of][and references therein]{somerville15} can more
successfully match observations than those which do not include such
effects, in which case too many stars are frequently formed.  This feedback can heat
and/or expel gas from galaxies, effectively reducing, or even
quenching further star-formation.  Such feedback can explain a
variety of observations.  For example, the mass-metallicity relation
observed at lower redshift \citep[e.g.,][]{tremonti04,erb06} can be
explained by supernova-driven winds preferentially removing metals from
lower-mass galaxies, while the increased potentials from higher mass
galaxies allows retention of these metals; \citep[e.g.,][]{dave11}.  

Given that these physical processes affect the shape of the luminosity
function, studying the evolution of this shape with redshift can
therefore provide information on the evolution of these processes.
Observations have shown that the abundance of bright
quasars decreases steeply at $z >$ 3 \citep[e.g.,][]{richards06}.
Although bright quasars do exist at $z >$ 6 \citep{fan06}, they are
exceedingly rare.  Thus, if AGN feedback was the primary factor
regulating the bright end of the luminosity function, one may expect a
decrease in the difference between the luminosity function and the
halo mass function at high redshift.  Likewise, if supernova feedback
became less efficient at higher redshift, one would expect the
faint-end slope to steepen at high redshift, approaching that of the
halo mass function ($\alpha \sim -$2).  In practice, this is more
complicated, as other effects are in play, such as
luminosity-differential dust attenuation \citep[e.g.,][]{bouwens14},
and perhaps a changing star-formation efficiency \citep[e.g.,][]{finkelstein15b}.  Nonetheless, the
shape of the luminosity function is one of the key probes that we can
now measure which can begin to constrain these processes.

The integral of the rest-frame UV luminosity function is also a
physically constraining quantity.  As the UV luminosity probes
recent star-formation \citep[on timescales of $\sim$100
Myr;][]{kennicutt98}, the integral of the UV luminosity function,
corrected for dust, measures the cosmic star-formation rate density
in units of solar masses per year per unit volume.  One can measure
this quantity as a function of redshift, and
such a figure shows that from the present day this quantity rises steeply into the past
\citep[e.g.,][]{lilly96,schiminovich05}, peaking at $z \sim$ 2--3
\citep[e.g.,][]{reddy09}, and declining at $z >$ 3
\citep[e.g.,][]{madau96,steidel99,bouwens07}.  A recent review of this
topic can be found in \citet{madau14}.
The extension of the cosmic SFR density to $z >$ 6 can provide
detailed constraints on the buildup of galaxies at early times.  If it
continues in a smooth trend, as observed from $z \sim$ 3 to 6, it
implies a smooth buildup of galaxies from very early times.
Alternatively, if the SFR density exhibits a steep dropoff at some
redshift, it may imply that we have reached the epoch of initial
galaxy formation.

Finally, another use of the integral of the rest-frame UV luminosity
function is as a constraint on reionization.  Galaxies are the leading
candidate for the bulk of the necessary ionizing photons for
reionization.  By assuming a (stellar-population dependent) conversion
between non-ionizing and ionizing UV luminosity, one can convert the
integral of the UV luminosity function (the specific luminosity
density, in units of erg s$^{-1}$ Hz$^{-1}$ Mpc$^{-3}$) to an ionizing
emissivity, in units of photons s$^{-1}$ Mpc$^{-3}$.
This can then be compared to models of the needed ionizing emissivity
to reionize the IGM, to assess the contribution of galaxies to
reionization.  As shown over a decade ago, galaxies much fainter than
the detection limit of {\it HST} are likely needed to complete
reionization \citep[e.g.,][]{bunker04,yan04}.  Thus, measuring an accurate
faint-end slope is crucial to allow a robust measurement of the total
UV luminosity density, and thus the total ionizing emissivity.  I will
cover this issue in \S 7.

\subsection{Observations at $z =$ 6--10}

In \S 3.1 and 3.2, I covered recent surveys for star-forming galaxies
at $z \geq$ 6.  Here I will discuss the measurements of the rest-frame
UV luminosity function from these surveys.  A number of recent papers
have studied this quantity at $z =$ 6
\citep{bouwens07,mclure09,willott13,bouwens15,finkelstein15,bowler15,atek15,livermore16}, $z =$ 7
\citep{ouchi09,castellano10,bouwens11,tilvi13,mclure13,schenker13,bowler14,bouwens15,finkelstein15,atek15,livermore16},
$z =$ 8 \citep{bouwens11,mclure13,schenker13,schmidt14,bouwens15,finkelstein15,atek15,livermore16},
$z =$ 9 \citep{mclure13,oesch13,oesch14,mcleod15,bouwens15c,ishigaki15} and $z =$ 10
\citep{oesch15b,bouwens15,bouwens15c}.  In the
interest of presenting constraints from the most recent studies in
a concise manner, I will focus on the studies of \citet{bowler14,bowler15},
\citet{finkelstein15}, \citet{bouwens15} at $z =$ 6--8, and
\citet{bouwens15c} and \citet{mcleod15} at $z =$ 9--10.

At $z =$ 6--8, \citet{finkelstein15} and \citet{bouwens15} used data
from the CANDELS and HUDF surveys, while \citet{bowler14,bowler15} used
ground-based imaging from the UltraVISTA and UKIDSS UDS surveys to
discover brighter galaxies.  \citet{finkelstein15} used only data from
the CANDELS GOODS-N and GOODS-S fields, which have deep {\it HST}
imaging in seven optical and near-infrared filters, versus only four
filters in the other three fields.  Specifically, only the CANDELS
GOODS fields have deep space-based $Y$-band imaging, which is necessary for robust removal of
stellar contaminants \citep{finkelstein15}.  \citet{bouwens15} used
all five CANDELS fields, making use of ground-based optical and
$Y$-band imaging to fill in the missing
wavelengths from {\it HST}.  Both studies use the HUDF and
associated parallels, while \citet{finkelstein15} also used the
parallels from the first year of the Hubble Frontier Fields program,
and \citet{bouwens15} used the BoRG/HIPPIES pure parallel program data \citep{schmidt14}.

In spite of different data reduction schemes, selection techniques,
data used, and completeness simulations, the results of
\citet{finkelstein15} and \citet{bouwens15} are broadly similar
(Figure~\ref{fig:schechter}).
Both studies find a characteristic magnitude M$^{\ast}$ which is
constant or only weakly evolving from $z =$ 6--8, and both find a
significantly evolving faint-end slope (to steeper values at higher redshift), and
characteristic number density $\phi^{\ast}$ (to lower values at higher
redshift).  This is a change from initial studies at $z >$ 6 (most
probing smaller volumes), which
found that M$^{\ast}$ significantly evolved to fainter values from $z
=$ 4 to 8, with less evolution in $\phi^{\ast}$
\citep[e.g.,][]{bouwens07,bouwens11,mclure13}.  The faint-end slope
$\alpha$ now appears to
match (or even exceed) the value from the halo mass function ($\alpha
\sim -$2) at the highest redshifts ($\alpha < -$2 can be possible due
to effects of baryonic physics).
The primary difference between these studies appears to be in the normalization, as the
\citet{bouwens15} data points are systematically slightly
higher/brighter than those of \citet{finkelstein15}, which can be
attributed to differences in the assumed cosmology ($\sim$5\%), aperture
corrections utilized to calculate the total fluxes in the photometry,
and differences in contamination.

\begin{figure*}
\begin{center}
\includegraphics[width=6in]{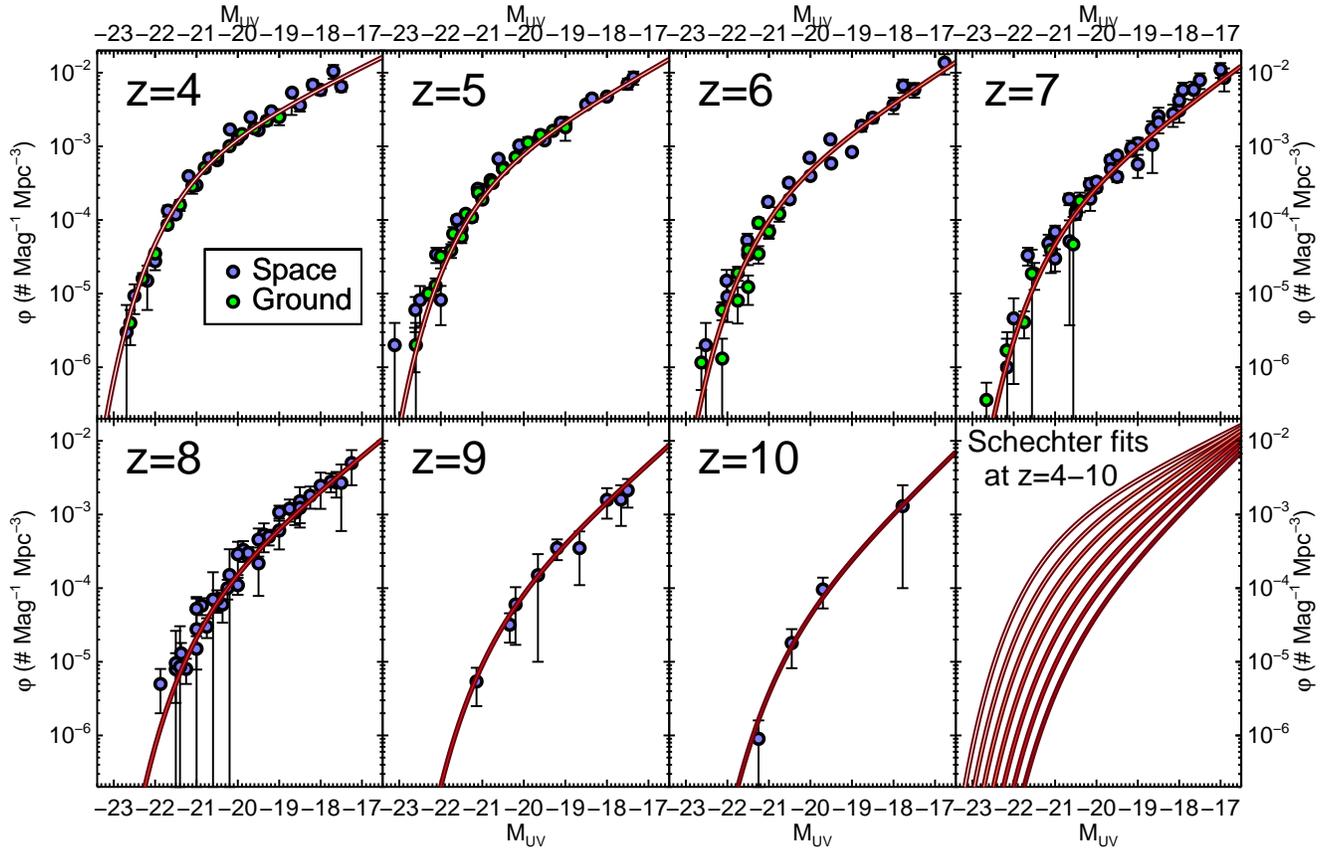}
\caption{A compilation of luminosity function data from the
  literature.  Data from space-based surveys are shown in
  blue, and ground-based surveys in green.  In each panel, I show the
  reference Schechter function fit (\S 5.3) to all
  available data points as the red curves.  The lower-right
  panel overplots the fiducial Schechter functions together at all five
redshifts.  These Schechter function values and associated
uncertainties are given in Table~\ref{tab:lf}.  The studies used in
the fitting are: \citet{bouwens15} at $z =$ 4--10;
\citet{finkelstein15} at $z =$ 4--8; \citet{vdburg10} at $z =$
4--5; \citet{mclure09} at $z =$ 5--6; \citet{mclure13} at $z =$ 7--9;
\citet{schenker13} at $z =$ 7--8; \citet{bouwens15c} at $z =$ 9--10; \citet{bowler15} at $z =$ 6;
\citet{castellano10}, \citet{tilvi13} and \citet{bowler14} at $z =$ 7;
\citet{schmidt14} at $z =$ 8; \citet{oesch13} and \citet{mcleod15} at
$z =$ 9; and
\citet{oesch14} and \citet{mcleod16} at $z =$ 10.}
\vspace{-10mm}
\label{fig:lf}
\end{center}
\end{figure*}

As \citet{bowler14,bowler15} used ground-based data to probe larger volumes,
they were thus sensitive to brighter galaxies than either
\citet{finkelstein15} or \citet{bouwens15}.  Broadly
speaking their results are consistent with the {\it HST}-based studies where
there is overlap.  However, there does seem to be a modest tension, in
that the \citet{bowler14,bowler15} ground-based results exhibit slightly
lower number densities than those from either of the space-based
studies, though the tension only exceeds 1$\sigma$ significance at the
faintest ground-based magnitude (M$_\mathrm{UV} = -$21.5).  This is
true compared to both {\it HST} studies at $z =$ 7, while
\citet{bowler15} and \citet{finkelstein15} are in agreement at $z =$ 6.  To fit a
Schechter function, \citet{bowler15} combine their data with data at
fainter luminosities from \citet{bouwens07} at $z =$ 6, while
\citet{bowler14} combine with fainter data from \citet{mclure13} at $z
=$ 7.  The combination of the deeper {\it HST} imaging with their much
larger volumes allows the ground-based studies to perhaps place
tighter constraints on M$^{\ast}$.  They do find more significant
evolution in M$^{\ast}$ than that found by \citet{finkelstein15} or
\citet{bouwens15}, with M$^{\ast}_{z=7} = -$20.56 $\pm$ 0.17, compared
to $-$21.03$^{+0.37}_{-0.50}$ from \citet{finkelstein15} and $-$20.87
$\pm$ 0.26 from \citet{bouwens15}.  Given these uncertainties, some
evolution in M$^{\ast}$ towards fainter luminosities at higher
redshift is certainly plausible (including the modest $dM/dz =$ 0.2
proposed by Bowler et al.), and there appears no strong
disagreement between these complementary studies.
The data points from these three studies, as well as a number of other
recent works, are shown in Figure~\ref{fig:lf}, and the fiducial
Schechter function parameters are shown in Figure~\ref{fig:schechter}.

Given the wide dynamic range now probed in luminosity, each of the
aforementioned studies understandably pay careful attention to
the shape of the luminosity function. Specifically, they investigate
whether a Schechter function shape is required by the data, or whether
another function, such as a double power law (where the bright-end
exponential cutoff is replaced by a second power law), or even a
single power law (where the exponential cutoff disappears altogether)
is demanded.  \citet{finkelstein15} considered all three functions.
While they found significant evidence that a Schechter function was
required at $z =$ 6 and 7, at $z =$ 8 the luminosity function was
equally well fit by a single power law as by a Schechter function.
This is intriguing, as this is exactly the signature one might expect
were AGN feedback to stop suppressing the bright end (changing dust attenuation
is likely not a dominant factor, as bright galaxies at $z =$ 6--8 have
similar levels of attenuation;
\citealt[e.g.,][]{finkelstein12a,bouwens14,finkelstein15b}).
\citet{bouwens15} found a similar result, claiming that there was no
overwhelming evidence to support a departure from a Schechter
function, but that the available data at $z >$ 6 made it difficult to
constrain the exact functional form.  

While \citet{bowler15} reach a
similar conclusion at $z =$ 6, in that either a double power law or a
Schechter functional form could explain the bright end shape, at $z
=$ 7 \citet{bowler14} find significant evidence that a double power
law is a better fit than Schechter.  With this result, they find that the shape of the
bright end closely matches that of the underlying halo mass function,
implying little quenching in bright galaxies at $z =$ 7.
\citet{bouwens15} discuss this difference in conclusions, and
attribute it to the differences in the measured number density at $M
= -$21.5; the higher number densities found by the {\it HST} studies
allow a Schechter function to be fit equally well.  In any case, at $z
>$ 6, there no longer remains overwhelming evidence to support a
Schechter function parameterization.  To robustly constrain the shape
of the luminosity function requires further data, in particular in
the overlap range between the ground-based and {\it HST} studies.

At higher redshifts of $z =$ 9--10, the data do not currently permit
such detailed investigations into the shape of the luminosity
function.  However, we have begun
to gain our first glimpse into the evolution of the Schechter
parameters to the highest redshifts (if, in fact, this function
ultimately does describe the shape of the luminosity functions).
Studies of galaxies at these redshifts are difficult with the data
used in the above studies, as at these redshifts the Lyman break
passes through the {\it HST}/WFC3 F125W filter, rendering most objects
only detectable in the F160W filter in the CANDELS fields.
Nonetheless, several groups have attempted to select F125W-dropout
galaxies, with some using {\it Spitzer}/IRAC data as a potential
secondary detection band \citep[e.g.,][]{oesch14}.  Additionally,
recent surveys have started adding observations in the {\it HST}/WFC3
F140W filter, which should show a detection for true $z \approx$ 9
galaxies, though galaxies at $z >$ 10 may still have most of their
flux attenuated by the IGM in that band.

\citet{mcleod15} probed the faint-end at $z \sim$ 9
using data from the
first year of the Hubble Frontier Fields (though they explicitly do
not include regions with high magnifications), which 
contains F140W imaging, allowing for robust
two-band detections.  They found 12 new $z
\sim$ 9 galaxy candidates in these data, which they combined with
previously discovered $z \sim$ 9 candidates in the HUDF
from \citet{mclure13} to constrain the luminosity function.  They do
not fit all Schechter parameters, choosing to leave the
faint-end slope fixed to $\alpha = -$2.02, and explore possible values
of the characteristic magnitude and normalization under the scenarios
of pure density or pure luminosity evolution.  They find only modest
evolution from $z =$ 8, consistent with a continued smooth decline in
the UV luminosity density.  \citet{bouwens15c} recently searched the
CANDELS fields for bright $z \sim$ 9 and 10 galaxies, finding nine $z
\sim$ 9 galaxies, and five $z \sim$ 10 galaxies (all
with $H <$ 27).  They do not attempt to derive Schechter function
parameters, rather they use the redshift evolution of said parameters
from \citet{bouwens15} to show that it is consistent with their
updated results at the bright end, as well as results from the
literature at the faint end.  They however conclude that the UV
luminosity density at $z \sim$ 9 and 10 is $\sim$2$\times$ lower than
would be expected when extrapolating the observed trend from $z \sim$
4--8.

These studies highlight the remarkable capability of the
modestly-sized {\it Hubble} to study galaxies to within 500 Myr after
the Big Bang.  However, the sizable uncertainties remaining,
especially at $z >$ 8, lead to fundamental disagreements about
the evolution of the UV luminosity density (\S 5.4) which will likely not be
resolved until the first datasets come in from {\it JWST}.

\subsection{A Reference Luminosity Function}
As discussed at the beginning of the previous subsection, many groups
have published measurements of rest-frame UV luminosity functions at
$z \geq$ 6.  There are a variety of datasets used, both deep
space-based, and relatively shallow ground-based, thus one study alone may not
have the dynamic range to fully constrain the full luminosity
function.  As a perhaps crude, yet illustrative attempt to shed light
on the evolution of this useful tool, in this subsection I combine
published results from a number of studies in an attempt to derive a
set of ``reference'' rest-frame UV luminosity functions at $z =$
4--10\footnote[2]{I extend down to $z =$ 4 to allow the use of this
  analysis in the subsequent sections, using data from
\citealt{vdburg10,bouwens15,finkelstein15} at $z =$ 4, and
\citealt{vdburg10,mclure09,bouwens15,finkelstein15} at $z =$ 5}.  I use the data from
all studies listed in the first paragraph of \S 5.2 where the
luminosity function data was available, with the
exception of the $z =$ 6 results from \citet{willott13} as
  \citet{bowler15} found that the Willott et al.\ sample did not
  include many true $z \sim$ 6 galaxies, possibly due to the
  shallowness of the earlier data.  Likewise, I do not include the $z =$ 7
  results from \citet{ouchi09}, as they applied a very high
  contamination correction of 50\%; see discussion in Appendix F.3 of
  \citet{bouwens15}.  I also do not include the data from
  \citet{bouwens07,bouwens11}, as they are superseded by
  \citet{bouwens15}.  Finally, I do not include the recent lensed
  luminosity functions of \citet{ishigaki15}, \citet{atek15}, or \citet{livermore16} due to
  the likely strong presence of Eddington bias (see discussion in
  \citealt{livermore16}).

I acknowledge that many of these studies use
  similar datasets in the same survey fields, thus the same galaxies
  may contribute to multiple data points, and this analysis does not include possible
  systematic effects which could broaden the error budget.  However, even studies
  which utilize the same surveys use a wide range of data
  reduction, photometry, and sample selection
  techniques which can and does result in differences in the measured
  number densities at a given magnitude (Figure~\ref{fig:lf}).
The results of this analysis are illustrative of the constraints possible
  when marginalizing over these differences, as well as when combining
  ground with space-based datasets.
However, I do acknowledge that the faintest data points used
  typically all come from the same field: the HUDF.  Therefore, during
  the MCMC analysis (described below), in magnitudes bins at $-$18 or fainter each step of the
MCMC chain randomly chooses one data point per magnitude bin.  This
should suppress any ``artificial" damping of the uncertainties on the
faint-end slope, as during each step we use only a single measurement
from the HUDF per magnitude bin.

From each of the studies used, I extracted the measured, completeness
corrected, number densities (and associated uncertainties) as a
function of UV absolute magnitude, shown in Figure~\ref{fig:lf}.
I then calculated the likelihood that the data were represented by a given Schechter function via
a Markov Chain Monte Carlo (MCMC) algorithm.  I used an IDL implementation (R.\
Ryan, private communication) of
the Python $\tt{Emcee}$ package \citep{foreman-mackey13}, which utilizes
an affine-invariant ensemble sampler to sample the parameter space.  
However, rather than fitting a Schechter function to each redshift
independently, I made the assumption that the Schechter function parameters linearly vary with
redshift.  This is an assumption often assumed in the literature,
though typically after the fact \citep[e.g.,][]{bouwens15,finkelstein15}.
The advantage of this method is that it allows the data from all
redshifts to be fit \emph{simultaneously}, solving for the posterior
distribution of the coefficients of the linear function which
describes $M^{\ast}(z)$, $\phi^{\ast}(z)$ and $\alpha(z)$.
In effect, this replaces 21 parameters (the three Schechter parameters at seven
redshifts) with six parameters: the slope and intercept of the three
redshift-dependent Schechter parameter trends.  An additional
advantage is that this method allows convergence of the chain even at
$z =$ 9 and 10, where there are few datapoints.  Without the
assumption of linear evolution, a prior would have to be placed on the
individual Schechter parameters at these highest redshifts to obtain a
result.

\begin{figure*}
\begin{center}
\includegraphics[width=6in]{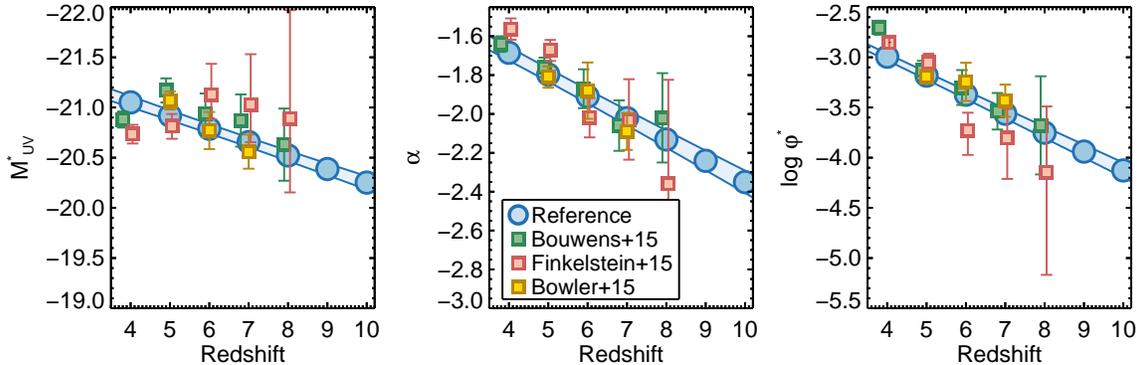}
\caption{The derived evolution of the three Schechter function
  parameters with redshift, derived by fitting all redshifts
  simultaneously.  The shaded blue regions
  denote the 68\% confidence range of the linear evolution of these
  parameters with redshift, while the circles denote the reference value at a
  given redshift. Evolution
  with increasing redshift in $M^{\ast}$ (to fainter values), $\alpha$ (to steeper values) and
  $\phi^{\ast}$ (to lower values) is detected at $>$10$\sigma$ significance.}
\vspace{-10mm}
\label{fig:schechter}
\end{center}
\end{figure*}

I ran the MCMC algorithm for 10$^6$ total steps, and checked each
parameter for convergence using the Geweke diagnostic, which compares
the average and the variance of the first 10\% to the last half of the
steps \citep{geweke92}.  For each of the six parameters of the fit, I
calculated the median value and the 68\% confidence range from the
median and central 68\% range of the posterior distribution,
respectively.  The results are:
\begin{align*}
M^{\ast}(z) &= -20.79_{-0.04}^{+0.05} + 0.13_{-0.01}^{+0.01}~ (z-6)\\
\alpha(z) &= ~\!\!-1.91_{-0.03}^{+0.04} - 0.11_{-0.01}^{+0.01}~ (z-6)\\
log~\phi^{\ast}(z) &= ~\!\!-3.37_{-0.04}^{+0.05} - 0.19_{-0.01}^{+0.01}~ (z-6)
\end{align*}
The fiducial Schechter functions from this exercise derived by
evaluating the above equations at a given redshift are shown alongside the data
points in Figure~\ref{fig:lf} and the values are tabulated with their
uncertainties in Table~\ref{tab:lf}.  In the lower-righthand panel of
Figure~\ref{fig:lf}, I show the fiducial Schechter functions from
this analysis at $z =$ 4--10 together, highlighting a relatively
smooth decline from $z =$ 4 to 10.  In Figure~\ref{fig:schechter}
I show the redshift-dependent Schechter parameter evolution
with the median values at each redshift, along with the results
of a variety of studies listed above.  
The reference results are for the most part consistent with results in
the literature within the uncertainties (with the exception of M$^{\ast}$
at $z =$ 4, which is brighter than the value found by
\citet{bouwens15} and \citet{finkelstein15}).

\begin{table}
\caption{Median Schechter Parameters to Compilation of Literature Data}
\begin{center}
\begin{tabular*}{0.45\textwidth}{@{}c\x c\x c\x c@{}}
\hline \hline
Redshift & M$^{\ast}$ & $\alpha$ & $log \phi^{\ast}$ \\
\hline
4&$-$21.05$^{+0.05}_{-0.06}$&$-$1.69$^{+0.03}_{-0.04}$&$-$2.99$^{+0.04}_{-0.04}$\\
5&$-$20.92$^{+0.05}_{-0.05}$&$-$1.80$^{+0.04}_{-0.04}$&$-$3.18$^{+0.04}_{-0.04}$\\
6&$-$20.79$^{+0.05}_{-0.04}$&$-$1.91$^{+0.04}_{-0.03}$&$-$3.37$^{+0.05}_{-0.04}$\\
7&$-$20.66$^{+0.06}_{-0.04}$&$-$2.02$^{+0.05}_{-0.03}$&$-$3.56$^{+0.05}_{-0.04}$\\
8&$-$20.52$^{+0.06}_{-0.04}$&$-$2.13$^{+0.05}_{-0.03}$&$-$3.75$^{+0.06}_{-0.04}$\\
9&$-$20.39$^{+0.07}_{-0.05}$&$-$2.24$^{+0.06}_{-0.04}$&$-$3.94$^{+0.07}_{-0.05}$\\
10&$-$20.25$^{+0.07}_{-0.06}$&$-$2.35$^{+0.06}_{-0.04}$&$-$4.13$^{+0.08}_{-0.06}$\\
\hline \hline
\end{tabular*}\label{tab:lf}
\end{center}
\tabnote{Median Schechter function values to a compilation of
  data from the literature.  These values are derived from a MCMC fit
  with a prior that the evolution of these parameters are linear with
  redshift.  The value of each parameter is the median value of that parameter's
  linear function evaluated at the given redshift.  The uncertainty
  listed is the 68\% confidence range.}
\end{table}

In the discussion in \S 5.2, I highlighted recent investigations into the evolution of the faint-end
slope $\alpha$ and the characteristic magnitude M$^{\ast}$.  As shown
in the left-hand panel of Figure~\ref{fig:schechter}, the results at
4 $< z <$ 7 are consistent with a significant dimming in M$^{\ast}$
with increasing redshift, with $dM^{\ast}/dz =$0.13.  The sign of
this effect is similar to that found in \citet{bowler15}, though they
found a more accelerated evolution, with $dM^{\ast}/dz =$0.20-0.25 over 5
$< z <$ 7.  Investigating $\alpha$ and $\phi^{\ast}$, the
reference luminosity functions
confirm the previous results of a steepening faint-end
slope and decreasing characteristic number density with increasing
redshift, with $d\alpha/dz =-$0.11 and $d\phi^{\ast}/dz =-$0.19.
These trends are qualitatively similar to those from
\citet{bouwens15} of d$\alpha$/d$z=-$0.10 and
dlog$\phi^{\ast}$/d$z=-$0.27, and from \citet{finkelstein15} of
d$\alpha$/d$z=-$0.19 and dlog$\phi^{\ast}$/d$z=-$0.31.  However, due to
the higher fidelity of the reference
Schechter fits, the significance of the trends derived here is higher,
such that a $>$10$\sigma$ significance
evolution is detected in both $\alpha$ and $\phi^{\ast}$, respectively
(though the partial correlation between some of the data points may be
responsible for some of the apparent improvement).  While I did not
include recent lensed-galaxy data from the Hubble Frontier Fields, the
faint-end slopes I derive at $z =$ 6--8 are in excellent agreement
with those from the Frontier Fields of $\alpha \sim -$2 \citep{atek15,livermore16}.

A general conclusion which can
be made from this exercise is that the current data can reasonably
constrain all three Schechter function parameters out to $z =$ 10 when
all redshifts are fit simultaneously, under the assumption that the
parameters evolve linearly with redshift.
By inspecting Figure~\ref{fig:lf}, one can see that the resultant parameterizations at
each redshift appear to excellently describe the data, thus it would
appear that this assumption is valid, at least at the limit of the data
presently in hand.  
By fitting all data simultaneously, I arrive at potentially more
precise values of the evolution of the rest-frame UV luminosity
function with redshift, which increases the constraints on reionization which can be
derived by integrating the UV luminosity function.  I explore this
further in \S 7.

\begin{figure*}
\begin{center}
\includegraphics[width=5.5in]{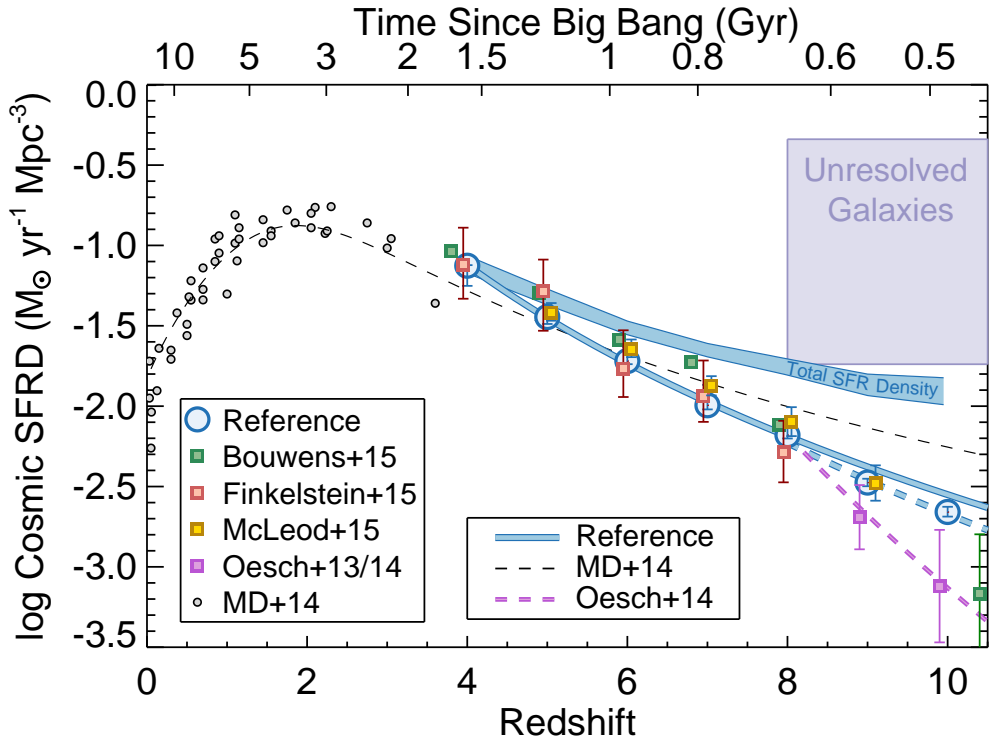}
\vspace{-4mm}
\caption{The evolution of the cosmic star-formation rate density,
  comparing the values from the integral of the reference luminosity
  function to those from the literature.  All points have been
  corrected to represent a lower limit on the luminosity function
  integration of M$_\mathrm{UV} < -$ 17, and have been corrected for dust
  attenuation (with the exception being the low-redshift far-infrared
  datapoints from \citealt{madau14}).  The solid blue curve shows a
  power-law fit to the reference data at 4 $< z <$ 8 ($\propto$[1$+z$]$^{-4.17 \pm 0.27}$), extrapolated to
  higher redshift, while the dashed line shows a fit only to the data
  at $z \geq$ 8 ($\propto$[1$+z$]$^{-5.10 \pm 0.69}$).
The results from the reference luminosity function
are consistent with a smooth decline in the SFR density at 4 $< z <$
10, with no significant evidence for an accelerated evolution at $z >$ 8.
However, the 68\% confidence range on the total SFR density (blue
shaded region; derived from integrating the
  reference luminosity functions to M$_\mathrm{UV} = -$ 13) is
  consistent with an even shallower decline in
the SFR density over $4 < z < 10$.  The light purple region denotes constraints on
the total luminosity density from unresolved background fluctuations
\citep{mitchell-wynne15}, which also imply a relatively shallow evolution
of the total SFR density.  The potential for a surprisingly high SFR density at $z >$ 8 will soon be settled by {\it JWST}.}
\vspace{-8mm}
\label{fig:sfrd}
\end{center}
\end{figure*}

\subsection{The Evolution of the Cosmic Star-Formation Rate Density}
A final quantity one can study with the UV luminosity function is that
of the evolution of the cosmic star-formation rate density.  This is
derived by integrating the UV luminosity function down to a specified
threshold magnitude, and then converting to a SFR using a conversion
based on stellar population modeling \citep[e.g.,][]{madau14}.  As
this is derived from the UV which is susceptible to dust reddening, an
extinction correction also needs to be estimated and applied.
As shown in a variety of previous studies, the evolution of this
quantity has been shown to smoothly vary with redshift at 3 $< z <$ 8,
such that the data are consistent with a single power-law $\propto
(1+z)^{-(3-4)}$ \citep[e.g.,][]{madau14,oesch14,finkelstein15,bouwens15}.  The
evolution to higher redshift has been less well constrained, with some
observations supporting a continued smooth decline with increasing redshift at $z >$ 8
\citep[e.g.,][]{ellis13,coe13,mcleod15}, while others report that
their data require a steeper decline
\citep[e.g.,][]{oesch14,bouwens15c,bouwens15}.

To explore whether the reference luminosity functions derived here can
distinguish between these two scenarios, in
Figure~\ref{fig:sfrd} I show the values of the SFR density as a
function of redshift from this reference analysis, alongside values at
$z <$ 4 from the compilation of \citet{madau14}.  These values were
derived by integrating the luminosity functions at each redshift to
$-$17 \citep[a value typically used;][]{bouwens15,finkelstein15}, to
calculate the observed UV luminosity density.  This was then converted
into a SFR density using the conversion factor of 1.15 $\times$
10$^{-28}$ M\sol~yr$^{-1}$/(erg s$^{-1}$ Hz$^{-1}$) from
\citet{madau14}.  Dust corrections were derived
using the redshift-dependent M$_\mathrm{UV}$ -- $\beta$ relation from
\citet[][see \S 6.1, including at 0.35 dex scatter]{bouwens14}
coupled with the relation between dust extinction and $\beta$ from
\citet{meurer99}.  Formally, the combination of these two relations
can result in negative values of A$_\mathrm{UV}$; these were set to
zero.  The total dust correction factors
(L$_\mathrm{intrinsic}$/L$_\mathrm{observed}$) were 3.0, 2.2, 1.7, 1.4
and 1.4 at $z =$ 4, 5, 6, 7 and 8, respectively.  Zero dust attenuation was assumed at $z =$ 9 and
10. 

I show my results in Figure~\ref{fig:sfrd} alongside several results
from the most recent studies at each redshift in the literature.
Those literature values that were integrated to brighter limits
(typically $-$17.7) were corrected down to $-$17 using the luminosity
functions from each paper.  Those which did not perform a dust
correction had a correction applied, using the value for a given
redshift derived here as discussed in the previous paragraph.  We note
that the reference results are consistent with those from
\citet{finkelstein15} and \citet{mcleod15}, while those from
\citet{bouwens15} are elevated at $z \leq$ 7.  The latter discrepancy can be
explained as the \citet{bouwens15} luminosity functions are slightly
elevated over those from other studies.

\citet{madau14} found that the
low-redshift evolution of the SFR density was proportional to
$(1+z)^{2.7}$, while at high redshift it was proportional to
$(1+z)^{-2.9}$ (fitting only to the previously available data at $z <$
8).  To explore the evolution of the SFR density implied by the reference
values derived here, I fit a single power law to the reference data at
4 $< z <$ 10, finding that the SFR density is $\propto$
(1$+z$)$^{-4.55 \pm 0.14}$.  Fitting to the data only at 4 $< z <$ 8,
the slope is slightly shallower, (1$+z$)$^{-4.17 \pm 0.27}$, in
excellent agreement with the slope of $-$4.3 found by
\citet{finkelstein15} over that same redshift range.  The fit to the 4 $< z <$ 8 data
is shown as the solid blue line in
Figure~\ref{fig:sfrd}.  We extrapolate this line out to $z =$ 10, and
find that it is generally in good agreement with the reference SFR
density values at those redshifts, although they do appear to be
slightly lower.

To explore this further, I fit a separate power law only to the data
at 8 $< z <$ 10, finding a power-law slope of $-$5.10 $\pm$ 0.69.  The
difference between this slope and that derived from the data at 4 $< z
<$ 8 is only 1.3$\sigma$.  Therefore, I conclude there is no
significant evidence for an accelerated evolution of the SFR density
at $z >$ 8 given the currently available data.
This is contrary to the evolutionary trend
proposed by \citet{oesch14} of $(1+z)^{-10.9}$, shown by the dashed
purple line, which becomes progressively more discrepant with
the reference shallower evolution with increasing redshift at $z >$ 8.

I note that by forcing the assumption that each of the
  Schechter function parameters evolve linearly with redshift, I
  effectively bias the reference SFR density results to evolve
  smoothly with redshift.  However, as shown by the excellent
  agreement between the data and the Schechter fits in
  Figure~\ref{fig:lf}, the reference results supporting a smooth
  evolution of the SFR density are fully consistent with the data out
  to those high redshifts.  To examine this quantitatively, I
  compared the fiducial results at $z =$ 9
and 10 to those from a single Schechter function fit at each redshift
(holding M$^{\ast}$ fixed at the $z =$ 8 value of $-$20.5 to allow the
fits to converge).  Comparing the goodness-of-fit via the Bayesian
Information Criterion \citep{liddle04}, I find no evidence that the
fiducial fit is worse than when fitting the $z =$ 9
and 10 luminosity functions without the assumption of linear
evolution.  Therefore, I conclude that this assumption is not biasing
these results.

As a final point here, I emphasize that a decline in this quantity -- the SFR
density for galaxies with M$_\mathrm{UV} \leq -$17 --
is not unexpected, as we are
comparing SFR densities above a fixed absolute magnitude.  As the
luminosity function evolves with redshift, a progressively smaller
fraction of the total UV luminosity density will come from galaxies
with UV absolute magnitudes fainter than $-$17 due to the steepening
of the faint-end slope.  Assuming that the luminosity function extends with the measured
faint-end slope down to M$_\mathrm{UV} = -$13, the SFR density derived at M$_\mathrm{UV} < -$17 is $\sim$60\%
of the \emph{total} SFR density at $z =$ 6, yet only $\sim$20\% at $z
=$ 10.  Thus, in the regime of an
evolving luminosity function, the choice of the limiting
magnitude directly affects the inferred evolution in the SFR density. 

To illustrate this point, the
shaded blue region in Figure~\ref{fig:sfrd} shows the 68\% confidence
range on the {\it total} SFR density from the reference UV luminosity
functions (integrated down to $M = -$13).  This is observed to evolve
only very shallowly from $z \sim$ 4 --
8, leaving open the intriguing possibility that there is significant
star formation hiding below our current detection limits at $z =$ 10.
A resolution of this issue will require a
more robust measurement of the \emph{total} SFR density at $z =$ 10, which we will
have in just a few years with the launch of {\it JWST}.   A {\it JWST}
deep field should be able to detect galaxies down to M$_\mathrm{UV} =
-$16 at $z =$ 10; more than 1.5 magnitudes fainter than the currently
available sample of galaxies.  However, the recent detection of
near-infrared background fluctuations in the CANDELS GOODS-S field \citep{mitchell-wynne15}
imply a total UV luminosity density at $z >$ 8 consistent with our
estimate of the total SFR density, thus substantial star-formation
activity at $z >$ 8 may yet be found.

\section{The Physical Properties of Distant Galaxies}

While the UV luminosity function can allow us to probe the global
mechanisms affecting the evolution of galaxies, we can obtain a deeper
insight into galaxy evolution by directly measuring galaxy physical
properties.  In this section I highlight two areas of significant
recent activity: the chemical enrichment of galaxies, and the growth of galaxy stellar
masses.

\subsection{Dust Attenuation and Chemical Enrichment}
One of the most direct, and straightforward measures of galaxy
evolution is the evolution of galaxy colors.  Although the
interpretation of these colors requires assumptions and/or modeling,
the direct measurement is unambiguous, and can be done with a simple
photometric catalog.  At lower redshift, simple color-color or
color-magnitude plots are immensely useful to probe the star-forming
properties of galaxies, as galaxy populations typically separate out
into the star-forming ``blue cloud'', and the relatively quiescent
``red sequence''.  At $z >$ 6, when the universe was less than 1 Gyr
old, our {\it HST} observations probe only the rest-frame UV, thus our color
diagnostics are somewhat more limited than at lower redshift (though
we don't expect a large fraction of quiescent galaxies, e.g.,
\citealt{muzzin13, nayerri14}).

However, the rest-frame UV can provide robust constraints on the dust
content in distant galaxies.  While one would prefer to directly probe
the metallicities of such systems to track chemical enrichment, in
practice this is difficult with current technology.  However, as dust
grains are made from the heavy elements which form in stars, the 
build-up of dust in the universe tracks the build-up of metals.
Additionally, although dust is not unique in reddening the rest-frame
UV colors of galaxies, the change in UV color with increased dust
content is larger than that with a comparable increase in metallicity
or stellar population age (particularly at $z >$ 6, as stars must
be younger than the age of the universe).  Thus, tracking the
evolution of the UV colors of distant galaxies provides an excellent
path for following the evolution of the dust content in these systems,
therefore providing a proxy to track the build-up of heavy metals in
the universe.

Rather than using a single pair of filters, the rest-frame UV color is
typically parameterized by the UV spectral slope $\beta$, which is
defined as $f_{\lambda} \propto \lambda^{\beta}$ \citep{calzetti94}.
The quantity $\beta$ is an excellent tracer of dust extinction, as it
has been found to be well correlated with the ratio of the
far-infrared to UV flux at both low
and moderate redshift \citep[e.g.,][]{meurer99,reddy12}, though the
exact shape of this correlation depends on the dust attenuation curve
\citep[e.g.,][]{capak15} and other factors
\citep[e.g.,][]{wilkins12,wilkins13}.  Earlier work showed that
$\beta$ became
substantially more negative (i.e., bluer UV colors) with increasing
redshift, from $\beta\sim-$1.5 at $z \sim$ 2 to $\beta\sim-$2 at $z \sim$ 6
\citep[e.g.,][]{bouwens09}.  

\subsubsection{Colors of Faint Galaxies}
Measurements at higher redshift required
the deep near-infrared imaging discussed in \S 3.1.  The earliest
results on the UV colors of $z \geq$ 6 galaxies focused on faint
galaxies at $z \sim$ 7, using the first year of data from the HUDF09
program (which, targeting a small field to a deep magnitude limit,
was well-suited to address faint galaxies).  \citet{bouwens10b}
measured $\beta = -$3.0 $\pm$ 0.2 for faint (M$_\mathrm{UV} = -$ 18)
galaxies at $z \sim$ 7.  They postulated that this extremely blue
color may imply very low metallicities, as even young, dust free
populations have $\beta \geq -$2.7 when nebular emission is accounted
for.  Using the same imaging dataset, \citet{finkelstein10} also found
$\beta = -$3.0, but with a significantly higher uncertainty, of $\pm$
0.5.  Given the larger uncertainty, they concluded there was no
evidence for ``exotic'' stellar populations, as the measured colors
were consistent with local blue galaxies within a significance of 2$\sigma$.

Further evidence for ``normal'' stellar populations in faint galaxies
at $z \sim$ 7 came from \citet{dunlop12}.  They ran simulations which
showed that faint galaxies exhibit a bias, providing measurements
which are bluer than the true colors of galaxies.  This occurs because
the colors which are used to measure $\beta$ are the same colors used
to select the galaxy sample.  In particular, at $z =$ 7, the {\it
  HST}/WFC3 $J$ and $H$ bands were used by these initial studies to
measure the colors, and they were two of the only three bands with
detections.  If one imagines a scenario of a faint galaxy with an intrinsic
$\beta = -$2 where the position of the galaxy in the $J$-band image
falls on a positive noise spike, the color will be measured to be
bluer than the true value.  There is an equal probability that 
a noise spike would occur at this position in the $H$-band image, providing a
color measured which is redder than the true value.  However, a redder
color would may push the galaxy out of the LBG selection box (or,
equivalently, push more of its redshift probability distribution
function to lower redshifts), thus, galaxies biased red may not make
it into the galaxy sample, providing a net blue bias.  Correcting for
this, \citet{dunlop12} found no significant evidence that faint $z =$
7 galaxies have $\beta < -$2, similar to blue star-forming galaxies at
lower redshifts, a result which was confirmed by \citet{rogers13}, who
ran further simulations as well as justifying the benefits of the eventual inclusion
of the F140W filter from the UDF12 program.  A similar result was found by \citet{wilkins11}.

Further progress was made with the addition of the CANDELS and UDF12
imaging.  In addition to much larger samples of galaxies, more
attention was put into the measurement of $\beta$ itself.
\citet{finkelstein12a} showed that measuring $\beta$ with a single
color resulted in much larger uncertainties than when using multiple
colors, obtaining $\beta$ either via a power-law fit to all available
colors \citep[e.g.,][]{castellano12,bouwens12,bouwens14}, or by fitting $\beta$
directly to the best-fitting stellar population model \citep{finkelstein12a}.
Both \citet{finkelstein12a} and \citet{bouwens12,bouwens14} made use of these
much larger datasets to increase the robustness of constraints on the
evolution of $\beta$.  Both studies found that the updated sample (as
well as measurement techniques) resulted in colors of faint $z =$ 7
galaxies which did not require exotic stellar populations ($\beta \sim -$2.3
to $-$2.4, correcting for any biases).  A qualitatively similar result
was found by \citet{dunlop13}, who obtained a bias-free measurement of
$\beta$ for faint galaxies by leveraging observations in a new WFC3/IR
filter (F140W), finding $\beta \sim -$2.1 for faint galaxies.  From
the combination of these above studies, one can thus conclude that
even the faintest galaxies we can see at $z \sim$ 7 have colors which
imply the presence of some metals, although not much dust, and
therefore do not represent truly primordial systems.  Such systems, if
they exist, remain to be found with {\it JWST}.

\subsubsection{Colors of Bright Galaxies}
The large dynamic range in luminosity and stellar mass of the galaxy
samples studied by \citet{finkelstein12a} and
\citet{bouwens12,bouwens14} allowed further redshift and
luminosity/stellar mass dependent trends to be explored.
\citet{bouwens12,bouwens14} explored the dependance of galaxy colors
on UV luminosity, and found a significant color-magnitude relation,
where brighter galaxies are redder than fainter galaxies at a given
redshift.  \citet{finkelstein12a} explored the correlation of colors
with stellar mass, and found that at a given redshift, more massive
galaxies are redder than less massive galaxies.  They concluded that
this was likely a manifestation of the mass-metallicity relation seen
at lower redshift \citep[e.g.,][]{tremonti04,erb06}, where more
massive galaxies have higher gas-phase metallicities.  The interpretation
is that lower-mass galaxies are more susceptible to losing their
metals (and dust) in outflows due to their shallower gravitational
potentials, thus a mass (or luminosity) dependent rest-frame UV color
would be an expected signature of this physical process was in place
at early times.  More generally, both \citet{finkelstein12a} and
\citet{bouwens12,bouwens14} found that the average dust attenuation in galaxies
rises with decreasing redshift, from near-zero at $z \sim$ 7, to
A$_\mathrm{V} \sim$ 0.5 at $z =$ 4.

Perhaps the most interesting result from these studies are the colors
of not the faintest galaxies, but the brightest/most massive
galaxies.  \citet{finkelstein12a} found that the most massive galaxies
have a roughly constant value of $\beta$ from $z =$ 4 to 7, of $\beta
\sim -$1.8, similar to what \citet{bouwens12,bouwens14} found for the
brightest galaxies at these redshifts.  \citet{rogers14} found a
similar result for the brighest galaxies at $z \sim$ 5.
Similarly, via SED fitting,
\citet{finkelstein15b} found that bright galaxies at $z =$ 4--7 have
similar values of $E(B-V)$, of $\sim$0.1-0.15 (A$_\mathrm{V} =$ 0.4--0.6
assuming a \citealt{calzetti00} attenuation curve).  The significant
dust content in these galaxies is consistent with the apparently
ubiquitous strong [O\,{\sc iii}] emission implied by {\it
  Spitzer}/IRAC colors in many distant galaxies
\citep[e.g.,][]{finkelstein13,smit14,oesch15}.  As shown by
\citet{finkelstein13}, galaxies with very strong [O\,{\sc iii}]
emission must have low enough stellar metallicities such that the
stellar emission is hard enough to excite the oxygen in
the ISM to this state, yet must have high enough gas-phase metallicities such
that enough oxygen is available in the ISM.  For the
particular galaxy studied with an [O\,{\sc
  iii}] EW of 600 \AA, they found that the stellar metallicity was
likely $\sim$0.2--0.4 Z\sol, consistent with the modest levels of dust
attenuation observed in these bright systems.

\subsubsection{Implications of Dust at Early Times}
The presence of dust in bright/massive galaxies at such early times implies that
whatever mechanism dominates the creation of dust in distant galaxies
is already in place by $z \sim$ 7, only $\sim$750 Myr after the Big
Bang, and only 500 Myr after $z =$ 15, a reasonable guess for the epoch
of the first galaxies.  \citet{finkelstein12a} postulated that the
presence of significant dust at $z \sim$ 7 implied that dust
production was dominated by supernovae which have been occurring
since the epoch of the first stars, assuming that stars with
masses of 2--4 M\sol, which are the most efficience dust producers when in the AGB phase,
have not yet had time to evolve \citep[][and references
therein]{gall11}.  However, low-metallicity stars can enter the AGB
phase much sooner \citep[e.g.,][]{maraston05,karakas14}, and grain growth in the
ISM can occur at arbitrarily early times, as soon as the ISM is
enriched \citep[e.g.,][]{michalowski10,mancini15}, thus the picture is
more muddled than originally thought.

However, if dust can be found to be present at even earlier times,
that can begin to constrain the mechanisms of dust production in the
early universe.  \citet{dunlop13},
using the UDF12 dataset, found $\beta = -$1.8 $\pm$ 0.6 for faint
galaxies at $z =$ 9.
While this is consistent with the red colors of bright/massive
galaxies at $z =$ 4--7, it is also quite uncertain, thus conclusions
are difficult.  \citet{wilkins15} recently leveraged {\it
  Spitzer}/IRAC detections of several $z =$ 9 and 10 candidate
galaxies to place the first constraints on UV colors out to $z =$ 10, finding
colors only slightly bluer than similarly luminous galaxies at lower redshift.
These results hint that even at $z \approx$ 10, only 500 Myr from the
Big Bang, and 200 Myr from $z =$ 15, dust may be present.
While a definitive conclusion will certainly require {\it
  JWST}, these initial hints of dust production at $z \gtrsim$ 10
provide a tantalizing glimpse into the early production of heavy elements.

\subsection{The Growth of Galaxies}

\subsubsection{Galaxy Stellar Masses}

The distribution function of galaxy stellar masses is also a useful
tool to probe the galaxy physics discussed in \S 5.1.   In
particular, comparing the slope of the low-mass-end of the stellar
mass function to that of the underlying dark matter halo mass function
can provide an impression of the impact of feedback (supernovae, or
stellar radiative) on the suppression of star formation.  This has the
advantage over the UV luminosity function in that it directly probes
an intrinsic galaxy property without the troubling effect of dust. 
However, the measurement of galaxy stellar masses is an inferred
quantity, therefore the galaxy stellar mass function is a less
direct quantity to measure than the UV luminosity function.  I will
not discuss here the detailed methods for the measurements of galaxy
stellar masses, but a recent review is available in \citet{conroy13}.
In general, with decent photometry, galaxy stellar masses can be
measured accurately to within a factor of a few, typically a higher
accuracy than other photometrically-determined
properties (though a changing initial mass function can result in
  a much larger systematic uncertainty in the stellar mass).

The optical and near-infrared imaging typically used to discover
distant galaxies probes only the rest-frame ultraviolet, and is thus
dominated by the most massive stars present in these galaxies.
Stellar masses measured using only these filters can be subject to the
``outshining'' phenomenon, where an older, perhaps more massive
population is ``outshone'' by a more recent episode of
star-formation at the observed wavelengths.  Robust stellar mass measurements thus require longer
wavelength imaging.  While rest-frame near-infrared imaging would be
desirable, this falls longward of 10 $\mu$m at $z >$ 6, and is thus
presently inaccessible.  However, rest-frame optical imaging can still
constrain moderately older populations, and is supplied by {\it
  Spitzer}/IRAC imaging at 3--5 $\mu$m at these redshifts.  Direct detections at these
wavelengths are not necessarily required, as if the IRAC upper limits
occur at similar flux levels as the optical/near-infrared detections,
useful constraints on the stellar mass can still be obtained (by,
e.g., ruling out a massive older population, which would exceed the
IRAC upper limits).

Individual galaxy stellar masses were determined for galaxies at $z =$
6 from a variety of studies.  In particular, \citet{yan05,yan06} and
\citet{eyles07} studied $i$-band dropout galaxies in the HUDF and/or GOODS fields
which had clear IRAC detections, and found that they had stellar
masses of $\gtrsim$10$^{10}$ M\sol, surprisingly high for only $\sim$1
Gyr after the Big Bang.  \citet{stark09} also measured stellar masses
for a sample of a few dozen $z \sim$ 6 galaxies.  They were not all
detected in IRAC, thus the typical
stellar masses were lower, $\sim$10$^9$ M\sol\ for galaxies with M$_\mathrm{UV} =
-$20 to $-$21, though there was a significant scatter with some galaxies as
massive as 10$^{10}$ M\sol.  These massive galaxies with IRAC
detections were fit with strong 4000 \AA\ breaks, indicative of a
dominant population of older stars ($\sim$500 Myr; e.g.,
\citealt{eyles07}).  At higher-redshift,
\citet{finkelstein10} measured the stellar masses for $\sim$30 $z =$
7--8 galaxies in the HUDF (though see \citealt{egami05} for an earlier mass
measurement of an individual lensed $z \sim$ 7 candidate galaxy).  As these
galaxies were typically faint,
they were mostly not detected in IRAC, and had correspondingly low
stellar masses of $\sim$10$^{9}$ M\sol, an order of magnitude less
massive than that of an L$^{\ast}$ galaxy at $z =$ 3, though
intriguingly similar to the masses of narrowband-selected LAEs at $z
=$ 3--6.  \citet{labbe10} found similar stellar mass results in a
stack of $z =$ 7--8 galaxies in the HUDF, but through a stacked
analysis was able to measure a significant average IRAC flux for these galaxies,
inferring a somewhat surprising age of 300 Myr, implying that
the majority of stars in these galaxies formed prior to $z =$ 10.

One complicating issue to the inclusion of IRAC photometry in these fits is the
potential for a significant contribution from nebular emission lines.
In particular, if the [O\,{\sc iii}]
$\lambda\lambda$4959,5007 and/or H$\alpha$ emission were strong enough, they could add
significantly to the bandpass-averaged flux in the IRAC bands.
This was initially not thought to be a problem, as the majority of low-redshift
star-forming galaxies do not show extremely strong lines, and thus
these lines were not accounted for in these initial studies.  However, a bevy
of observational results have since shown that high-EW [O\,{\sc iii}] and
H$\alpha$ emission is common, if not ubiquitous amongst distant
star-forming galaxies, possibly indicating changes in the physical environments
of typical star-forming regions
\citep[e.g.,][]{schaerer09,schaerer10,finkelstein09c,finkelstein11a,shim11,stark13,labbe13,song14,song15,salmon15,smit15,kfinkelstein15}.
These strong lines boost the IRAC flux, mimicking the presence of a strong 4000 \AA\ break,
causing one to infer a higher mass and/or older population than may be correct.
\citet{schaerer09,schaerer10} showed that by including
nebular lines in the stellar population modeling process much lower
stellar masses and stellar population ages can be obtained, with ages
as young as 4 Myr possible for the stacked photometry of faint $z
\sim$ 7 galaxies from \citet{labbe10}.  \citet{salmon15} quantified
this further, finding that when including nebular emission
lines, the stellar masses of $z \sim$ 6 galaxies in the
CANDELS/GOODS-S field were systematically measured to be
$\sim$20--30\% lower than when emission lines were ignored.

Turning to the distribution of stellar masses, the first full,
completeness-corrected stellar mass functions at $z \geq$ 6 were
published by \citet{gonzalez11}.  They found low-mass-end slopes which
were surprisingly shallow, with $\alpha = -$1.44 at $z =$ 6 and
$-$1.55 at $z =$ 7, compared to simulations at the time, which often
predicted $\alpha < -$2 \citep[e.g.,][]{jaacks12a}.  The observations
thus implied that the implementation of feedback in the simulations
was possibly too weak.  However, these initial results were based on
small samples of only $\sim$60 galaxies total at $z =$ 6 and 7
combined, and also did not include the impact of nebular lines.  Recently, three separate studies on the stellar mass
function have been completed using a combination of different subsets
of the CANDELS data with the HUDF.  
\citet{duncan14} used the CANDELS
GOODS-S field to find much steeper faint-end slopes of $\sim -$1.9 at
these redshifts.  \citet{grazian15} used both the CANDELS GOODS-S and
UDS fields, to find a similarly steep slope at $z =$ 7, though they
found $\alpha = -$1.55 at $z =$ 6.  The most recent study, by
\citet{song15} also used two CANDELS fields (GOODS-N and GOODS-S), but
made use of deeper
IRAC data newly available from the S-CANDELS survey (PI Fazio).  By using simulations
to verify that their methods minimized systematic offsets in their
results, \citet{song15} found that the low-mass-end slope $\alpha$
steadily becomes steeper, from $-$1.53 at $z =$ 4 (similar to
\citealt{gonzalez11}) to $-$2.05 at $z =$ 7.  At the highest
redshifts, this slope is similar to that of the dark matter halo mass
function, implying that feedback may be less efficient at suppressing
star formation towards higher redshift.

\begin{figure}
\begin{center}
\includegraphics[width=3.5in]{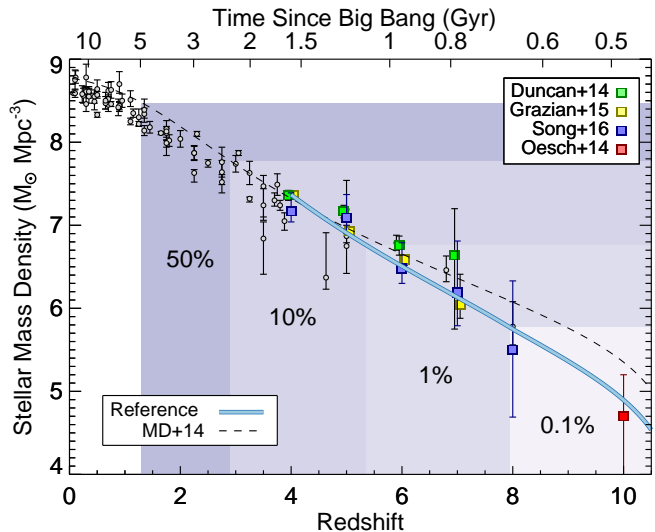}
\caption{The evolution of the total stellar mass density in the
  universe, all derived assuming a Salpeter IMF.  The low redshift
  measurements have a range of definitions, but I note that all high
  redshift measurements were obtained by integrating
  stellar mass functions from 8 $< log M/M$\sol $<$ 13 (the flatter
  slope of the low-mass end of the stellar mass function at $z <$ 4
implies that the lower limit of the integration is less important).  The gray points
show the data from the compilation of \citet{madau14}, while the other
symbols come from recent estimates of the stellar mass function at
high redshift, defined in the legend.  The blue and dashed black curves
show the stellar mass density obtained by integrating the SFR density
evolution from Figure~\ref{fig:sfrd}.  The left side of the shaded
regions denote the redshift at which the total stellar mass density
formed is equal to the listed percentage of the $z =$ 0 value.}
\vspace{-10mm}
\label{fig:smd}
\end{center}
\end{figure}

The integral of the stellar mass function provides the stellar mass
density of the universe at a given epoch.  As measurements at very
low-redshift from, e.g., the SDSS, are quite robust, it is interesting
to compare the stellar mass density at high redshift to low, to
investigate what fraction of the total stellar mass of the universe
exists at a given epoch.  This has been done in a variety of studies,
\citep[e.g.,][]{dickinson03,rudnick03,duncan14,grazian15,oesch15}, which are
highlighted in Figure~\ref{fig:smd}.  In this figure I also show the
stellar mass density obtained by integrating the evolution of the SFR
density from Figure~\ref{fig:sfrd}, showing both the updated reference
trend from this work, as well as the values from \citet{madau14}.  At $z <$
2, the stellar mass density derived in this way is systematically
higher by a small amount than that obtained directly from stellar mass
measurements.  \citet{madau14} discussed a number of possible causes
for this offset, including overestimation of the star-formation rates
(which could be the case if, for example, galaxies have a dust
attenuation curve similar to the Small Magellanic Cloud),
or underestimation of the stellar masses (due to lack of constraints
on older stars, and/or an evolving IMF).  However, at $z >$ 3, our
integrated SFR density shows excellent agreement with the observations of
\citet{grazian15}, \citet{song15} and \citet{oesch14}.  The values
from \citet{duncan14} show a modest significant difference at $z \geq$
5 due to their much steeper low-mass end slope (see discussion in
\citealt{song15}, and also \citealt{graus15}).  By comparing the values of my model at a given
redshift to that at $z =$ 0, I find that the universe formed 50\% of its stellar mass
by $z \sim$ 1.3, with $\sim$10\%, 1\%, 0.1\% and 0.01\% of the stellar
mass in place by $z =$ 2.9, 5.4, 8.0 and 10.2, respectively.  
Observations have therefore tentatively inferred the presence of
99.99\% of all the stellar mass which has ever formed in the universe,
though much of it remains to be directly detected.

\subsubsection{Star-Formation Histories}
While the measurement of the instantaneous stellar mass is relatively
straightforward, teasing out the growth of that stellar mass with time
is more difficult.  When measuring stellar population properties via
SED fitting, the star-formation history (SFH) is typically assumed to follow
some functional form.  Initial studies assumed a SFH which declined
exponentially with time (so-called tau models), which successfully
works at late times when many galaxies are in the process of 
quenching and gas inflow is likely less than at higher redshifts
\citep[e.g.,][and references therein]{maraston10}.  

However, several years ago it was noted that this assumption may not be valid at
higher redshifts.  \citet{maraston10} found that exponentially
declining models produced extremely young ages for a sample of
$z \sim$ 2 star-forming galaxies, while ``inverted-tau''
models (where the SFR increases exponentially with time) produced more
realistic results.  \citet{papovich11} also found that, by linking galaxies
from $z =$ 8 to $z =$ 3 using a constant number density tracking
technique \citep[e.g.,][]{vandokkum10,leja13,behroozi13c,jaacks15},
galaxies on average have rising star-formation histories (see also \citealt{salmon15}).
\citet{reddy12} reached a similar conclusion by comparing SED-fitting
derived SFRs to UV$+$IR SFRs, finding that when a declining SFH was
assumed, the SED-fitting derived SFR was too low.  A bevy of
theoretical studies have reached similar conclusions: at $z \gtrsim$ 3,
the average SFRs of galaxies grow roughly steadily with time \citep[e.g.,][]{finlator07,finlator11,jaacks12b}.

Such growth cannot continue unchecked.  In particular, at $z \lesssim$ 3
galaxies begin to quench, thus clearly such a rising SFH is not
appropriate for all galaxies at all times.  In an advanced 
MCMC analysis of galaxy evolution from $z =$ 8 to 0,
\citet{behroozi13b} found that while galaxies at $z >$ 3 appeared to
have SFRs which increased with time as a power law \citep[similar to
][though with a steeper exponent]{papovich11}, at $z \sim$ 2-3
there existed a mass-dependent redshift where the SFHs began to
turnover and decline.  This turnover occurs as late as $z =$ 0.5 for
lower-mass galaxies, thus even at relatively low redshift there may
be a mix of rising and falling SFHs.  They proposed either a double power law SFH
(Eqn. 23 in \citealt{behroozi13b}), or a combination of a power-law
and declining exponential, either of which should be appropriate for galaxies at
any epoch.

Finally, I note that the majority of the discussed studies are
considering ensembles of galaxies, therefore the SFHs which were derived
represent the average growth of stellar mass in these galaxies.  The
SFH for any individual galaxy may be quite stochastic, depending on a
variety of effects, including environment, merger activity, gas
accretion, etc.  Thus, caution should be used when considering the
evolution of individual galaxies.

\subsubsection{Evolution of Galaxy Specific Star-Formation Rates}
The specific star-formation rates (sSFR; SFR/stellar mass) of galaxies
are an excellent tracer of galaxy growth, as this quantity normalizes the growth of
new stars by the current stellar mass.
Theoretical simulations predict that the sSFR at fixed stellar mass
should be rising with increasing redshift, as such models predict that
star formation is governed by cold gas accretion, and should therefore
increase with increasing redshift \citep[e.g.,][]{neistein08,dave11,krumholz12}.
It was thus surprising when the first measurements of the evolution of the sSFR at high
redshift showed it to be roughly flat from $z \sim$3 to 7
\citep{gonzalez10}.  Subsequent work by \citet{stark13} considered the
impact of nebular emission, which was not accounted for in
\citet{gonzalez10}, and found hints that the sSFR may be rising from
$z =$ 5 to 7.  \citet{gonzalez14} performed an updated analysis with a
larger sample, including nebular lines and varying SFHs (including
those which increase with time) and also found evidence for a rising
sSFR with increasing redshift, though with a redshift dependance much
weaker than simulations.  

Most recently, \citet{salmon15} found a
steeper increase in sSFR with increasing redshift from $z =$ 4--6.
Their fiducial analysis tracking galaxies at constant mass was not a
great match to simulations, but when they instead considered tracking
galaxies with an evolving cumulative number density to more accurately
track progenitors and descendants \citep{behroozi13c}, they found an
excellent match to both \citet{neistein08} and \citet{dave13} in both
normalization and slope.  Therefore, at present there is no
significant discrepancy with models at high redshift.  However,
current results at $z >$ 4 have large uncertainties, and, at the
highest redshifts, are based on small samples, thus there is room for
improvement.  Additionally, there is a significant discrepancy between
models and observations at 0.5 $< z <$ 2 (see Fig.\ 9 of
\citealt{gonzalez14}), where observed sSFRs evolve shallowly down to $z
\sim$ 1 then fall off quickly towards lower redshift, while simulations predict a smoother
evolution.  Thus, the evolution of the sSFR will likely
remain an active area of inquiry at all redshifts.

\subsubsection{Evolution of the Stellar-Mass to Halo-Mass Ratio}
As a final consideration into the growth of galaxies, I consider the
dependance of galaxy growth on the underlying dark matter halo mass.
Because the growth of dark matter halos is well understood via
simulations such as the Millenium, Bolshoi, and Illustris projects \citep[e.g.,][]{springel05,klypin11,vogelsberger14}, linking galaxies to these halos can
allow us to gain insight into the efficiency with which galaxies turn
gas into stars.  A common way to do this is to calculate the
stellar mass to halo mass (SMHM) ratio.  The halo mass can best be observationally
inferred via galaxy clustering, which allows one to
estimate the dark matter halo mass for a specific sample of galaxies
by comparing their spatial distribution to the underlying dark
matter halo distribution.  In practice, this requires large
samples of galaxies, particularly if one wishes to split their sample
into several bins by, for example, stellar mass (see
\citealt{overzier06} and \citealt{baronenugent14}  for initial measurements at $z
\geq$ 6).  Lacking sufficient galaxy numbers, an alternative method
has been derived known as abundance matching
\citep[e.g.,][]{moster10,behroozi10}.  This technique assumes that a
galaxy property (for example, the stellar mass, or UV luminosity) is a
monotonic function of the halo mass, such that the most
massive/luminous galaxies live in the most massive halos.  One can
then match a galaxy distribution function (i.e., the stellar mass or
UV luminosity function) to a dark matter halo mass function, linking
galaxies to halos at a constant cumulative number density, to derive
the halo mass for a given stellar mass or UV luminosity.

\citet{behroozi13b} used abundance matching (as part of their MCMC
analysis considering a wide range of observables) to constrain the
evolution of the SMHM relation from $z =$ 0 to 8.  They found the
interesting result that this relation seems roughly constant from $z
=$ 0 to 4, with a peak SMHM ratio of $\sim$1-2\% at halo masses of
$\sim$10$^{12}$ M\sol.  There are two interesting points to take away
from this.  First, the apparent constant halo mass where the SMHM reaches
its peak value implies that there is a characteristic halo mass where galaxies
form stars most efficiently.  At lower and higher halo masses, the
feedback effects discussed in \S 5 are likely stronger (at least at $z <$ 4).
Second, the peak ratio of $\sim$1-2\% is much lower than
the cosmic fraction of available baryons in halos, which should be
$\sim$17\% (for WMAP7 cosmology, \citealt{komatsu11}).  Thus, galaxies
are not efficient star formers, as the majority of gas does not
make it into stars.  \citet{behroozi13b} continued this analysis to
$z >$ 4, finding that the SMHM relation began to evolve, in that at a constant halo
mass, a $z >$ 4 galaxy will have a higher stellar mass than at lower
redshift, perhaps implying a change in the efficiency governing star
formation.

This result was based on very early observational
results at $z >$ 6, including pre-2013 luminosity and
stellar mass functions, which differ 
significantly from the most recent results.  \citet{finkelstein15b}
performed an updated, albeit much simpler, analysis, performing abundance
matching using their updated luminosity
functions.  They derived the halo mass at
M$_\mathrm{UV} = -$21 at $z =$ 4--7 (approximately L$^{\ast}$ at these
redshifts), finding that the halo masses were lower (by
0.6 dex) at $z =$ 7 compared to $z =$ 4.  Measuring the stellar masses
for these bright galaxies, they found them to remain roughly constant
across this redshift range (log M/M\sol\ $\sim$9.6--9.9), thus the SMHM
ratio of galaxies at this constant luminosity/stellar mass increased
towards higher redshift.  Normalizing by the cosmic abundance of
baryons, they found that the fraction of baryons converted into stars
in these galaxies was roughly 2.3$\times$ higher at $z =$ 7 than at $z
=$ 4, again implying that higher redshift galaxies are more efficient
star-formers.  This provides further evidence that the physics shaping
galaxy evolution may be evolving at $z >$ 4.  The explanation need not
be exotic; rather, such changes may be expected, as a changing gas
density, black hole accretion, and/or impact of SNe feedback may be
expected as the expansion of the universe plays back in reverse.
Further theorietical and observational work is needed to clarify
exactly the physical processes which may be changing.  In particular,
these results are based entirely on the measurement of luminous
matter.  A direct measurement of the evolution of the gas reservoirs
in these galaxies will provide a much clearer insight into how these distant
galaxies are converting their gas into stars.

\section{Reionization}

At the time that CMB photons began their long journey to our
telescopes, the universe had expanded and cooled enough for hydrogen
gas to become neutral.  This was the state for the next few hundred
million years --- a period known as the ``dark ages''.  Eventually, baryonic
matter cooled and condensed in dark matter halos to form the first
stars and galaxies in the universe.  These objects provided a key
source of ionizing photons, ionizing
the cosmic haze of neutral intergalactic gas in a process known as reionization.
Understanding how reionization proceeds, both the evolution of the
neutral fraction of the IGM with time as well as its spatial
variation provides key constraints on
the nature of these first luminous sources in the universe.
In this section I present current observational constraints on reionization, focusing on
those derived directly from galaxies at $z >$ 6, with a discussion of
the current limiting factors.

\subsection{The Timeline of Reionization}

\subsubsection{Ly$\alpha$ Forest}

The most robust constraints on the end of reionization comes
from the measurement of the Gunn-Peterson trough in the spectra of
distant quasars.  By measuring the transmission in the Ly$\alpha$
forest of $z >$ 6 quasars, the neutral fraction of the IGM has been
constrained to to be $\ll$1\% at $z \approx$ 6
\citep[e.g.,][]{fan06}.  As individual quasars probe only single sightlines, there
are bound to be variations, and in fact \citet{becker15} recently
measured a significant trough in a $z =$ 6.0 quasar, implying a
partially neutral IGM all the way down to $z \sim$ 5.5 on that line-of-sight.  Additionally,
from the variations in transmission from quasar spectra,
\citet{becker15} find that the variations in the UV background at $z
>$ 5.6 require significant variations in the IGM neutral fraction.
The observations do not become consistent with an IGM opacity arising
from only the matter density field until $z \sim$ 5.5.  Thus, while
reionization is likely mostly complete by $z \sim$ 6, current
observations do not constrain it to be fully complete until $z \lesssim$ 5.5.  Quasars are
somewhat less useful to probe deep into the epoch of reionization,
however, as this measurement saturates at a relatively small neutral fraction
($\sim$10\%), and only a single $z \geq$ 7 quasar is currently known \citep{mortlock11}.

\subsubsection{Cosmic Microwave Background}

Reionization is expected to be a prolonged process, with the overdense
regions which formed stars first beginning the process at 
early times, with neutral gas lingering in the densest filaments until
late times \citep[e.g.,][]{finlator09}.  While it is difficult to obtain strict constraints on the beginning or
end of reionization, measurements of the optical depth due to electron scattering from the
cosmic microwave background (CMB) does provide observational constraints
on the total number of free electrons along the line of sight to the
CMB, which can be used to estimate the midpoint of reionization.
Initial results from the {\it Wilkinson Microwave Anisotropy Probe}
({\it WMAP}) measured a large optical depth, implying a midpoint for
reionization at $z =$ 17 $\pm$ 5 \citep{spergel03}, which led to a
bevy of work imposing significant early star formation.  However,
subsequent analysis with additional {\it WMAP} data resulted in lower
optical depths, and the latest results from {\it Planck} find
$\tau_\mathrm{es} = $0.066 $\pm$ 0.016, implying $z_\mathrm{reion} =$
8.8$_{-1.1}^{+1.2}$ \citep{planck15}, implying a later start to reionization.
Thus, it appears likely that reionization started not
much before $z \sim$ 10, and was predominantly completed by $z \sim$
6.  

\subsubsection{Ly$\alpha$ Emission}

Another probe into reionization is the measurement of
Ly$\alpha$ emission from galaxies, as the resonant scattering nature of
Ly$\alpha$ photons means that their detection is impaired by a
significantly neutral IGM.  This was initially done with
Ly$\alpha$ emitters discovered via narrowband imaging surveys.  The
latest studies do find evidence of a drop from $z =$ 5.7 to 6.5, where
Ly$\alpha$ emitters are less luminous (and/or less common) at higher redshift \citep[e.g.,][]{ouchi10}.
Attempts to continue these narrowband studies to higher redshift have met
with difficulty, as very few reliable candidates have been discovered
\citep[e.g.,][; see \S 3.3]{tilvi10,krug12,faisst14}.

A more recent method involving Ly$\alpha$ is to perform
spectroscopic follow up of continuum-selected galaxies to measure their
Ly$\alpha$ emission.  This is powerful as it relies on a known sample
of robustly selected galaxies, and even non-detections are very
constraining as one can model upper limits in the data and thus study
the non-detections on a galaxy-by-galaxy basis (unlike
narrowband studies, where the high-redshift galaxies are only found if
they have strong Ly$\alpha$ emission).  This method is possible
because the fraction of continuum-selected galaxies
with detected Ly$\alpha$ emission rises from $\sim$30\% at $z =$ 3 to
60-80\% at $z =$ 6 \citep[e.g.,][]{shapley03,stark10,stark11}.
This increase is likely due to decreasing dust attenuation (\S 6.1), which should continue to
higher redshifts, thus Ly$\alpha$ \emph{should} be nearly ubiquitous
amongst star-forming galaxies at $z >$ 6.  An observed drop in the
fraction of galaxies exhibiting Ly$\alpha$ emission (typically
parameterized by comparing the EW distribution at higher redshift to
that at $z =$ 6)
may signal a rapidly evolving IGM neutral fraction.

As discussed in \S 4.2, such a drop has been exactly what has been
seen.  While it may be that a larger fraction of the samples are
interlopers at these redshifts, that seems unlikely given the high
confirmation fraction of similarly-selected $z \sim$ 6 galaxies \citep[e.g.,][]{pentericci11}.
Converting this observed evolution of the Ly$\alpha$ EW
distribution into constraints on reionization is less direct,
requiring advanced modeling with a number of assumptions.  
In particular, one needs to know the emergent Ly$\alpha$ profile,
which depends on the ISM H\,{\sc i} column density and geometry, as
well as the kinematics of the ISM.  Reasonable assumptions on these
properties can be made based on observations at $z \sim$ 2
\citep[e.g.,][]{chonis13,hashimoto13,song14}.  One then needs to
model the IGM, which can be done with semi-numerical models, or full-on
cosmological hydrodynamic simulations
\citep[e.g.,][]{dijkstra11,jensen13}.  These models include not only
diffuse neutral gas in the epoch of reionization, but also attenuation
post-reionization from dense ionized gas.  When such
models are compared to observations they
imply a volume-averaged neutral fraction of $\sim$50\% at $z =$ 7 \citep[e.g.,][]{pentericci14}.
This is a drastic change from the apparently fully ionized universe at
$z \sim$ 6, and is too fast compared to what is seen in theoretical
reionization models (see also \citealt{treu13} and \citealt{tilvi14} for
qualitatively similar results at $z \sim$ 8).  I direct the reader to the recent
review on this topic by \citet{dijkstra14} for more details on the
modeling process.

As discussed by \citet{dijkstra14}, there are a number of alternatives
which may allow the current observations to be consistent with a lower
volume-averaged neutral fraction at $z \sim$ 7.  One likely candidate
is that the observed evolution is primarily due to an increase in the
opacity of self-shielding Lyman limit systems.  
Prior to the completion of reionization, the ionizing background
is smaller than just after reionization, which makes it easier for these systems to
self-shield \citep[e.g.,][]{bolton11,mesinger15}.  At this time, dense
filaments will ``thicken'', increasing the optical depth to
Ly$\alpha$ photons.  The galaxies we attempt to observe in Ly$\alpha$ likely lie in
overdense environments, increasing the likelihood that a line-of-sight
to a given galaxy will pass through one (or more) nearby filaments.
\citet{bolton13} found that when including these systems in their
models, a volume-averaged neutral fraction of only 10-20\% was needed
to explain the observed Ly$\alpha$ evolution (though see \citealt{mesinger15}).

A final possibility is that at least some of the observed drop in
Ly$\alpha$ detectability is caused by the properties of galaxies
themselves.  While at lower redshift dust can significantly
attenuate Ly$\alpha$, dust is unlikely to provide a higher Ly$\alpha$
attenuation at $z \geq$ 7 than at $z =$ 6 (see \S 6.1).  However,
Ly$\alpha$ is also highly sensitive to the column density of neutral
hydrogen, and if the potential evolution of the SMHM relation
discussed in \S 6.3 is due to increased gas reservoirs in galaxy, then it may be
that at higher redshift more Ly$\alpha$ photons suffer additional
resonant scattering within the galaxy, reducing the Ly$\alpha$ observability with current
telescopes (see also \citealt{papovich11} and \citealt{finkelstein12b}).  
Finally, it may be that galaxies at higher redshift
have higher escape fractions of ionizing photons; every ionizing
photon which escapes is one less photon which is converted into
Ly$\alpha$, lowering the intrinsic Ly$\alpha$ luminosity
\citep[e.g.,][]{dijkstra14b}.  However, although relatively high
escape fractions have occasionally been observed
\citep[e.g.,][]{steidel01,shapley06,iwata09,vanzella10,nestor11}, some
may be due to superimposed foreground objects \citep[e.g.,][]{vanzella12}, and
the majority of models predict that the currently observable galaxies
leak essentially zero ionizing photons \citep[e.g.,][]{paardekooper15,ma15}.

Regardless of the changing IGM or ISM, the few
detections of Ly$\alpha$ which we do have at $z >$ 7 implies that some
combination of global or local IGM state combined with ISM geometry
and kinematics will allow
focused observational studies with future facilities to use
Ly$\alpha$ as a spectroscopic indicator deep into the epoch of reionization.

\subsection{Constraints from Galaxies}

The declining quasar luminosity function at $z >$ 3 (e.g., Richards et
al.\ 2006) has led to the conclusion that 
star-formation in galaxies produced the bulk of ionizing photons
necessary for reionization.  A key measure is thus a
cosmic ``census'' of distant galaxies, measuring their rest-UV light,
then inferring the likely ionizing photon emissivity to see if the
observed star-formation at a given epoch can sustain an ionized IGM.
If one assumes that galaxies do indeed provide all the needed ionizing
photons, this analysis provides an independant measure into the
timeline of reionization.

There are a number of assumptions which need to be made to convert
rest-frame UV observations of galaxies into a constraint on
reionization.
\begin{itemize}
\item Stellar population: As one observes non-ionizing UV photons, 
  that observable needs to be converted to the emergent ionizing
  radiation, which requires assuming a stellar
  population.  The primary considerations here are the initial mass
  function and the metallicity.  The conversion factors do not vary
  wildly if one assumes a Salpeter IMF and non-zero metallicity
  ($\sim$ 20-30\%; e.g., \citealt{finkelstein12b}).  However, a
  changing IMF, particularly one which results in more massive stars,
  could produce a much higher number of ionizing photons per unit UV
  luminosity making it easier for galaxies to reionize the IGM.
\vspace{1mm}

\item Escape fraction: Only ionizing photons which
  escape a galaxy are available to contribute to reionization, thus
  one needs to assume an escape fraction of ionizing photons
  (f$_{esc}$).  This quantity cannot be directly observationally
  ascertained in the epoch of reionization due to attenuation along
  the line of sight.  Studies at $z <$ 4 have found that the majority
  of galaxies have little-to-no escaping ionizing radiation \citep[e.g.,][]{siana10}, though
  examples of escape fractions on the order of 20\% have been found
  \citep[e.g.,][]{nestor11}.  \citet{finkelstein12b} showed that
  values greater than 13\% at $z =$ 6 would violate the ionizing photon
  emissivity obtained from observations of the Ly$\alpha$ forest.
  Most simulations predict that
  galaxies massive enough to be currently observed have very low
  escape fractions, and values $>$20\% only occur in the very lowest
  mass halos \citep[e.g.,][]{paardekooper15}.  Values of $\sim$20\%
  are frequently assumed in the literature when obtaining reionization
  constraints from galaxies, which may thus be optimistically too high \citep[e.g.][]{finkelstein10,robertson13,robertson15}.
\vspace{1mm}

\item Limiting Magnitude: When integrating the UV luminosity function,
  one needs to assume a limiting magnitude.  An absolute limit would
  be the magnitude of a single O star ($\sim -$5); however, galaxies likely have
  their star formation suppressed in more massive halos, due to
  photoionization of their gas from star formation in nearby halos.
  Galaxies are commonly observed at $z \sim$ 6--8 down to magnitudes of
  M$_\mathrm{UV}$ $= -$17, thus it may
  be reasonable to assume fainter values of $-$13 or $-$10, as is
  often done in the literature.  Recent simulations have found that
  the luminosity function may break at M$=-$14 to
  $-$16 due to a combination of physical processes
  \citep{jaacks13,oshea15}, or at $-$13 by
  simulating the progenitors of galaxies in the Local Group
  \citep[e.g.,][]{mbk15}.  However, initial results from the Hubble
  Frontier Fields lensing survey show that the luminosity function
  continues steeply to M$_\mathrm{UV} = \sim$13 at $z =$ 6 \citep{livermore16}, and
  future studies with the full Frontier Fields dataset should produce
  even better constraints.
\vspace{1mm}

\item Clumping Factor of IGM: The density of gas in the IGM is a key
  parameter, as it is more difficult to keep dense gas ionized than
  diffuse gas, as the recombination time is shorter.  This is
  typically parameterized by the clumping factor (C), where higher
  values of $C$ require more ionizing photons to sustain
  reionization.  While $C$ was originally thought to be quite high
  ($\sim$ 30; \citealt{gnedin97}), more recent models, with a better
  understanding of the interface between galaxies and the IGM, find
  lower values ($<$5; e.g., \citealt{pawlik09,finlator12,pawlik15}).
  The most recent models of \citet{pawlik15} find a clumping factor
  which is $<$3 at $z >$ 8, rising to $C=$4 at $z =$ 6.  Most studies
  below assume $C =$ 3.
\end{itemize}

With these assumptions, one can use a model of reionization to
determine whether the galaxy population at a given redshift can
sustain an ionized IGM.  The model of \citet{madau99} is commonly
used, which calculates a redshift-dependent limiting UV
luminosity density to sustain an ionized IGM, assuming a stellar
population term, and values for f$_{esc}$ and $C$.  One can then
integrate an observed luminosity function (to an assumed limiting
magnitude), and compare to this model to explore the contribution of
galaxies to reionization.  This has been done in a variety of studies;
here I will review a few recent results.

Using new samples of $z =$ 6, 7 and 8 galaxies from the CANDELS and
HUDF surveys, \citet{finkelstein12b} examined the contribution to reionization
solely from observed galaxies (i.e., with no extrapolation to fainter
magnitudes).  They found that if the typical galaxy had an escape
fraction of $\sim$30\%, then the observed population of galaxies could
sustain an ionized IGM by $z =$ 6, in contrast to previous studies
which found that fainter galaxies were required
\citep[e.g.,][]{bunker04,yan04}.  By using limits on the ionizing
emissivity from observations of the $z \approx$ 6 Ly$\alpha$ forest,
\citet{finkelstein12b} found that \emph{if} the UV luminosity function
continues with the observed steep slope to $M_\mathrm{UV} = -$13, then the
typical escape fraction could be no higher than $\sim$13\%.  However,
even with this modest escape fraction, galaxies could still sustain an
ionized IGM by $z =$ 6 if the luminosity function extended down to
such faint magnitudes.  \citet{robertson13} performed a complementary
analysis, using not only the UV luminosity function, but also
measurements of the stellar mass density and the electron scattering
optical depth from the {\it WMAP} measurements of the CMB as
constraints.  They found similar
results, in that by assuming f$_{esc} =$ 20\%, galaxies can 
sustain an ionized IGM by just after $z =$ 6, but that fainter galaxies would be
required to complete reionization at earlier times.

Over the next few years, the advent of {\it Planck} coupled with the
large improvement on the precision of the UV luminosity function from the completed
CANDELS survey allowed these constraints to be tightened.
\citet{finkelstein15} used their updated luminosity functions to
revisit the constraints placed on reionization from galaxies.  Assuming
a fiducial model with $C =$ 3 and f$_{esc} =$ 13\%, they found that
galaxies suggest that reionization was largely complete by $z =$ 6,
and that the universe was primarily neutral by $z >$ 10.  Their
observations constrained a midpoint of reionization (defined as the
epoch when the volume-averaged ionized fraction reached 50\%) of 6.7
$< z <$ 9.4 (at 68\% confidence).  Although this analysis did not use
the recent observations from {\it Planck} as constraints, the electron
scattering optical depth derived from this reionization history is in
excellent agreement with the {\it Planck} observations, implying that
a significant contribution to reionization from galaxies at $z >$ 10
is not needed.  \citet{robertson15} updated their previous analysis
with the {\it Planck} constraints and the latest luminosity function
results, found a similar result in that a reionization process which
starts at $z \approx$ 10 and completes by $z \approx$ 6 is consistent
with both the observations of galaxies and the {\it Planck} observations of the CMB.
Finally, \citet{bouwens15b} recently performed a complementary
analysis, eschewing constraints from the luminosity functions of galaxies in their analysis to
derive independant constraints on the ionizing sources.  As
constraints, they combined observations of the Ly$\alpha$ forest from
quasar spectra with the {\it Planck} optical depth measurements, as
well as observations of the evolving Ly$\alpha$ emission from
galaxies.  They derived an evolution in the ionizing background which
was consistent with the ionizing emissivity derived from observations
of galaxies (with $f_{esc} \sim $10-20\%) even though galaxies were
{\it not} used as constraints.  This again suggests that galaxies are
the major producers of the ionizing photons which drove the
reionization of the IGM, provided such ionizing escape fractions are achievable.

\subsection{Reference Reionization Constraints}
The studies discussed above uniformly arrive at a similar conclusions:
galaxies were responsible for reionization, and reionization completed
by $z \sim$ 6.  However, the time-evolution of the neutral fraction
during reionization is less well constrained.  For example, the constraints on the mid-point
of reionization from \citet{finkelstein15} of $\Delta z =$ 2.7 are
largely due to the faint-end slope uncertainties from the luminosity
function.  As discussed in \S 5.3, we can improve upon these
constraints with my derived reference UV luminosity function.  For
example, the fractional uncertainty on the faint-end slope
($\sigma_\alpha/\alpha$) from \citet{finkelstein15} and
\citet{bouwens15} was 10\% and 6\%, respectively, while from my
reference UV luminosity function at $z =$ 7 we I find
$\sigma_\alpha/\alpha =$ 2\%.  Therefore, to conclude this
discussion of reionization I explore the improvements in the
constraints on the contribution of galaxies to reionization available
when I use luminosity functions derived in \S5.

\begin{figure}
\begin{center}
\includegraphics[width=3.5in]{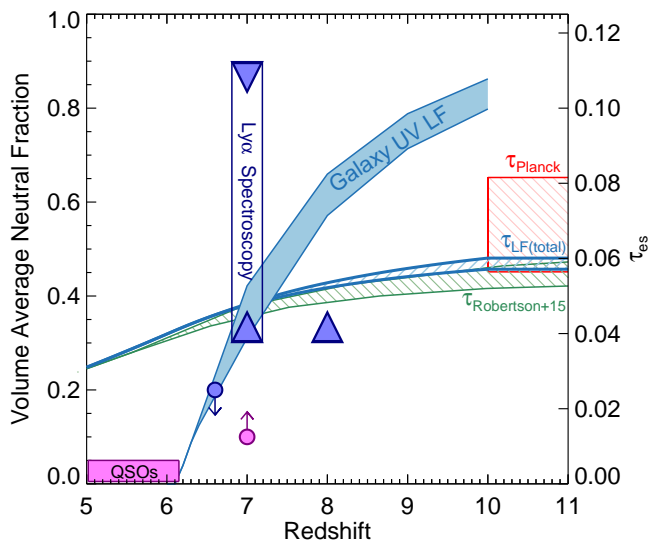}
\caption{The volume-averaged neutral fraction as a function of
  redshift, with the 68\% confidence range on constraints from the
  integral of the reference UV luminosity functions shown by the
  shaded blue region.  An upper limit from the Ly$\alpha$ narrowband survey
  by \citet{ouchi10} is shown by the blue circle, while constraints
  via Ly$\alpha$ spectroscopy at $z =$ 6.5--7 by \citet{pentericci14}
  and at $z =$ 7.5--8 by \citet{tilvi14}
  are shown by the blue arrows.  Constraints via quasars are shown by
  the magenta regions (from \citealt{fan06} at $z <$ 6, and \citealt{bolton11}
  at $z =$ 7).  The Thomson optical depth to electron scattering
  derived from the reference UV luminosity function constraints is
  shown by the hatched blue region, and is consistent
with constraints from {\it Planck}, shown by the hatched red region
(both corresponding to values on the right-hand vertical axis), as
well as the recent study by \citet{robertson15} shown in green.
This fiducial reionization history constrains reionization to be
$<$10\% complete by $z \sim$ 10, and $>$85\% complete by $z \sim$ 7.}
\vspace{-10mm}
\label{fig:reion}
\end{center}
\end{figure}

To derive this reionization history, I follow the methodology of
\citet{finkelstein15}, assuming $C =$ 3 and f$_{esc} =$ 13\% (which does
not violate constraints on the Ly$\alpha$ forest at $z =$ 6).  I then
integrate the fiducial reference UV luminosity functions from \S 5.3
at each redshift to a limiting magnitude of $M_\mathrm{UV} = -$13 to derive the UV
luminosity density, using the results from the MCMC chains to derive
uncertainties on these values.  Using the model of \citet{madau99} with the above
assumptions, I then calculate the volume-averaged neutral fraction
(x$_\mathrm{HI}$) at each redshift, propagating through the uncertainties on the UV
luminosity density to derive a 68\% confidence range on
x$_\mathrm{HI}$.  The resultant 68\% confidence reionization history
is shown in Figure~\ref{fig:reion}.

The tighter constraints on the luminosity functions available when
using all available data lead to more constraining results on the
evolution of the volume averaged neutral fraction x$_{HI}$ than
previous studies utilizing only galaxy data.  This analysis
constrains x$_\mathrm{HI} =$ 0\% at $z =$ 6 and x$_\mathrm{HI} >$ 80\% at $z =$ 10,
with the 68\% confidence constraint on the midpoint of reionization being 7.3 $< z <$ 7.7.
I calculated the Thomson optical depth to electron scattering from both my upper and
lower 68\% confidence constraints on the neutral fraction, finding
$\tau_{es} =$ 0.057 -- 0.060 (blue hatched region in Figure~\ref{fig:reion}).  This is
in consistent with the low end of the observational constraints from {\it
  Planck} of $\tau_{es} =$ 0.066 $\pm$ 0.012.  This is particularly
striking as the {\it Planck} measurements were not used as an input constraint on the galaxy
luminosity function-based reionization history.  Rather, the galaxy
observations, combined with assumptions on the limiting magnitude,
clumping factor, and escape fraction, produce a result which is
consistent with the {\it Planck} observations.  The results found here
are highly consistent with those from \citet{robertson15}, as shown by
the green shaded region in Figure~\ref{fig:reion}.  If the true value
of $\tau_{es}$ is closer to the higher end of the {\it Planck}
constraints, it could leave room for modest star-formation activity at
$z >$ 10, while if it is closer to the value implied by our
observations at the lower end of the {\it Planck}
constraints, it leaves
little room for a significant number of free electrons at higher redshift.

Comparisons to other complementary observations are also shown in Figure~\ref{fig:reion}.
Current results from quasars imply a fully ionized universe at $z <$ 6
\citep[e.g.,][]{fan06}, while the single known quasar at $z =$ 7
implies a volume averaged neutral fraction $>$10\%.  On the other hand, \citet{ouchi10}
found x$_\mathrm{HI} <$ 20\% at $z =$ 6.6, using the observed
evolution of the Ly$\alpha$ luminosity function via narrowband
imaging.  Using the incidence of Ly$\alpha$ emission in spectroscopic
followup of known high-redshift sources, \citet{pentericci14} and
\citet{tilvi14} constrained 30\% $< x_\mathrm{HI} <$ 90\% at $z =$ 7,
and x$_\mathrm{HI} >$30\% at $z =$ 8, respectively.  
While both the quasar and Ly$\alpha$ luminosity function-derived
constraints are fully consistent with my inferred
reionization history, taken at face value the $z =$ 7 Ly$\alpha$ results are in modest tension.  These studies imply a
significant neutral fraction of $>$30\% by $z =$ 7, while the luminosity
functions seemingly allow at most 20\% of the IGM to be neutral.

However, there is presently enough uncertainty in all of these
measures such that the tension can be easily resolved.  For example,
reducing the assumed escape fraction to 10\% increases the neutral
fraction at $z =$ 7 to be consistent with the lower limit from
\citet{pentericci14}, while still generating enough free electrons to
satisfy the {\it Planck} constraint.  Likewise, the constraints via
Ly$\alpha$ spectroscopy are relatively poor, due both to small sample
sizes as well as modeling uncertainties, both which should improve
significantly in the coming years as observers build up more robust
samples of $z >$ 7 spectroscopic observations.  The escape fraction is
likely the biggest wild card, and while here I impose an upper limit
set by Ly$\alpha$ forest constraints at $z =$ 6, we have no reasonable
constraint on the {\it lower} limit of this quantity.  Future studies
at $z =$ 2--3, as well as more advanced modeling, should shed light on
this issue in the coming years.  Should the escape fractions be
universally lower than currently assumed, it may be that galaxies play
less of a role in reionization than is currently thought, opening the
door for a larger contribution from low-luminosity AGN \citep[e.g.,][]{giallongo15,madau15}.

\section{Prospects over the coming decade}

In this review I have focused on the extensive knowledge we have
gained about the distant universe via observations.  Galaxy surveys can now
discover galaxies out to redshifts greater than 10, with spectroscopic
redshifts now possible out to $z >$ 8.  The shapes of
the galaxy rest-frame UV luminosity functions
at $z =$ 4--7 have become highly constrained, enabling their
use as tools to study galaxy evolution.  Deep mid-infrared photometry
allows robust stellar masses to be measured out to $z >$ 8, perhaps accounting
for the first 0.01\% of the stellar mass to ever have formed in the
Universe.  A better understanding of the dust attenuation in galaxies
allows us to not only better track the build-up of heavy elements in
the distant universe, but also, combined with robust estimates of the
luminosity function faint-end slope, to constrain the total amount of
star-formation at each epoch, which may remain relatively flat out to
$z =$ 8 with a possible decline at higher redshift.

The exquisite observational results which have been summarized here
are representative of the truly outstanding amount of resources, both
human and technological, which have been applied to this epoch in the universe.
However, in many respects, we have only touched the tip of the
iceberg, which makes the coming decades ripe for a new era of
discovery.  To conclude this review, I will focus on what improvements
we can expect with the coming generations of new observational facilities.

\subsection{Probing the Dark Side with ALMA}
Much of the observational studies above have focused on observations
of the starlight from galaxies.  While this is a direct probe into
galaxy evolution, it leaves out a key component, which is the
ubiquitous gas which is fueling the star formation we observe.
Observations at lower redshifts imply that galaxies in the distant
universe are substantially more gas-rich than at low redshift
\citep[e.g.,][]{tacconi10,papovich11,papovich15}.  This could have strong
implications on the evolution of such distant galaxies, and indeed
there are hints, discussed above, that the star-forming properties of
distant galaxies may be different than expected, perhaps due to large
gas reservoirs.  The advent of ALMA now enables direct studies into the gas properties of
distant ($z >$ 6) galaxies.  While semi-direct probes of molecular gas
in normal galaxies such as CO emission may observationally highly
expensive \citep[e.g.,][]{tan13}, more indirect probes such as the
dust emission \citep[e.g.,][]{scoville15} and/or dynamical modeling
via resolved [C\,{\sc ii}] emission can shed light on gas at high
redshift.  While ALMA observations have only just begun, recently
improved programs such as the mapping of the HUDF (PI Dunlop), and of a portion of
the GOODS-S field (PI Elbaz) should begin to allow us to probe the dark side of
the distant universe.  A better understanding of the gas properties of
these distant galaxies will shed light on the physical mechanisms
fueling galaxy evolution, perhaps explaining observed trends in galaxy
sSFRs and the possibly evolving SMHM relation.

\subsection{Action Items for {\it JWST}}
The launch of {\it JWST} is now less than three years away, thus the
time is now for identifying the key issues for study immediately
following the launch.  While there are clearly many important issues
that {\it JWST} will tackle at $z >$ 6, here I focus on
a few key issues.  The first issue is the faint-end slope of the
galaxy UV luminosity function.  Although the reference luminosity
functions derived here have improved the constraints on the faint-end
slope at $z =$ 7, there remains a large uncertainty in the shape (and
extent) of the faint end at 8 $< z <$ 10, which leaves possible a wide
range of possible IGM neutral fractions, as shown in Figure~\ref{fig:reion}.  A
deep {\it JWST} survey with a time investment similar to the HUDF, will reach UV
absolute magnitudes as faint as $-$16 ($-$15.5) at $z =$ 10 ($z =$ 7), two
magnitudes fainter than the current {\it HST} observations (without
the addition of magnification uncertainties).  Not only
will this allow much better constraints on the slope of the faint end
of the luminosity function, but it will also probe whether the single
power law slope is valid down to such faint magnitudes, which is not
the case in some models \citep{jaacks13,oshea15}.

Such a survey will also be capable of detecting galaxies out to $z >$
12, and perhaps $z =$ 15 (depending on their brightness, and on the
filters used).  A uniform sample of galaxies from 7 $< z \lesssim$ 12
will allow a robust investigation into the evolution of the cosmic SFR
density.  This analysis will answer the question of whether there is a
steep decline at $z =$ 8, as not only will it discover larger numbers
of moderately bright galaxies at 8 $< z <$ 10, but it will probe much
further down the luminosity function, allowing us to gain a much
greater confidence on the amount of \emph{total} star formation
occurring, constraining galaxy formation at the earliest epochs.

The infrared sensitivity of {\it JWST} will allow a more
robust investigation into the stellar mass density of the distant
universe.  The high-redshift data shown in Figure~\ref{fig:lf} are all based on
rest-frame UV selected samples, thus the stellar mass measured is only
for galaxies which have been forming stars at some point in the
$\sim$100 Myr prior to observation.  If the duty cycle of star formation is not near unity, it
is possible that there are populations of galaxies between bursts
of star-formation which are not accounted for in the current
observations.  Such galaxies, however, still host lower mass stars,
and thus would be visible in a longer wavelength survey.
A deep survey at observed 5 $\mu$m would measure rest-frame 0.6 $\mu$m light from
$z =$ 7 galaxies, allowing the detection of such massive UV-faint
systems.  The stellar mass function computed from such a survey would
represent a more robust census of the total stellar mass formed at
such redshifts, and will also allow a truly direct investigation into
the duty cycle of star formation at such redshifts.

Finally, spectroscopy with {\it JWST} will clearly be an enormous
discovery space.  {\it JWST} will observe Ly$\alpha$ 
to much fainter line flux levels than is currently possible, both increasing the
number of spectroscopically confirmed distant galaxies, and also
providing tighter constraints on reionization via Ly$\alpha$
emission.  Secondly, observations of rest-optical emission lines such
as [O\,{\sc iii}], which are already implied to be quite bright from
strong {\it Spitzer} colors, will provide an easier route to
spectroscopic confirmation, and also allow probes into the chemical
enrichment and ionization conditions inside these galaxies.

\begin{figure}
\begin{center}
\includegraphics[width=3.5in]{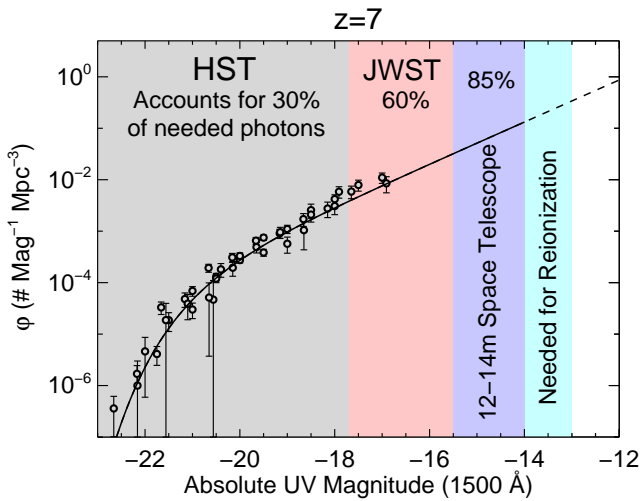}
\caption{The reference $z =$ 7 luminosity function, highlighting the absolute
  magnitudes reached in deep surveys by HST, and future deep surveys
  by {\it JWST} and a 12-14m space telescope.  Assuming that the $z
  =$ 7 luminosity function integrated down to M$=-$13 (with f$_{esc} =$
  13\%) can complete reionization, current {\it HST} observations only
account for 30\% of the needed ionizing photons.  While an ultra-deep survey
with {\it JWST} would only double this, a similar
allocation of resources with a 12-14m class space observatory, such as
HDST or ATLAST, would account for 85\% of the needed photons.}
\vspace{-10mm}
\label{fig:atlast}
\end{center}
\end{figure}

\subsection{Future Space-Based Telescopes}

The next decade will bring the Wide-Field Infrared Survey Telescope
(WFIRST), which will sport a 0.28 deg$^2$ camera housed on a 2.4m space
telescope.  While such a telescope will have a similar flux
sensitivity as {\it HST}, the wide field will allow a number of unique
investigations into the distant universe.  The planned 2227 deg$^2$
High Latitude Survey (HLS) to a near-infrared depth of 26.5 AB mag will allow a truly
unique view into the evolution of the bright end of the galaxy UV
luminosity function at very high redshifts. When paired with optical
data from the Large Synoptic Survey Telescope (LSST), this survey will
enable the selection of galaxies at 4 $< z <$ 10 over this enormous
area, approximately 10$^4$ times the size of the CANDELS
survey.   WFIRST will also have a General Observer component, and as this telescope
is similar in size to {\it HST}, a unique program would be to perform
an extremely deep survey over a single square degree, which would
encompass an area $\sim$720$\times$ that of the HUDF.  While deep {\it
  JWST} surveys will likely provide tight constraints on the faint-end
slopes at 4 $< z <$ 10 by this time, they will still be subject to
significant cosmic variance uncertainties due to the small
field-of-view of the {\it JWST} instruments.  By probing to 29 AB mag
over a very wide field, these uncertainties can be mitigated.  

In Table~\ref{tab:wfirst} I list the expected number counts and source
densities based on the reference UV luminosity functions, for both the
WFIRST HLS and a hypothetical 1 deg$^2$ m$_\mathrm{AB} =$ 29 WFIRST
survey.  At $z =$ 7, over one hundred thousand $m <$ 27 galaxies will be discovered,
$>$1000$\times$ the number of bright $z =$ 7 galaxies currently known.  This will
allow the relatively rough constraints on the bright end of the
high-redshift UV luminosity functions to be dramatically improved,
allowing much more detailed investigations into the issues discussed
in \S 5.1, such as the evolution of the impact of feedback on distant galaxies.
In particular, the cosmic variance which currently plagues the bright
end of the galaxy UV luminosity function
\citep{bouwens15,finkelstein15} will be mitigated.  Combining the
cosmic variance estimator of \citet{newman02}
for dark matter halos with the measured bias at $z =$ 7 from
\citet{baronenugent14}, \citet{finkelstein15} found an uncertainty in
the number of bright $z =$ 7 galaxies to be $\sim$20\%.  Using these
same methods, the WFIRST HLS will have a fractional uncertainty due
to cosmic variance of only 0.2\%.  For faint galaxies,
\citet{finkelstein15} found a cosmic variance uncertainty of
$\sim$40\%, while the analysis here shows that a deep 1 deg$^2$ survey
(discovering galaxies to $m =$ 28.5)
would reduce this uncertainty to only 8\%.  We will get a taste of
this science early in the next decade, when the smaller \emph{Euclid}
launches, which will perform a 40 deg$^2$ deep survey $\sim$1 mag
shallower than the \emph{WFIRST} HLS.

\begin{table*}
\caption{Predicted Source Counts and Densities for WFIRST HLS and Deep GO Surveys}
\begin{center}
\begin{tabular*}{0.8\textwidth}{@{}c\x c\x c\x c\x c\x c@{}}
\hline \hline
Redshift&Limiting&N(HLS)&$n$(HLS)&N(Deep deg$^2$)&$n$(Deep deg$^2$)\\
$ $&Magnitude&$ $&(arcmin$^{-2}$)&$ $&(arcmin$^{-2}$)\\

\hline
4&26.5&2.7 $\times$ 10$^7$&3.3&8.3 $\times$ 10$^4$&23\\
5&26.5&1.4 $\times$ 10$^7$&1.7&5.4 $\times$ 10$^4$&15\\
6&26.5&3.1 $\times$ 10$^6$&0.4&2.1 $\times$ 10$^4$&5.7\\
7&25.7&1.2 $\times$ 10$^5$&1.6 $\times$ 10$^{-2}$&1.2 $\times$ 10$^4$&3.4\\
8&26.0&4.3 $\times$ 10$^4$&5.3 $\times$ 10$^{-3}$&3.6 $\times$10$^3$&1.0\\
9&26.0&1.4 $\times$ 10$^4$&1.8 $\times$ 10$^{-3}$&2.0 $\times$ 10$^3$&0.5\\
10&26.0&4.3 $\times$ 10$^3$&5.4 $\times$ 10$^{-4}$&1.0 $\times$10$^3$&0.3\\
\hline \hline
\end{tabular*}\label{tab:wfirst}
\end{center}
\tabnote{The reference luminosity functions derived in this review
  were used for these predictions.  Columns 3 and 4 refer to the planned
  2227 deg$^2$ WFIRST High Latitude Survey,
while columns 5 and 6 refers to a hypothetical deep 1 deg$^2$ survey.  
To derive these predictions, I assumed a depth of the WFIRST
HLS of 26.5 AB in the $Y$, $J$ and $H$ filters.  Assuming a desired detectable
Lyman break of at least 0.5 mag this implies a limiting magnitude for
discoverable $z =$ 8, 9 and 10 galaxies of m$_\mathrm{AB} =$ 26 for the
HLS.  Galaxies at $z =$ 7 will be limited by the depth
of the LSST $z^{\prime}$-band, which is estimated to be
26.2\footnote[3]{http://www.lsst.org/scientists/scibook}, limiting $z
=$ 7 galaxy studies in the WFIRST HLS to m $<$ 25.7.  The bluer LSST
bands will go much deeper, thus I use a limiting magnitude of 26.5 for
$z =$ 4--6, such that these galaxies are still detected in the WFIRST HLS.}
\end{table*}

The somewhat more distant future (likely in the 2030s) should see the
advent of a 12-14m class space telescope.  Several science case and
design studies are underway for concept such as the High Definition
Space Telescope \citep[HDST;][]{dalcanton15} and the Advanced
Technology Large Aperture Space Telescope
\citep[ATLAST;][]{postman10}.  Such a telescope will likely be
designed for the discovery of habitable worlds around other stars,
but the technical requirements will result in an observatory with
directly applications to the distant universe.  One particular goal of
such an observatory would be to account for all of the ionizing
photons needed for reionization.  As discussed above, the current
observations likely only account for a tiny fraction of the galaxies
which power reionization.  As shown in Figure~\ref{fig:atlast}, even with a steep luminosity function, {\it
  JWST} deep surveys will still only account for 63\% of the necessary
photons.  A 12-14m space telescope, reaching a
limiting UV magnitude of $-$14 at $z =$ 7, will account for 85\% of
the total ionizing photon budget (albeit indirectly, through
observations of non-ionizing UV light).  More crucially, these studies
will allow a direct investigation into the shape of the UV luminosity
function, removing the need for the currently applied extrapolations
to levels well below that which we can observe.

\subsection{The Next Generation of Ground-Based Telescopes}

Much as todays 8-10m class telescopes provide detailed spectroscopic
followup for sources discovered by {\it HST}, the next generation of
25-40m class ground-based telescopes will allow us to study in detail
the new discoveries made by {\it JWST} and future space-based
telescopes.  There are presently three such telescopes in development:
the 25m Giant Magellan Telescope (GMT), the Thirty Meter Telescope
(TMT) and the 39m European Extremely Large Telescope (E-ELT).  These
telescopes will be complementary in many ways.  For example, the E-ELT will
clearly have the largest light-gathering power, while the GMT will
have the largest simultaneous field-of-view.
These facilities will enable a number of new observational analyses for the $z >$ 6
universe; here I will focus on two promising future lines of inquiry
directly relevant to the topics in this review.

First, high resolution spectroscopic followup of very high redshift
gamma ray bursts (GRBs) offers the potential to study both galaxy ISMs and
the state of the IGM at 6 $< z <$ 10.  Among the most intense
explosions in the universe, GRBs are predicted
to occur early in the Universe's history, and have been
observed at $z =$ 8.2 \citep{tanvir09} and likely at $z =$ 9.4
\citep{cucchiara11}, thus GRBs provide extremely bright pencil-beam
flashlights into gas in the distant universe.  By examining metal absorption
line features in the otherwise featureless spectra of GRBs, one can
directly study the abundance of heavy metals in the host galaxy of the
GRB \citep[e.g.,][]{prochaska07}.  Likewise, GRBs offer a chance to
probe the ionization state and temperature of the IGM at $z >$ 7.
While quasars have been utilized for these analyses previously
\citep[e.g.,][]{fan06}, there is only a single quasar currently known
at $z =$ 7, and none at higher redshifts.  Should a space-based gamma
ray observatory capable of discovering such sources be in place by the
next decade, instruments such as GMTNIRS \citep{jaffe14} available at
first light on the GMT, or NIRES on the TMT and SIMPLE \citep{origlia10} on the
E-ELT, will allow high-resolution spectroscopic followup into the
state of the ISMs of galaxies and the IGM at early times.

A second promising future analysis is the study of Ly$\alpha$ emission
at $z >$ 6 to much fainter flux levels than currently possible.
Current studies (primarily with MOSFIRE on Keck) can only reach low
limiting Ly$\alpha$ EWs for the brightest known $z >$ 7 galaxies,
rendering it presently very difficult to place constraints on
Ly$\alpha$ in galaxies fainter than 27 AB mag, which comprise the
majority of the known $z >$ 7 galaxy population.  The 5--15$\times$
gain in light gathering power by these future telescopes will allow
much deeper line flux depths to be probed.  For example, deep
spectroscopy on the GMT is expected to reach 5$\sigma$ limiting line
fluxes of a few $\times$ 10$^{-19}$ erg s$^{-1}$ cm$^{-2}$, with the
TMT and E-ELT reaching even deeper. 

A direct future application of such a facility would be to map
Ly$\alpha$ across both ionized and neutral regions in the 6 $< z <$ 10 universe.
A future multi-object spectrograph on the GMT should have a field-of-view of
$\sim$8$^{\prime}$, comparable to the angular size of ionized bubbles
during reionization, and $>$5$\times$ larger in area than the
equivalent spectrograph on {\it JWST}.  While the imaging coverage
needed to select galaxies over an area covering multiple such bubbles is
likely beyond the realm of {\it HST} or {\it JWST} galaxy surveys, the {\it WFIRST} deep survey source densities in
Table~\ref{tab:wfirst} show that such an instrument can simultaneously
observe hundreds of 6 $< z <$ 10 galaxies down to $m =$ 28.5, allowing the
use of Ly$\alpha$ emission to probe a wide range of IGM states during
the epoch of reionization.  This will be highly complementary with
21cm line surveys by, e.g., the Square Kilometer Array, as the
combination of both surveys will allow a full picture of both the
bright and dark side of reionization.

\subsection{Final Words}
The future is very bright for observational studies of galaxies in the
first billion years after the Big Bang.  As a community, we have made
tremendous progress in not only discovering sources at such redshifts,
but beginning to probe the physical processes dominating their
evolution and gaining a better picture of how they affect the
universe around them during reionization.  These tantalizing glimpses
into the nascent stages of our Universe leave us yearning for answers
to the most pressing questions.  When did the first stars form, and
what were their characteristics?  When did the true first galaxies
form, and were their stellar populations substantially different than
expected?  Is reionization truly over at $z =$ 6, and was the neutral
fraction substantial at $z =$ 7, or is there some other effect
hindering our view of Ly$\alpha$?  Finally, what is the interplay between
gas and stars in such distant galaxies, and does star formation
proceed differently?  Although we may never truly be
able to answer all of these questions, the next generation of
astronomical observatories will allow us to make great progress in
beginning to tease out the truth of the physics of the $z >$ 6
universe.  Combining the pursuit of these studies with the new
questions we will uncover with these future facilities promises to make
the coming decades an extremely exciting
time for observational studies of the early universe.

\begin{acknowledgements}
I am highly grateful to Casey Papovich and Mark Dickinson for their
time to provide feedback on an early manuscript of this
review, and my editor Dawn Erb for useful advice throughout the
process.  I thank Russ Ryan for providing his IDL MCMC code.
I also thank Elizabeth Stanway, Mark Dickinson, Ross McLure and Vithal
Tilvi for providing various data products from their work.  We thank
the referee whose detailed report greatly improved this paper.
\end{acknowledgements}


\end{document}